\documentclass[twocolumn]{aastex62}
\turnoffeditone
\usepackage[english]{babel}
\usepackage{blindtext}
\usepackage{enumitem}
\usepackage{mathtools}
\usepackage[flushleft]{threeparttable}

\received{}
\revised{}
\accepted{}
\submitjournal{}

\shorttitle{Lensing dispersion of gravitational waves}
\shortauthors{Oguri and Takahashi}

\begin{document}

\title{Probing Dark Low-mass Halos and Primordial Black Holes with Frequency-dependent Gravitational Lensing Dispersions of Gravitational Waves}

\email{masamune.oguri@ipmu.jp}

\author[0000-0003-3484-399X]{Masamune Oguri}
\affiliation{Research Center for the Early Universe, University of Tokyo, Tokyo 113-0033, Japan}
\affiliation{Department of Physics, University of Tokyo, Tokyo 113-0033, Japan}
\affiliation{Kavli Institute for the Physics and Mathematics of the Universe (Kavli IPMU, WPI), University of Tokyo, Chiba 277-8582, Japan}

\author[0000-0001-6021-0147]{Ryuichi Takahashi}
\affiliation{Faculty of Science and Technology, Hirosaki University, 3 Bunkyo-cho, Hirosaki, Aomori 036-8588, Japan}

\begin{abstract}
We explore the possibility of using amplitude and phase fluctuations
of gravitational waves due to gravitational lensing as a probe of the
small-scale matter power spectrum. The direct measurement of the
small-scale matter power spectrum is made possible by making
use of the frequency dependence of such gravitational lensing
dispersions originating from the wave optics nature of the
propagation of gravitational waves. We first study the small-scale
behavior of the matter power spectrum in detail taking the so-called
halo model approach including effects of baryons and subhalos. We find
that the matter power spectrum at the wavenumber 
$k\sim 10^6h{\rm Mpc}^{-1}$ is mainly determined by the abundance of
dark low-mass halos with mass 
$1h^{-1}M_\odot \la M \la 10^4h^{-1}M_\odot$ and is relatively
insensitive to baryonic effects. The matter power spectrum at this
wavenumber is probed by gravitational lensing dispersions of
gravitational waves at frequencies of $f\sim 0.1-1$~Hz with predicted
signals of $\mathcal{O}(10^{-3})$. We also find that primordial black
holes (PBHs) with $M_{\rm PBH}\ga 0.1~M_\odot$ can significantly
enhance the matter power spectrum at $k \ga 10^5h{\rm Mpc}^{-1}$ due
to both the enhanced halo formation and the shot noise from PBHs.
We find that gravitational lensing dispersions at $f\sim 10-100$~Hz 
are particularly sensitive to PBHs and can be enhanced by more than an
order of magnitude depending on the mass and abundance of PBHs.
\end{abstract}

\section{Introduction}
\label{sec:intro}

The nature of dark matter remains one of the central problems in cosmology.
While the cold dark matter (CDM) model
\citep{peebles82,blumenthal84,davis85} is successful in explaining a
variety of cosmological observations including the cosmic microwave
background \citep{planck16} and the large-scale structure of the
Universe \citep{boss17}, there are several competing candidates of
cold dark matter, including weakly interacting massive particles,
ultra-light dark matter, and primordial black holes 
\citep[see e.g.,][]{uscv17}.  A number of experiments are ongoing and
planned to detect dark matter and to discriminate these different
candidates. 

Cosmological and astrophysical observations provide an important means
of studying the property of dark matter particles, because different
dark matter candidates can predict quite different small-scale
distributions of dark matter. In observations, there have been debates
about the validity of the simplest collisionless CDM model at the
dwarf galaxy scale. For example, core-like radial density profiles of
many dark-matter dominated dwarf galaxies and the small number of
satellite dwarf galaxies in the Milky Way and the Local Group 
\citep[see][for a review]{bullock17} may hint the particle nature of
dark matter, although it has been argued that the modification of dark
matter distributions by complex baryon physics such as star formation
and supernova feedback may well explain the observed properties of
dwarf galaxies even in the context of the simplest collisionless CDM
model. 

One way to settle the debate is to study dark low-mass halos.
Since the galaxy formation theory predicts that halos with masses 
$\la 10^7M_\odot$ contain very little or no star, properties of such
dark low-mass halos are barely affected by the complex baryon physics,
which makes them an ideal site for testing various dark matter
candidates. However, detecting such dark low-mass halos is quite
challenging. Several ideas include flux ratio anomalies in
gravitationally lensed quasars \citep[e.g.,][]{mao98,inoue15,gilman20},
perturbations in galaxy-galaxy strong lensing
\citep[e.g.,][]{inoue03,koopmans05,vegetti12,ritondale19},
perturbations in stellar streams in the Milky Way
\citep[e.g.,][]{ibata02,bovy17,banik19}, pulsar timing array 
\citep[e.g.,][]{kashiyama18,dror19},
astrometric weak gravitational lensing 
\citep[e.g.,][]{vantilburg18,mondino20}, caustic crossings in
massive clusters \citep[e.g.,][]{kelly18,dai18a,dai20}, and
diffraction effects in gravitational lensing \citep[e.g.,][]{dai18b}.
Some of these techniques already place interesting constraints on the
abundance of low-mass halos down to $\sim 10^8M_\odot$ that is roughly
consistent with the standard CDM prediction. Thus pushing such
constraints to even lower halo masses is anticipated. 
 
One of the candidates of CDM includes Primordial Black Holes (PBHs)
that formed in the early Universe 
\cite[see e.g.,][for reviews]{sasaki18,carr20}. The PBH dark matter
scenario has attracted a lot of attention given the discovery of
gravitational waves from a binary black hole merger \citep{ligo16}.
The abundance of PBHs around the mass scale of such binary black hole
mergers has been constrained by e.g., quasar microlensing
\citep{mediavilla17}, caustic crossings \citep{oguri18}, and
gravitational lensing of Type Ia supernovae \citep{zumalacarregui18}.
Tighter constraints on their abundance are needed to check whether PBHs
can account for observed gravitational wave events.

In this paper, we explore the possibility of using gravitational
lensing dispersions of gravitational waves as a new probe of dark
low-mass halos and PBHs. The gravitational lensing dispersion refers
to the dispersion of the brightness of a distant source due to
gravitational lensing (de)magnification caused by intervening matter
along the line-of-sight. Such gravitational lensing dispersions have
been detected from samples of Type Ia supernovae 
\citep{jonsson07,jonsson10,kronborg10,karpenka13,smith14}, and
contain information on the matter power spectrum at small scales 
\citep{bernardeau97,metcalf99,hamana00,dodelson06,quartin14,fedeli14,bendayan14,bendayan16,hada16,hada19,agrawal19}. 

This approach can easily be extended to gravitational waves from
compact binary mergers given their standard siren nature
\citep{schutz86,holz05}. A notable difference of gravitational
lensing of gravitational waves from that of supernovae is that 
wave optics effects can play an important role in some situations 
\citep[see e.g.,][for reviews]{nakamura99,oguri19}. 
For instance, gravitational lensing magnifications are significantly
suppressed due to diffraction when the wavelength of gravitational
waves is larger than the Schwarzschild radius of the lens. In the case
of gravitational lensing dispersions, density fluctuations below the
so-called Fresnel scale, which depends on the frequency of
gravitational waves, do not contribute to the dispersion due to
diffraction \citep{macquart04,takahashi05,takahashi06}. Taking
advantage of this effect, \citet{takahashi06} proposed to use
amplitude and phase changes as a function of the frequency of
gravitational waves to probe the matter power spectrum at the Fresnel
scales. In this paper we extend this idea and explore the
detectability of dark low-mass halos and primordial black holes with
frequency dependent gravitational lensing dispersions of gravitational
waves. For this purpose, we study the behavior of the matter power
spectrum at very small scales ($k\gg 1$~Mpc$^{-1}$) in detail,
including the modification of the power spectrum due to baryon
physics, taking the halo model approach. 

This paper is organized as follows. 
We present our halo model including effects of baryons and subhalos in
Section~\ref{sec:halo_model}. We then present results of gravitational
lensing dispersions of gravitational waves in Section~\ref{sec:dispersions}.
Some discussions are given in Section~\ref{sec:discussion}.
Finally we conclude in Section~\ref{sec:conclusion}.
Throughout the paper we adopt the $\Lambda$-dominated CDM model with
the matter density $\Omega_{\rm m}=0.3089$, the baryon density 
$\Omega_{\rm b}=0.0486$, the cosmological constant $\Omega_\Lambda=0.6911$,
the dimensionless Hubble constant $h=0.6774$, the spectral index
$n_{\rm s}=0.9667$, and the normalization of the density fluctuation
$\sigma_8=0.8159$, which are cosmological parameters adopted in the
IllustrisTNG cosmological hydrodynamical simulations 
\citep{nelson18,springel18,pillepich18,marinacci18,naiman18}. 
Throughout the paper we always assume a flat Universe for calculating distances.
   
\section{Halo Model}
\label{sec:halo_model}

\subsection{Standard Calculation}
\label{sec:halo_standard}

The halo model \citep[see e.g.,][ for a review]{cooray02} provides a
powerful means of studying nonlinear gravitational clustering. It
assumes that all the matter is confined in dark matter halos. With
this assumption, the matter density field $\rho(\boldsymbol{x})$
is written as
\begin{equation}
\rho(\boldsymbol{x})=\sum_i M_i u\left(\boldsymbol{x}-\boldsymbol{x}_i|M_i\right),
\label{eq:def_rhom_1}
\end{equation}
where $i$ labels dark matter halos, $M_i$ and $\boldsymbol{x}_i$
are the mass and the spatial position of $i$-th halo, and
$u(\boldsymbol{x}|m)$ denotes the normalized density profile of a halo
with mass $M$ that satisfies $\int
d\boldsymbol{x}\,u(\boldsymbol{x}|M)=1$. 
Usually the \citet[][hereafter NFW]{navarro97} density profile is
adopted as the density profile of each halo. 
From Equation~(\ref{eq:def_rhom_1}), it is found that the matter power
spectrum is described by the sum of the so-called 1-halo and 2-halo
terms (see Appendix~\ref{app:halo_model}  for the derivation) 
\begin{equation}
P(k)=P^{\rm 1h}(k)+P^{\rm 2h}(k),
\end{equation}
\begin{equation}
P^{\rm 1h}(k)=\int dM\frac{dn}{dM}\left(\frac{M}{\bar{\rho}}\right)^2u^2(k|M),
\label{eq:pk_1h}
\end{equation}
\begin{equation}
P^{\rm 2h}(k)=\left[\int dM \frac{dn}{dM}\left(\frac{M}{\bar{\rho}}\right)b(M)\,u(k|M)\right]^2P_{\rm lin}(k),
\label{eq:pk_2h}
\end{equation}
where $dn/dM$ denotes the halo mass function, $\bar{\rho}$ is the
mean comoving matter density, $u(k|M)$ is the Fourier transform of the
normalized density profile, $b(M)$ is a linear halo bias, and 
$P_{\rm lin}(k)$ is the linear matter power spectrum.

\subsection{Modifications of 1-halo Term}
\label{sec:halo_modified}

The baryon cooling and star formation modify the matter distribution
in each halo, and thereby affect the matter power spectrum
\citep[see e.g.,][for a review]{chisari19}. Such baryonic effects have
been studied using the halo model, mostly focusing on their impact on
cosmic shear cosmology
\citep[e.g.,][]{white04,zhan04,semboloni11,fedeli14a,fedeli14b,debackere20}.
In addition, substructures or subhalos in dark matter halos may
affect the matter power spectrum at very small scales, as studied in
\citet{giocoli10}. 

Following the literature, we consider effects of the stellar
component and subhalos, both of which can be important at very small
scales, and ignore the effect of the gas component. In presence of
these components, Equation~(\ref{eq:def_rhom_1}) is rewritten as
\begin{equation}
\rho(\boldsymbol{x})=\sum_i M_i 
\left[(1-f_{\rm s})u_{\rm h}\left(\boldsymbol{x}-\boldsymbol{x}_i|M_i\right)
+f_{\rm s}u_{\rm s}\left(\boldsymbol{x}-\boldsymbol{x}_i|M_i\right)\right],
\label{eq:def_rhom_mod}
\end{equation}
where
\begin{eqnarray}
u_{\rm h}\left(\boldsymbol{x}-\boldsymbol{x}_i|M_i\right)
&=&\frac{1-f_*-f_{\rm s}}{1-f_{\rm s}}u\left(\boldsymbol{x}-\boldsymbol{x}_i|M_i\right)\nonumber\\
&&+\frac{f_*}{1-f_{\rm s}}u_*\left(\boldsymbol{x}-\boldsymbol{x}_i|M_i\right),
\end{eqnarray}
\begin{equation}
u_{\rm s}\left(\boldsymbol{x}-\boldsymbol{x}_i|M_i\right)
=\frac{1}{f_{\rm s}M_i}\sum_j m_ju_{\rm sub}
(\boldsymbol{x}-\boldsymbol{x}_j|M_i,m_j,\boldsymbol{x}_j-\boldsymbol{x}_i),
\end{equation}
$f_*$ and $f_{\rm s}$ are mass fractions of stellar and subhalo
components, respectively, $m_j$ is the mass of the $j$-th subhalo, and
$u_*$ and $u_{\rm sub}$ denote normalized density profiles of stellar
and subhalo components, respectively. For simplicity, throughout the
paper we ignore the dependence of $u_{\rm sub}$ on the position within
a halo by setting $u_{\rm sub} (\boldsymbol{x}-\boldsymbol{x}_j|
M_i,m_j,\boldsymbol{x}_j-\boldsymbol{x}_i)=u_{\rm sub}
(\boldsymbol{x}-\boldsymbol{x}_j|M_i,m_j)$. Repeating the similar
calculation as done for the standard halo model case 
(see Appendix~\ref{app:halo_model}), we obtain
\begin{equation}
P^{\rm 1h}(k)=P^{\rm 1h,hh}(k)+P^{\rm 1h,hs}(k)+P^{\rm 1h,ss}(k),
\label{eq:pk_1h_mod}
\end{equation}
\begin{equation}
P^{\rm 1h,hh}(k)=\int
dM\frac{dn}{dM}\left(\frac{M}{\bar{\rho}}\right)^2(1-f_{\rm
  s})^2u_{\rm h}^2(k|M),
\label{eq:pk_1h_hh}
\end{equation}
\begin{equation}
P^{\rm 1h,hs}(k)=\int
dM\frac{dn}{dM}\left(\frac{M}{\bar{\rho}}\right)^22(1-f_{\rm s})u_{\rm
  h}(k|M)I(k|M),
\label{eq:pk_1h_hs}
\end{equation}
\begin{equation}
P^{\rm 1h,ss}(k)=\int
dM\frac{dn}{dM}\left(\frac{M}{\bar{\rho}}\right)^2\left[I^2(k|M)+J(k|M)\right],
\label{eq:pk_1h_ss}
\end{equation}
where
\begin{equation}
I(k|M)=\int dm\frac{dN_M}{dm}\left(\frac{m}{M}\right)u_{\rm sub}(k|M,m) U(k|M,m),
\end{equation}
\begin{equation}
J(k|M)=\int dm\frac{dN_M}{dm}\left(\frac{m}{M}\right)^2u^2_{\rm sub}(k|M,m),\nonumber\\
\end{equation}
$dN_M/dm$ is the subhalo mass function within a halo with mass $M$ and
$U(k|M)$ is the Fourier transform of the spatial distribution of
subhalos $U(\boldsymbol{x}|M,m)$. 

The stellar component $u_*$ actually consists of stars, which
indicates that the shot noise due to the discrete nature of the stellar
component may be important at very small scales. Following the
calculation in Appendix~\ref{app:halo_model}, we include the shot
noise from stars by replacing $u_*^2(k|M)$ to
\begin{equation}
u_*^2(k|M) \rightarrow u_*^2(k|M)+\frac{1}{N_*},
\label{eq:shot_star}
\end{equation}
where $N_*=f_*M/m_{\rm star}$ denotes the total number of 
stars in each halo with mass $M$ and $m_{\rm star}$ is the mass of
each star. Here we assume that all stars share the same mass
for simplicity.

It is instructive to approximate the expressions above further to
understand their behavior. Simulations suggest that the spatial
distribution of subhalos approximately follows that of the smooth
matter component. If we simply assume $U(k|M,m)\approx u_{\rm h}(k|M)$,
and use the fact that $u_{\rm sub}(k|M,m)\sim 1$ when $U(k|M,m)$ takes
large values, we obtain $I(k|M)\approx f_{\rm s} u_{\rm h}(k|M)$. 
Under this approximation the 1-halo power spectrum is further
simplified as
\begin{equation}
P^{\rm 1h}(k)\approx\int dM\frac{dn}{dM}\left(\frac{M}{\bar{\rho}}\right)^2
\left[u_{\rm h}^2(k|M)+J(k|M)\right],
\label{eq:p1h_approx}
\end{equation}
where the first term of the right hand side of Equation~(\ref{eq:p1h_approx}) 
corresponds to the 1-halo power spectrum without any subhalo, and the
dominant effect of the subhalo is given by the second term of the right
hand side of Equation~(\ref{eq:p1h_approx}), which represents the
auto-correlation of the matter distribution within each subhalo.

\subsection{Model Ingredients}

We adopt a smoothly truncated NFW profile studied by
\citet[][hereafter BMO]{baltz09} for the density profile of main
halos. Specifically we adopt the following density profile 
\begin{equation}
\rho_{\rm BMO}(r)=\frac{\rho_{\rm s}}{(r/r_{\rm s})(1+r/r_{\rm
    s})^2}\left(\frac{r_{\rm t}^2}{r^2+r_{\rm t}^2}\right)^2,
\label{eq:bmo}
\end{equation}
which we parametrize by the virial mass $M=M_{\rm vir}$, the
concentration parameter $c=c_{\rm vir}=r_{\rm vir}/r_{\rm s}$, and the
truncation radius $\tau=r_{\rm t}/r_{\rm s}$. We adopt a fitting form
of the mass-concentration relation presented by \citet{diemer15} with
updates of fitting parameters by \citet{diemer19} and the conversion
from $c_{\rm 200c}$ to $c_{\rm vir}$ assuming an NFW profile. 
We then compute $\rho_{\rm s}$ and $r_{\rm s}$ for a given mass and
redshift from the mass-concentration relation and in a standard manner
assuming the NFW profile (i.e., ignoring the effect of the
truncation). We determine $\tau$ such that the total mass of the BMO
profile matches $M$ i.e.,   
\begin{equation}
f_{\rm BMO}(\tau)=f_{\rm NFW}(c),
\end{equation}
\begin{equation}
f_{\rm BMO}(\tau)=\frac{\tau^2\left[(3\tau^2-1)(\pi\tau-\tau^2-1)+2\tau^2(\tau^2-3)\ln\tau\right]}{2(\tau^2+1)^3},
\end{equation}
\begin{equation}
f_{\rm NFW}(c)=\ln(1+c)-\frac{c}{1+c}.
\end{equation}
For typical values of $c$, we obtain $\tau\sim (1.4-1.6)c$
from this condition. The Fourier transform  $u(k|M)$ of the
normalized BMO profile is given in Appendix~B of \citet{oguri11}. The
procedure above ensures $u(k|M)\rightarrow 1$ at
$k\rightarrow 0$. Since the BMO profile is smoothly truncated, an
oscillating feature in $u(k|M)$, which is seen in the Fourier
transform of the NFW profile truncated at $r=r_{\rm vir}$
\citep[e.g.,][]{cooray02}, is suppressed.  

For the mass function $dn/dM$ and halo bias $b(M)$, we adopt a model
of \citet{sheth99} that is reasonably accurate for wide mass and
redshift ranges.

We need to specify the stellar mass fraction $f_*$ and the density
profile $\rho_*(r)=f_*Mu_*(r)$ as a function of the halo mass $M$ in
order to address baryonic effects. We adopt the stellar mass--halo
mass relation for all central galaxies presented by \citet{behroozi19}
as $f_*$. Note that we adopt the mean stellar mass--halo mass
relation, which is computed from the median relation in
\citet{behroozi19} and assuming the log-normal distribution with the
scatter of $0.3$~dex. We adopt the \citet{hernquist90} profile as the
density profile of the stellar component 
\begin{equation}
\rho_*(r)=\frac{f_*M}{2\pi(r/r_{\rm b})(r+r_{\rm b})^3},
\end{equation}
where $r_{\rm b}$ is related with the effective radius as $r_{\rm
  b}=0.551r_{\rm e}$. Since it has been shown that galaxy sizes
are proportional to virial radii of their host halos for a wide halo
mass range \citep[e.g.,][]{kravtsov13,kawamata15,huang17,kawamata18,kravtsov18,zanisi20},
in this paper we simply assume 
\begin{equation}
r_{\rm e}=0.006r_{\rm vir}, 
\end{equation}
at $z=0$, and it evolves with redshift with $\propto (1+z)^{-1}$ that
roughly matches the observed redshift evolution of galaxy sizes.
The Fourier transform of the normalized density profile is given by
\begin{equation}
u_*(k|M)=1-x{\rm Ci}(x)\sin x-\frac{1}{2}x\left[\pi-2{\rm Si}(x)\right]\cos x,
\label{eq:u_hern}
\end{equation}
where $x=kr_{\rm b}(1+z)$ and ${\rm Si}(x)$ and ${\rm Ci}(x)$ are sine and
cosine integrals, respectively. 
The shot noise from stars (equation~\ref{eq:shot_star}) is computed
assuming the star mass of $m_{\rm star}=0.5~M_\odot$.

We also need a model of subhalos. We adopt a simple analytic model
presented in Appendix~\ref{app:subhalo_model} to compute the mass
function and the density profile of subhalos. We assume that their
radial distribution within each halo follows that of the smooth dark
matter distribution i.e., the BMO profile $U(k|M,m)=u(k|M)$. We adopt
the BMO profile also for the mass distribution of each subhalo but
with different model parameters from those of main halos, as detailed
in Appendix~\ref{app:subhalo_model}. Similarly to main
halos, we consider baryonic effects for subhalos as well using 
the mean stellar mass--halo mass relation for all satellite galaxies
presented by \citet{behroozi19}. We estimate subhalo masses before
tidal stripping $m_{\rm f}$ (see Appendix~\ref{app:subhalo_model} 
for more details) as a proxy of the peak mass in \citet{behroozi19} to
derive the stellar mass fraction of subhalos, $f_*^{\rm s}$. The
Fourier transform of the normalized subhalo density profile is given by
\begin{equation}
u_{\rm sub}(k|M,m)= (1-f_*^{\rm s})u(k|m,M)+f_*^{\rm s}u_*(k|m_{\rm ext}),
\end{equation}
where $u(k|m,M)$ is the Fourier transform of the normalized BMO
profile with total mass $m$, the concentration parameter 
$c_{\rm sub}$, and truncated at $r_{\rm t}^{\rm ave}$, and
$u_*(k|m_{\rm ext})$ is given by Equation~(\ref{eq:u_hern}) with
$r_{\rm b}$ computed from $m_{\rm f}$. 

\begin{figure}
\begin{center}
 \includegraphics[width=1.0\hsize]{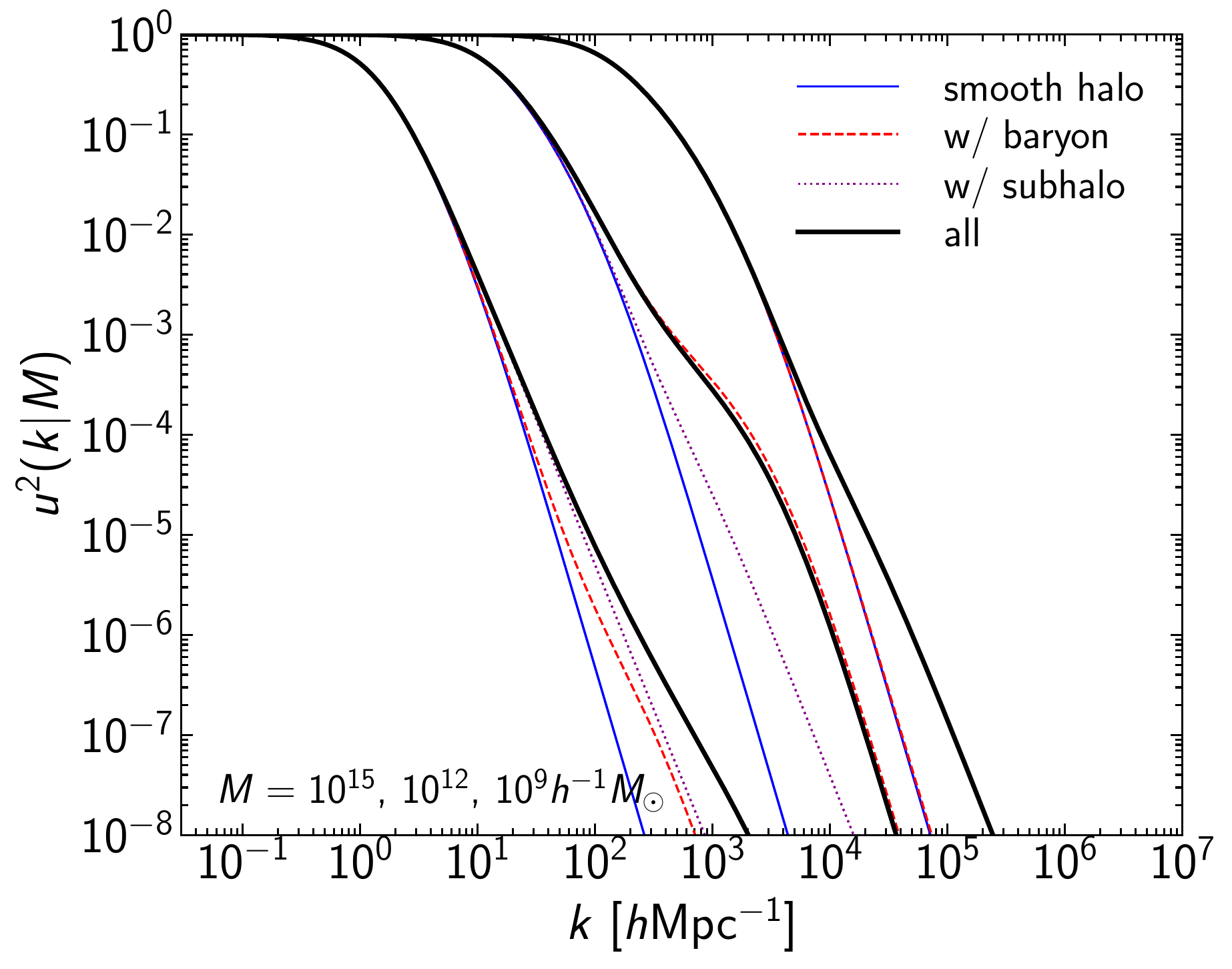}
\end{center}
\caption{Comparison of the Fourier transform of the halo density
  profile $u^2(k|M)$ at $z=0$ with and without effects of baryon and
  subhalos. From left to right, we show results for halos with mass
  $M=10^{15}h^{-1}M_\odot$, $10^{12}h^{-1}M_\odot$, and 
  $10^{9}h^{-1}M_\odot$, respectively.
  Thin solid lines corresponds to the case with only the smooth main
  halo, dashed and dotted lines are after adding stellar components
  and subhalos, respectively, and thick solid lines show the case with
  both stellar components and subhalos are included.
  Here the shot noise from stars is not included.
\label{fig:comp_u2}}
\end{figure}

\begin{figure}
\begin{center}
 \includegraphics[width=1.0\hsize]{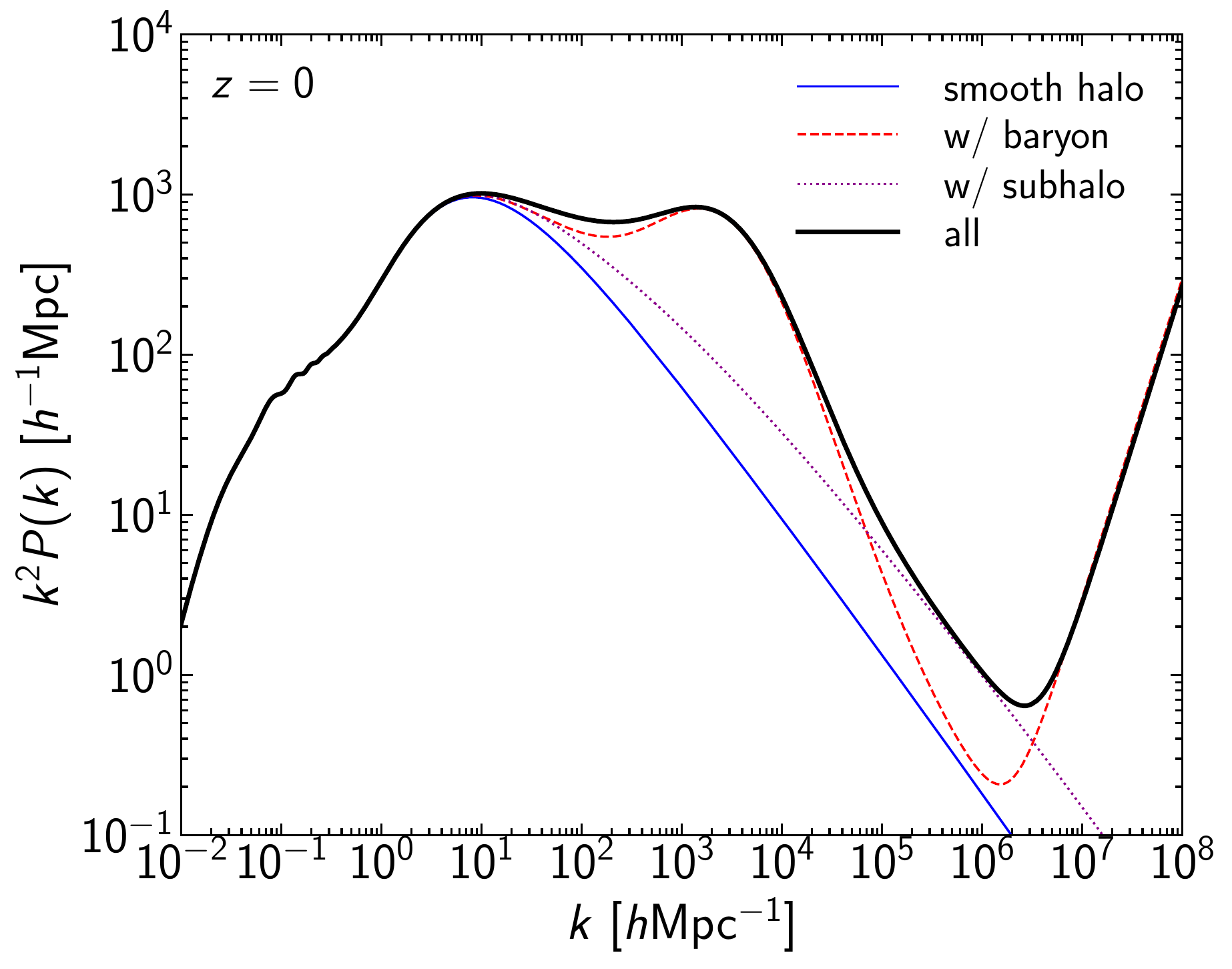}
\end{center}
\caption{Comparison of matter power spectra $P(k)$ at $z=0$
  with and without effects of baryon and subhalos. Lines are same as
  in Figure~\ref{fig:comp_u2}.
\label{fig:pk_barsub}}
\end{figure}

\begin{figure}
\begin{center}
 \includegraphics[width=1.0\hsize]{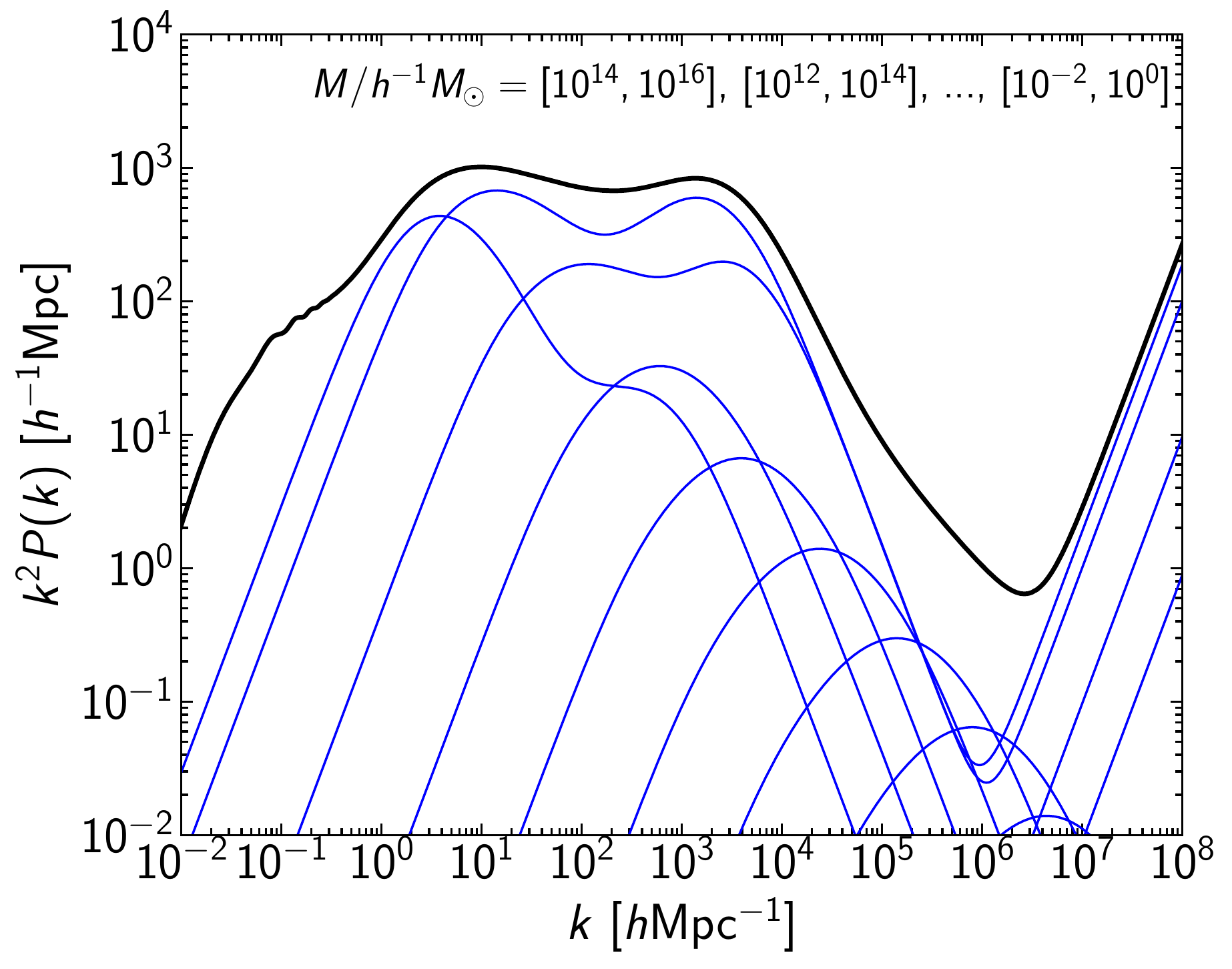}
\end{center}
\caption{Contributions from halos with different masses to the matter
  power spectrum $P(k)$ at $z=0$. From left to right thin lines, we show
  contributions in the 2 dex mass range from higher to lower masses of
  halos. Here we show contributions from main halos only i.e., without
  subhalos but including baryonic effects. 
\label{fig:break_pk_main}}
\end{figure}

\begin{figure}
\begin{center}
 \includegraphics[width=1.0\hsize]{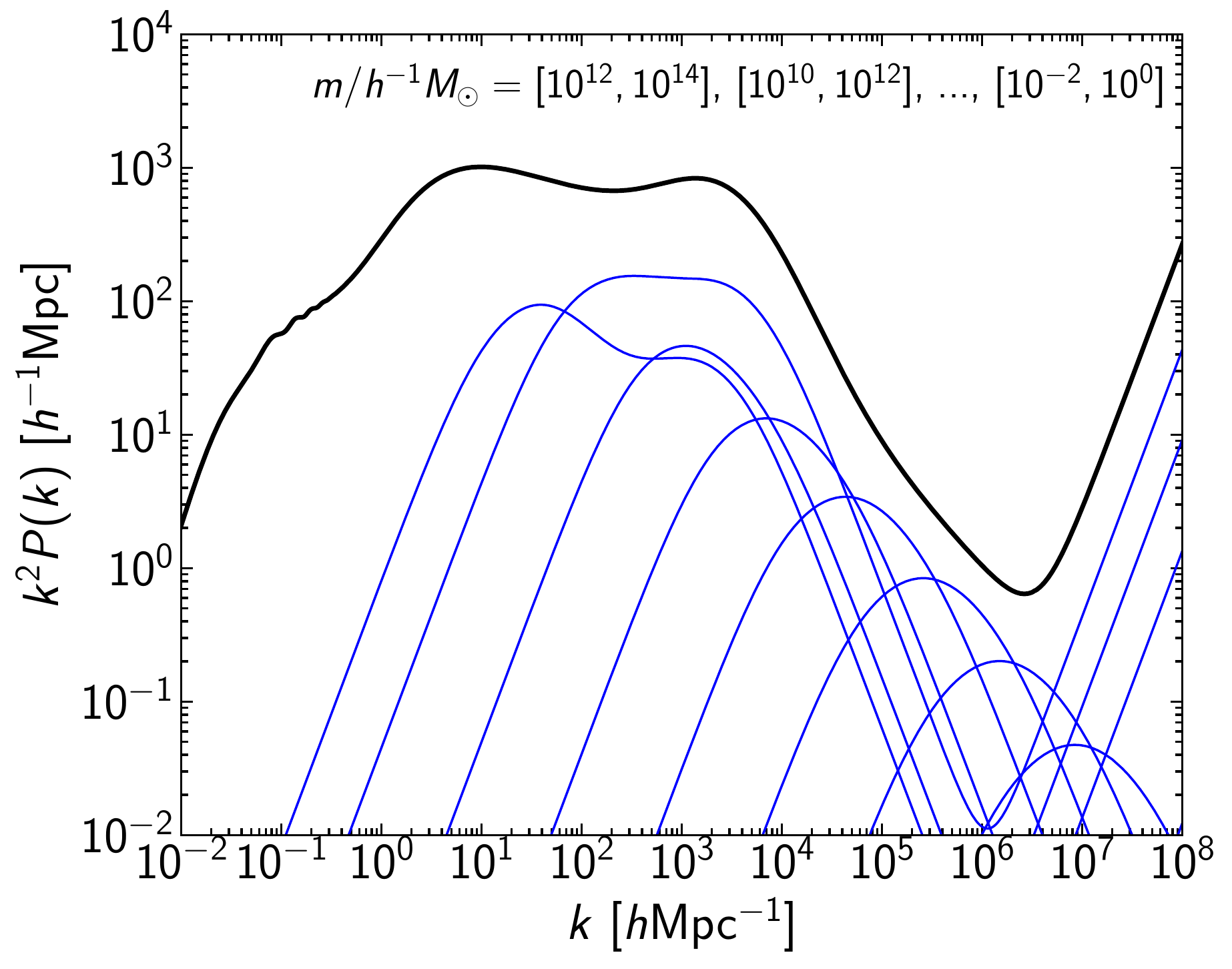}
\end{center}
\caption{Similar to Figure~\ref{fig:break_pk_main}, but contributions
  from subhalos with different masses are shown. To isolate effects
  of subhalos, here we include only the $J(k|M)$ term in
  Equation~(\ref{eq:pk_1h_ss}) to compute these contributions.
\label{fig:break_pk_sub}}
\end{figure}

\subsection{Some Examples}

Before presenting examples of calculations of matter power spectra, in
Figure~\ref{fig:comp_u2} we show the Fourier transform of the halo
density profile with and without effects of baryon and subhalos. The
Figure indicates that both subhalos and baryonic stellar components
significantly enhance the small scale power of individual halo density
profiles. Baryonic effects depend sensitively on the halo mass,
reflecting the halo mass dependence of the stellar mass--halo mass
relation. While the effects of baryon are more pronounced at 
$M\sim 10^{12}h^{-1}M_\odot$, the effects of subhalos are more
significant for very high and low-mass halos.

Figure~\ref{fig:pk_barsub} shows matter power spectra at $z=0$
computed from the halo model presented above. We show $k^2P(k)$
because it represents contributions to gravitational lensing
dispersions per $\ln k$. We find that effects of baryon and subhalos
are significant at $k\ga 10 h{\rm Mpc}^{-1}$. In our model, the effects
of baryon (stellar components) are dominated at $10 h{\rm Mpc}^{-1} \la
k \la 10^5 h{\rm Mpc}^{-1}$ and  $k\ga 10^7h{\rm Mpc}^{-1}$, and
interestingly the effects of subhalos dominates at 
$k \sim 10^6 h{\rm Mpc}^{-1}$. This indicates that observations of the
matter power spectrum at $k \sim 10^6 h{\rm Mpc}^{-1}$ would probe
dark low-mass halos. We note that the increase of 
$k^2P(k)$ at $k\ga 10^7h{\rm Mpc}^{-1}$ is due to the shot noise from
stars as described in Equation~(\ref{eq:shot_star}).

To check the possibility of studying dark low-mass halos 
more explicitly, we study contributions of the matter power spectrum
from different halo and subhalo masses. 
The results shown in Figures~\ref{fig:break_pk_main} and 
\ref{fig:break_pk_sub} indicate that halos and subhalos with masses
$1h^{-1}M_\odot \la M \la 10^4h^{-1}M_\odot $ most contribute to
the matter power spectrum at $k \sim 10^6 h{\rm Mpc}^{-1}$. For such
low-mass halos and subhalos there is virtually no star given the
current knowledge of the stellar mass--halo mass relation.

\begin{figure}
\begin{center}
 \includegraphics[width=1.0\hsize]{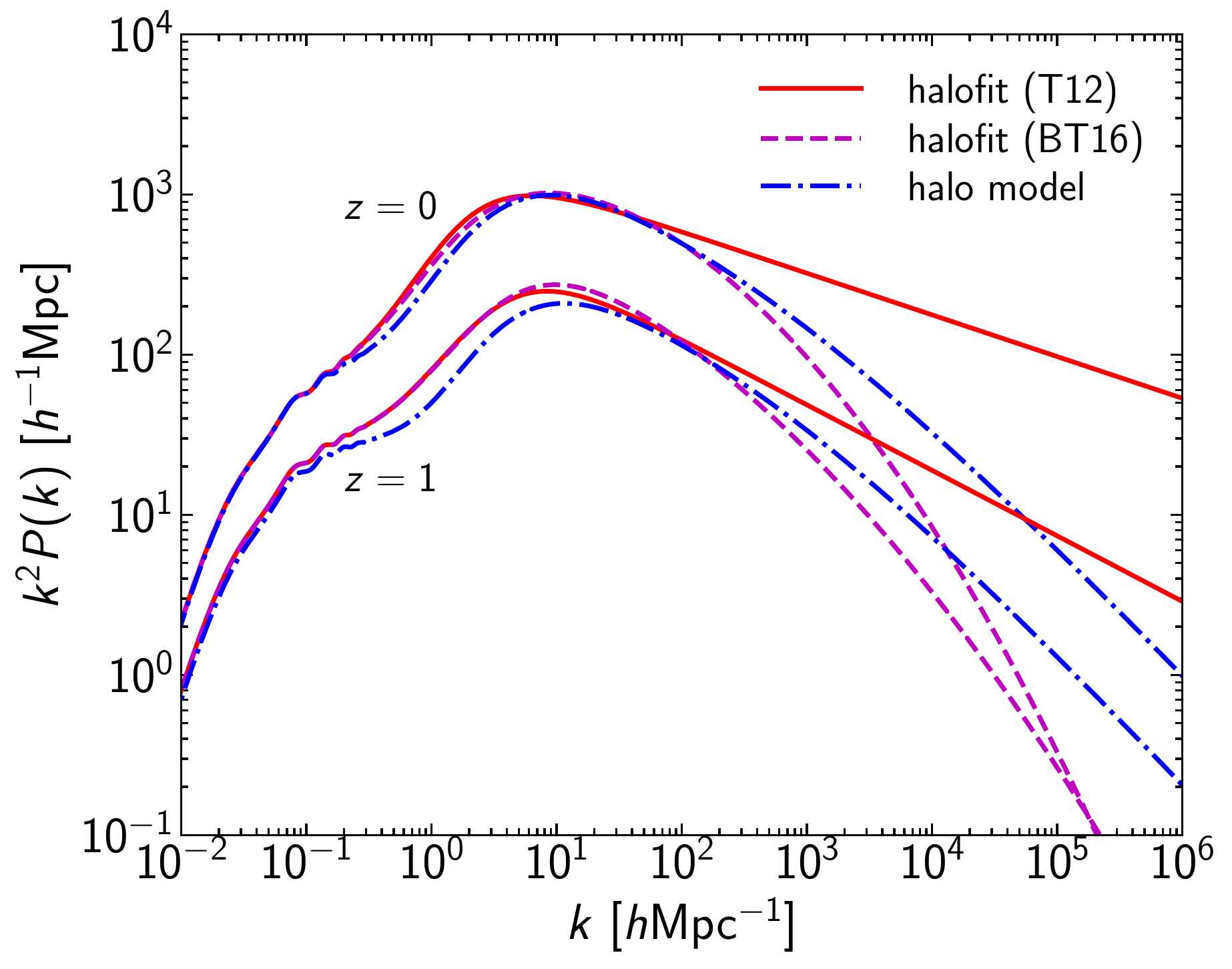}
\end{center}
\caption{Comparison of matter power spectra $P(k)$ without the baryonic
  effects computed from the halo model presented in this paper ({\it
    dash-dotted}) with the halofit models of
  \citet[][T12]{takahashi12} ({\it solid}) and
  \citet[][BT16]{bendayan16} ({\it dashed}). 
  The comparisons are made at redshift $z=0$ ({\it upper}) 
  and $z=1$ ({\it lower}). 
\label{fig:comp_pk}} 
\end{figure}

\begin{figure}
\begin{center}
 \includegraphics[width=1.0\hsize]{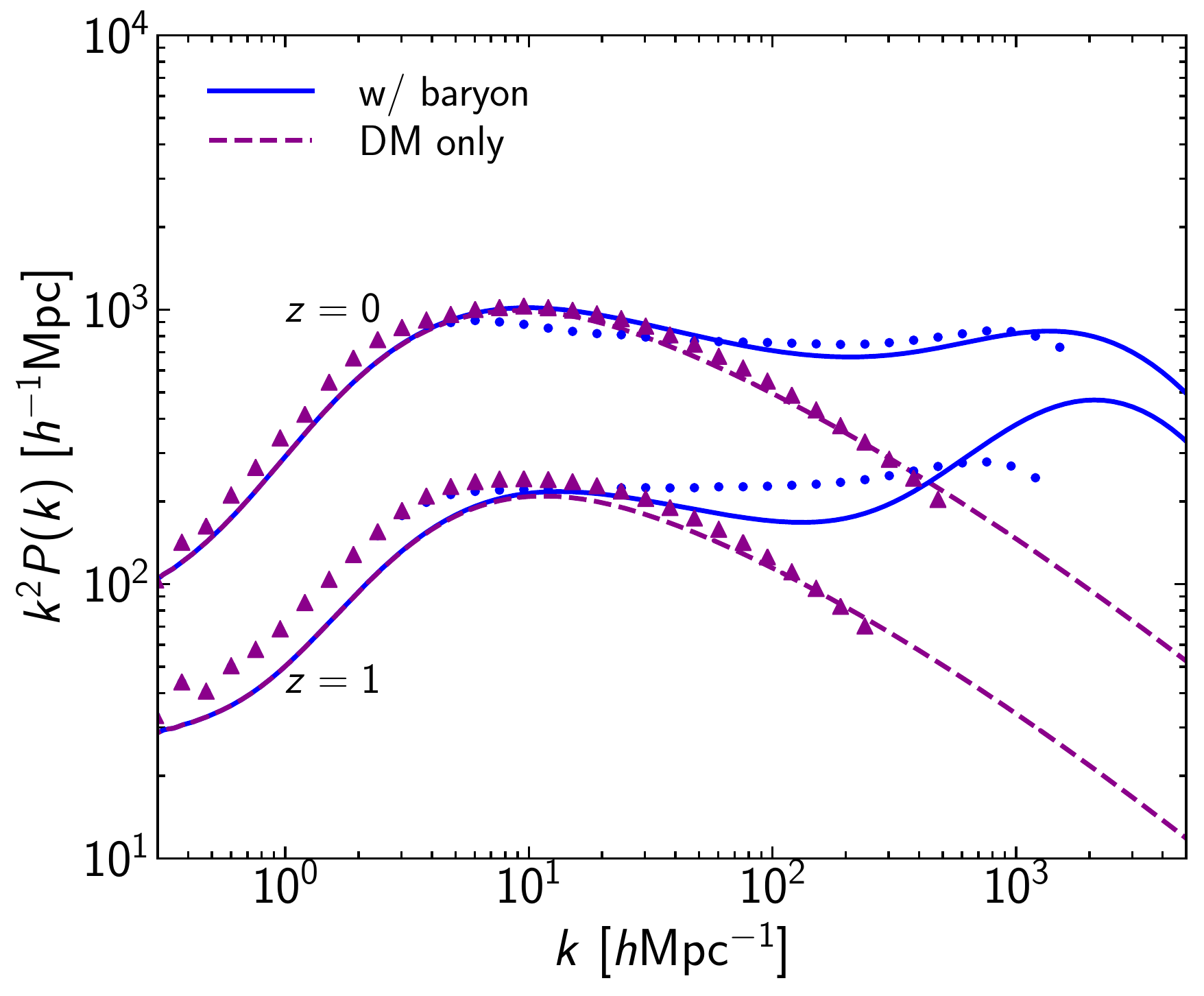}
\end{center}
\caption{Comparison of matter power spectra $P(k)$ computed from the
  halo model presented in this paper with IllustrisTNG cosmological
  hydrodynamical simulations 
 \citep{nelson18,springel18,pillepich18,marinacci18,naiman18}. 
 Filled circles and filled triangles show matter power spectra
 measured in TNG100-1 and TNG100-1-Dark, respectively, whereas 
 solid and dashed lines show halo model predictions with and without
 baryonic effects, respectively. The comparisons are made at redshift
 $z=0$ ({\it upper}) and $z=1$ ({\it lower}).
\label{fig:comp_pk_tng}}
\end{figure}

\subsection{Comparisons with Other Results}

We compare our halo model calculations with other results of the matter
power spectrum to check their validity. One of the most popular models
of the matter power spectrum without baryonic effects is the
so-called the halofit model, which is originally proposed by
\citet{smith03} and later improved by \citet{takahashi12}. The halofit
model is essentially a fitting formula whose functional form is
motivated by the halo model. In \citet{takahashi12}, model parameters
are calibrated by $N$-body simulation results at 
$k<30h{\rm Mpc}^{-1}$. \citet{bendayan16} updated the halofit model at
high wavenumber further by recalibrating model parameters with
$N$-body simulation results at $k<300h{\rm Mpc}^{-1}$. In
Figure~\ref{fig:comp_pk}, we compare our halo model calculations
without baryonic effects with the halofit models of both
\citet{takahashi12} and \citet{bendayan16}. We find that at  
$k<30h{\rm Mpc}^{-1}$ and $k<300h{\rm Mpc}^{-1}$ our halo model
results agree well with the halofit models of \citet{takahashi12} and
\citet{bendayan16}, respectively. At higher wavenumber $k$, however,
disagreements between different models get quite large. Since our halo
model is built on well-known properties of halos, we believe our halo
model predicts the matter power spectrum at high $k$ much more
accurately than the halofit models for which calculations of matter
power spectra at high $k$ have to rely on extrapolations of the
fitting formulae. 

To check the validity of our halo model including the baryonic stellar
components, we need to compare our results with matter power spectra
measured in cosmological hydrodynamical simulations. For this purpose,
we measure matter power spectra from IllustrisTNG cosmological
hydrodynamical simulations 
\citep{nelson18,springel18,pillepich18,marinacci18,naiman18}. 
Specifically, we measure matter power spectra for both TNG100-1 (with
baryonic effects) and TNG100-1-Dark (without baryonic effects)
that are publicly available and have the box size of 
$(110.7~{\rm Mpc})^3$. Figure~\ref{fig:comp_pk_tng} compares $P(k)$
from the halo model and IllustrisTNG cosmological hydrodynamical
simulations. We find that halo model predictions agree reasonably well
with matter power spectra from IllustrisTNG up to 
$k\sim 10^3h{\rm Mpc}^{-1}$. We thus conclude that our halo model is
suited form studying the behavior of the matter power spectrum at very
high $k$.

\begin{figure}
\begin{center}
 \includegraphics[width=1.0\hsize]{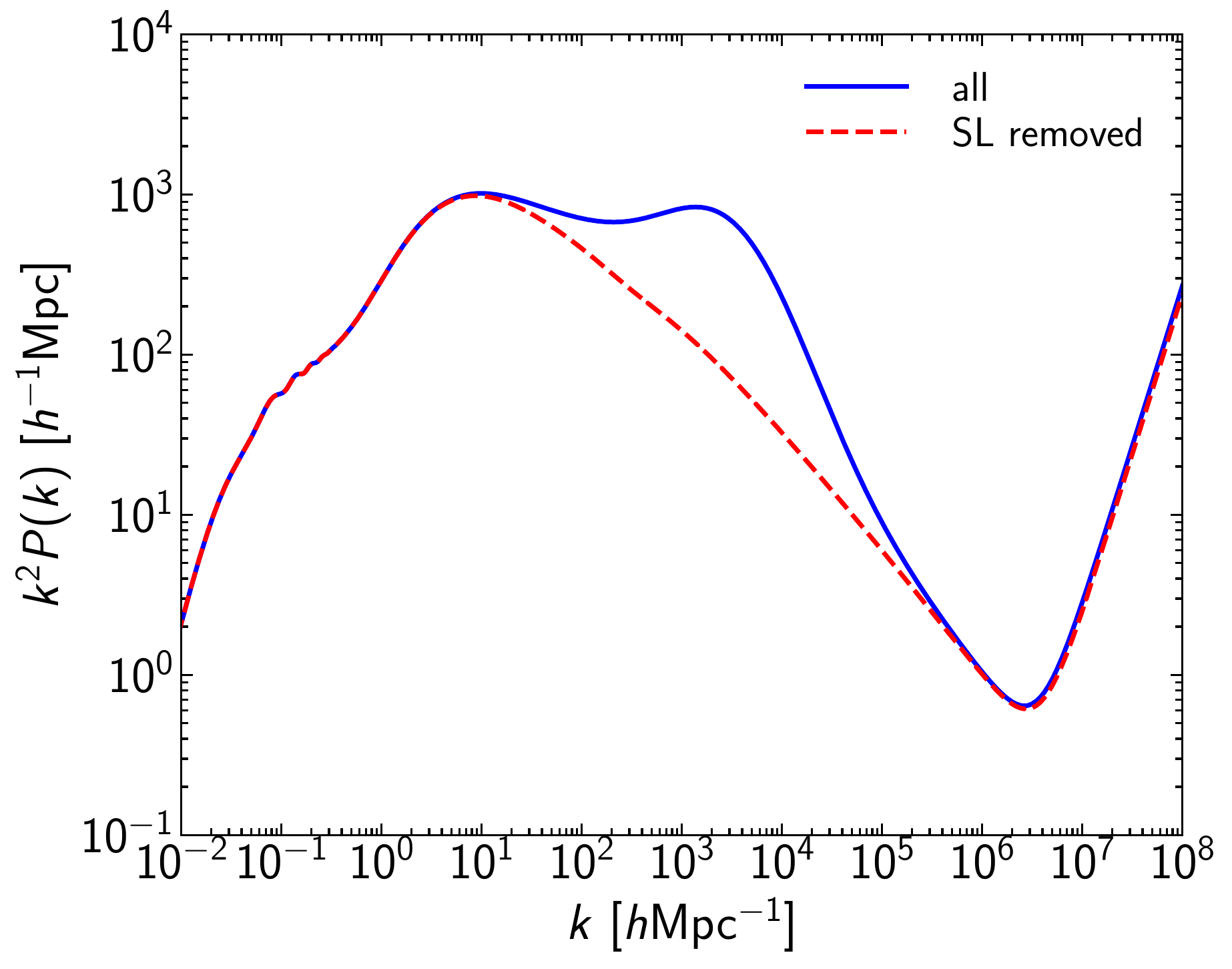}
\end{center}
\caption{Effects of removing central regions of halos that can produce
  strong lensing. The solid line shows the original halo model matter
  power spectrum at $z=0$, whereas the dashed line shows the result
  after removing those central regions following
  Equation~(\ref{eq:sl_remove}). 
\label{fig:pk_eli_sl}}
\end{figure}

\subsection{Contribution of Strong Lensing}
\label{sec:sl_remove}

Figure~\ref{fig:break_pk_main} suggests that the matter power spectrum
at $k\sim 10^3h{\rm Mpc}^{-1}$ is dominated by the stellar mass
components of halos with $M\sim 10^{13}h^{-1}M_\odot$ or so. The
central region of such halos is known to be a typical site for strong
gravitational lensing. Thus gravitational lensing dispersions caused
by such component must be highly non-Gaussian i.e., only a tiny
fraction of strong lensing events dominate the signal, which
complicates the analysis in observations. For instance, that
highly non-Gaussian component does not contribute to the gravitational 
lensing dispersion once strongly lensed events, which can be
identified in observations relatively easily, are removed from the
sample to derive the dispersion \citep[e.g.,][]{hada16,hada19}.

We evaluate the significance of such highly non-Gaussian contributions
corresponding to strong lensing as follows. We compute the matter
power spectrum including a suppression of centers of halos within
Einstein radii of a typical strong lensing configuration. 
Specifically, we modify the Fourier transform of the halo density
profile as
\begin{equation}
u(k)\rightarrow u(k)\exp\left(-k^2R_{\rm Ein,fid}^2\right),
\label{eq:sl_remove}
\end{equation}
where $u$ refers to both $u_{\rm h}$ and $u_{\rm sub}$, 
$R_{\rm Ein,fid}$ indicates the comoving Einstein radius of a singular
isothermal sphere with a fixed lens redshift $z=0.5$ and source
redshift $z_s=\infty$
\begin{equation}
R_{\rm Ein,fid}=4\pi \left(\frac{\sigma_v}{c}\right)^2\chi(z=0.5),
\end{equation}
with $\sigma_v$ being the velocity dispersion that is estimated from
the stellar mass $M_*$ (i.e., $f_*M$ for halos and $f^{\rm s}_*m$ for
subhalos) using the observed scaling relation
$\log(\sigma_v[{\rm km\,s^{-1}}])=-1.4+0.33\log(M_*[M_\odot])$ 
\citep{quimby14}. In this model, the Einstein radius reduces to zero
when halos contain no star, which is reasonable approximation because
the NFW profile alone has a negligibly small Einstein radius in the
low-mass limit \citep[see e.g.,][]{oguri19}.

We show the result in Figure~\ref{fig:pk_eli_sl}. As expected, the
matter power spectrum at 
$10^2 h{\rm Mpc}^{-1} \la k \la 10^5 h{\rm Mpc}^{-1}$ is
significantly affected by removing strong lensing regions of halos. 
In contrast, the matter power spectrum $k \sim 10^6 h{\rm Mpc}^{-1}$
is mostly unaffected, indicating that contributions from such highly
non-Gaussian strong lensing regions are not dominant. 

Since the contribution of strong lensing is not drastic in the
wavenumber range of our interest, in what follows we compute $P(k)$
without removing strong lensing regions unless otherwise stated. We
give additional  discussions on effects of strong lensing in
Section~\ref{sec:wl_approx}.  

\subsection{Effects of Primordial Black Holes}
\label{sec:pbh}

Primordial Black Holes (PBHs) are black holes generated in the early
Universe and are a viable candidate of dark matter \cite[see
  e.g.,][for reviews]{sasaki18,carr20}. Here we investigate effects of
PBHs on the small-scale matter power spectrum.

First, as in the case of stars, the shot noise from PBHs affects the
power spectrum. Denoting the total mass fraction of PBHs to dark
matter as $f_{\rm PBH}=\Omega_{\rm PBH}/\Omega_{\rm DM}$ where
$\Omega_{\rm DM}=\Omega_{\rm m}-\Omega_{\rm b}$ and the mass of each
PBH as $M_{\rm PBH}$, the comoving number density of PBHs is written as
\begin{eqnarray}
\bar{n}_{\rm PBH}
&=&7.224\times 10^{10}(h{\rm Mpc}^{-1})^3\nonumber\\
&&\times f_{\rm PBH}\left(\frac{\Omega_{\rm DM}}{0.26}\right)
\left(\frac{M_{\rm PBH}}{1h^{-1}M_\odot}\right)^{-1}.
\label{eq:nbar_pbh}
\end{eqnarray}
The contribution of the shot noise to the matter power spectrum is
simply given by
\begin{equation}
\Delta P_{\rm shot}(k)=\frac{f_{\rm PBH}^2}{\bar{n}_{\rm PBH}}.
\label{eq:pk_pbh_shot}
\end{equation}
Previous studies suggest that the Poisson fluctuation of the PBH
number density can be interpreted as an isocurvature perturbation
\citep[e.g.,][]{afshordi03,gong17,inman19}. We write the isocurvature
power spectrum due to the Poisson fluctuation as
\begin{equation}
\Delta P_{\rm iso}(k)=\left\{D_+(z)T_{\rm iso}(k)\right\}^2
\frac{f_{\rm PBH}^2}{\bar{n}_{\rm PBH}},
\label{eq:pk_pbh_iso}
\end{equation}
where the transfer function is approximated by 
\begin{equation}
T_{\rm iso}(k)=
\begin{cases}
  \frac{3}{2} (1+z_{\rm eq}) & (k_{\rm eq}<k<0.1k_*),\\
  0 & ({\rm otherwise}),
\end{cases}
\end{equation}
where $z_{\rm eq}$ is the redshift at the matter-radiation equality and 
$k_{\rm eq}=c^{-1}H(z_{\rm eq})(1+z_{\rm eq})^{-1}$ is the inverse of
the comoving Hubble horizon size at $z=z_{\rm eq}$. We truncate the
transfer function at $k\ga k_*$, where
$k_*=(2\pi^2\bar{n}_{\rm PBH})^{1/3}$ is the inverse of the length
scale within which there is on average one PBH and is given by
\begin{eqnarray}
k_*
&=&1.126\times 10^{4} h{\rm Mpc}^{-1}\nonumber\\
&&\times f_{\rm PBH}^{1/3}\left(\frac{\Omega_{\rm DM}}{0.26}\right)^{1/3}
\left(\frac{M_{\rm PBH}}{1h^{-1}M_\odot}\right)^{-1/3},
\end{eqnarray}
because the discreteness effects of PBHs become important as such small
scales. For instance, the fluid approximation used for the
calculation of the evolution of the isocurvature density fluctuations
clearly breaks down at $k\ga k_*$. In addition, halos containing the small 
($\la 10^3$) number of PBHs may be evaporated due to the relaxation
\citep{afshordi03}. We approximately take account of these effects by
truncating the transfer function at $k>0.1k_*$.

\begin{figure}
\begin{center}
 \includegraphics[width=1.0\hsize]{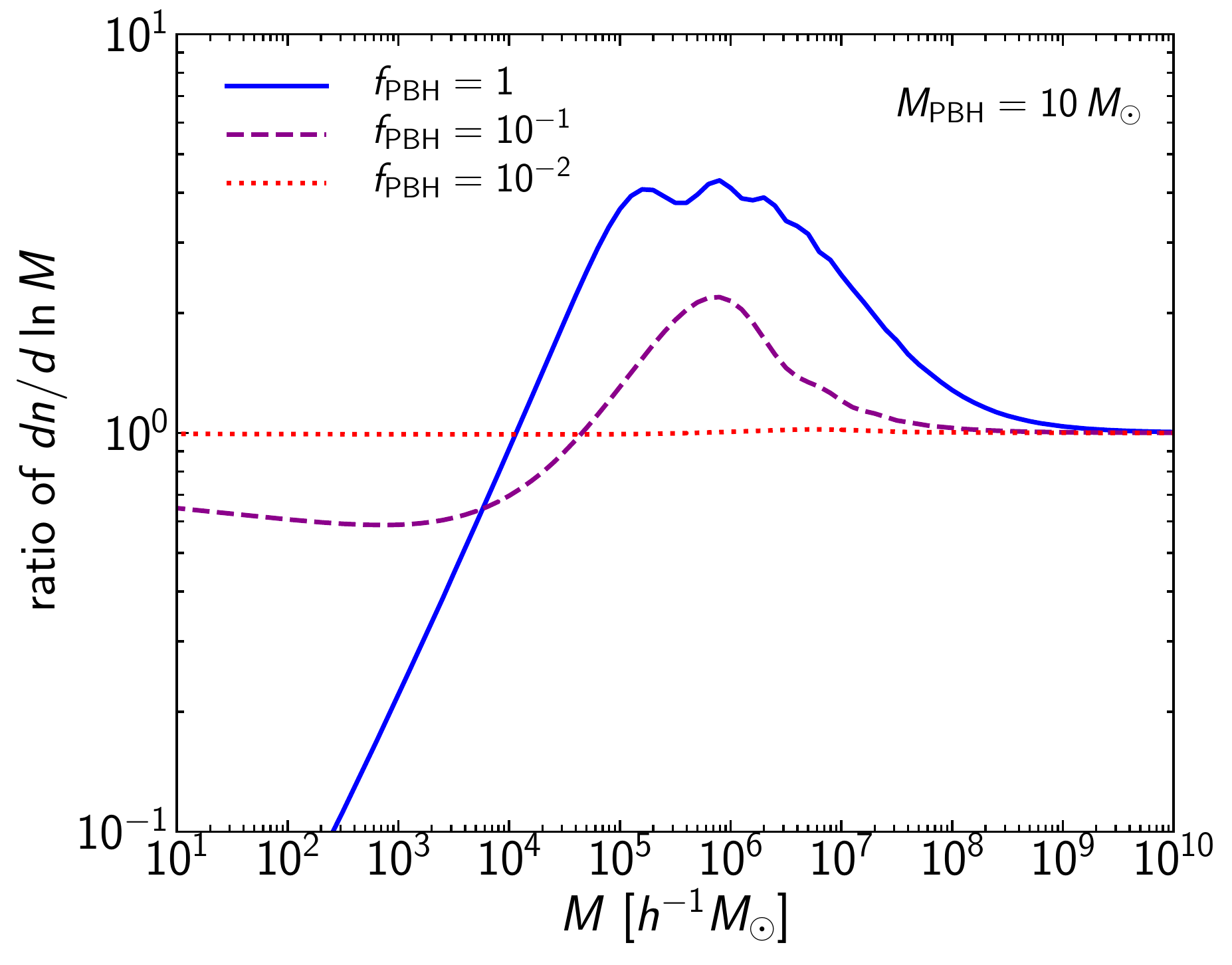}
\end{center}
\caption{The enhancement of the halo mass function $dn/d\ln M$ due to
  PBHs i.e., the ratio of $dn/d\ln M$ with PBHs to $dn/d\ln M$ without
  PBHs. The mass of PBHs is fixed to $M_{\rm PBH}=10~M_\odot$, whereas
  the total mass fractions are $f_{\rm PBH}=1$ ({\it solid}), 
  $10^{-1}$ ({\it dashed}), and $10^{-2}$ ({\it dotted}).
\label{fig:mf_pbh}}
\end{figure}

We include the effects of PBHs in our calculation of the nonlinear
matter power spectrum as follows. First, we add the contribution of
the isocurvature perturbation (equation~\ref{eq:pk_pbh_iso}) to the standard 
adiabatic linear matter power spectrum to compute the square root of
the mass variance $\sigma(M)$. By doing so the effect of PBHs is
included in the mass functions of main halos and subhalos, as well
as the concentration parameter of main halos. We show examples of
modifications of the halo mass function due to the isocurvature
perturbation in Figure~\ref{fig:mf_pbh}. After computing the
nonlinear matter power spectrum using the halo model, we add the shot
noise contribution (equation~\ref{eq:pk_pbh_shot}) to the nonlinear
matter power spectrum. 

\begin{figure}
\begin{center}
 \includegraphics[width=1.0\hsize]{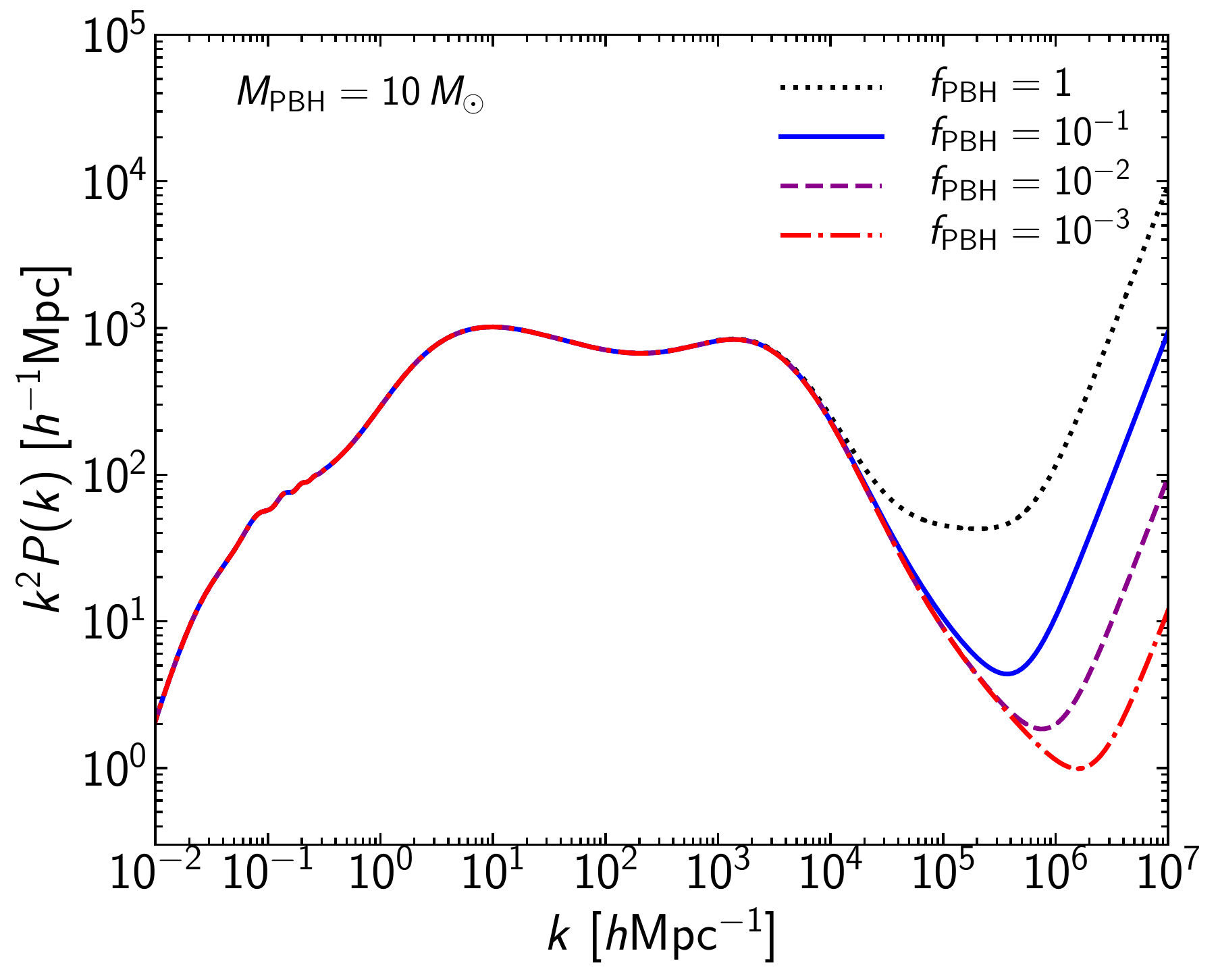}
\end{center}
\caption{Effects of PBHs on the matter power spectrum $P(k)$ at $z=0$
  assuming $M_{\rm PBH}=10~M_\odot$. The total mass fractions are 
  $f_{\rm PBH}=1$ ({\it dotted}), 
  $10^{-1}$ ({\it solid}), $10^{-2}$ ({\it dashed}), and $10^{-3}$ 
  ({\it dash-dotted}).
\label{fig:pk_pbh_f}}
\end{figure}

\begin{figure}
\begin{center}
 \includegraphics[width=1.0\hsize]{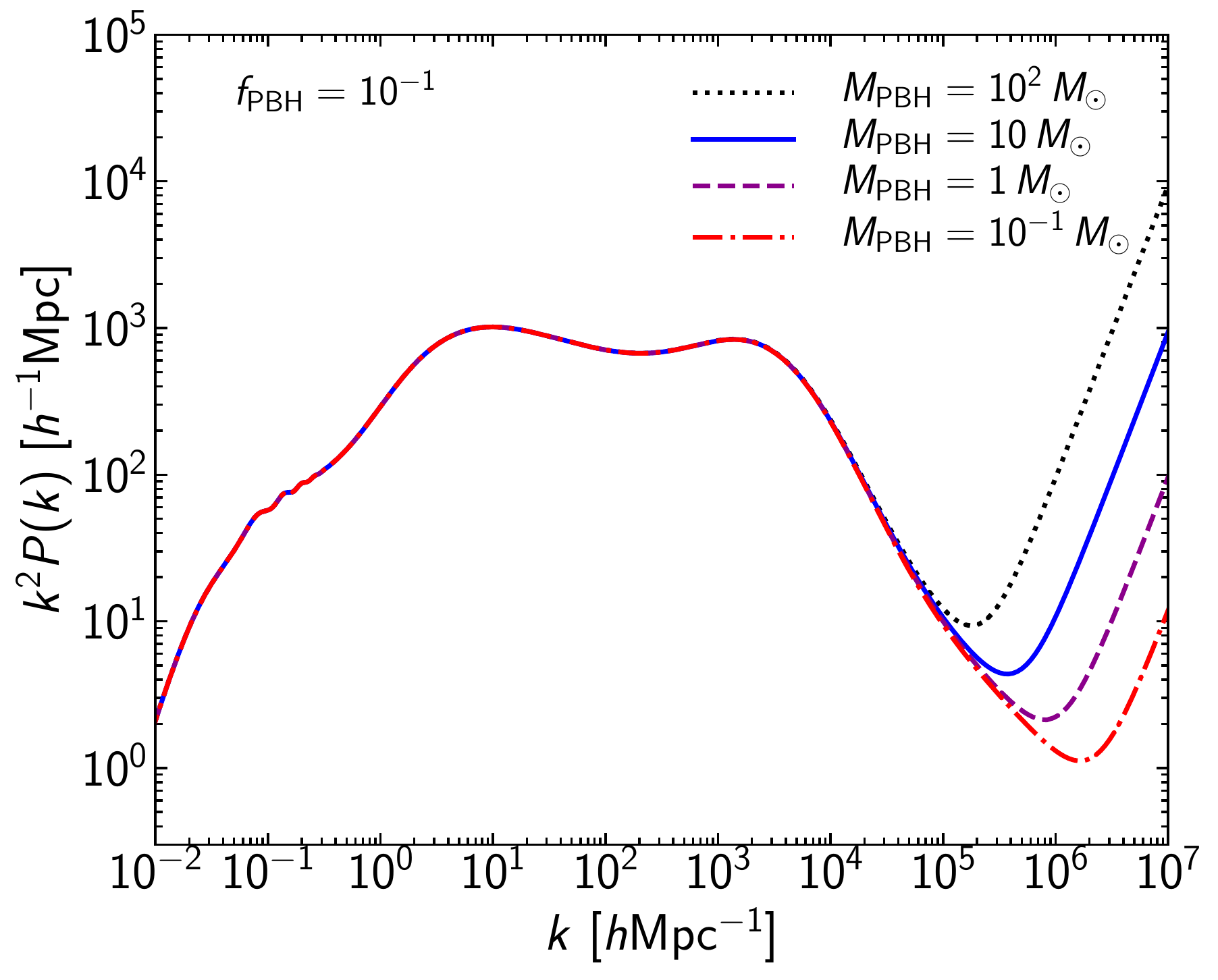}
\end{center}
\caption{Similar to Figure~\ref{fig:pk_pbh_f}, but the total mass
  fraction is fixed to $f_{\rm PBH}=10^{-1}$ and masses of PBHs of 
  $M_{\rm PBH}=10^2~M_\odot$ ({\it dotted}), 
  $M_{\rm PBH}=10~M_\odot$ ({\it solid}), $M_{\rm PBH}=1~M_\odot$ ({\it dashed}), 
  and $M_{\rm PBH}=10^{-1}~M_\odot$ ({\it dash-dotted}) are considered. 
\label{fig:pk_pbh_m}}
\end{figure}

We show several examples in Figures~\ref{fig:pk_pbh_f} and 
\ref{fig:pk_pbh_m}. We find that PBHs can significantly enhance the
small scale matter power spectrum at $k\ga 10^5 h{\rm Mpc}^{-1}$ due
to the enhanced number of dark low-mass halos as well as the shot 
noise from PBHs. Thus observations of small scale matter power spectra
not only directly probe dark low-mass halos but also can
constrain the mass and abundance of PBHs.

\section{Gravitational Lensing Dispersions}
\label{sec:dispersions}

\begin{figure}
\begin{center}
 \includegraphics[width=1.0\hsize]{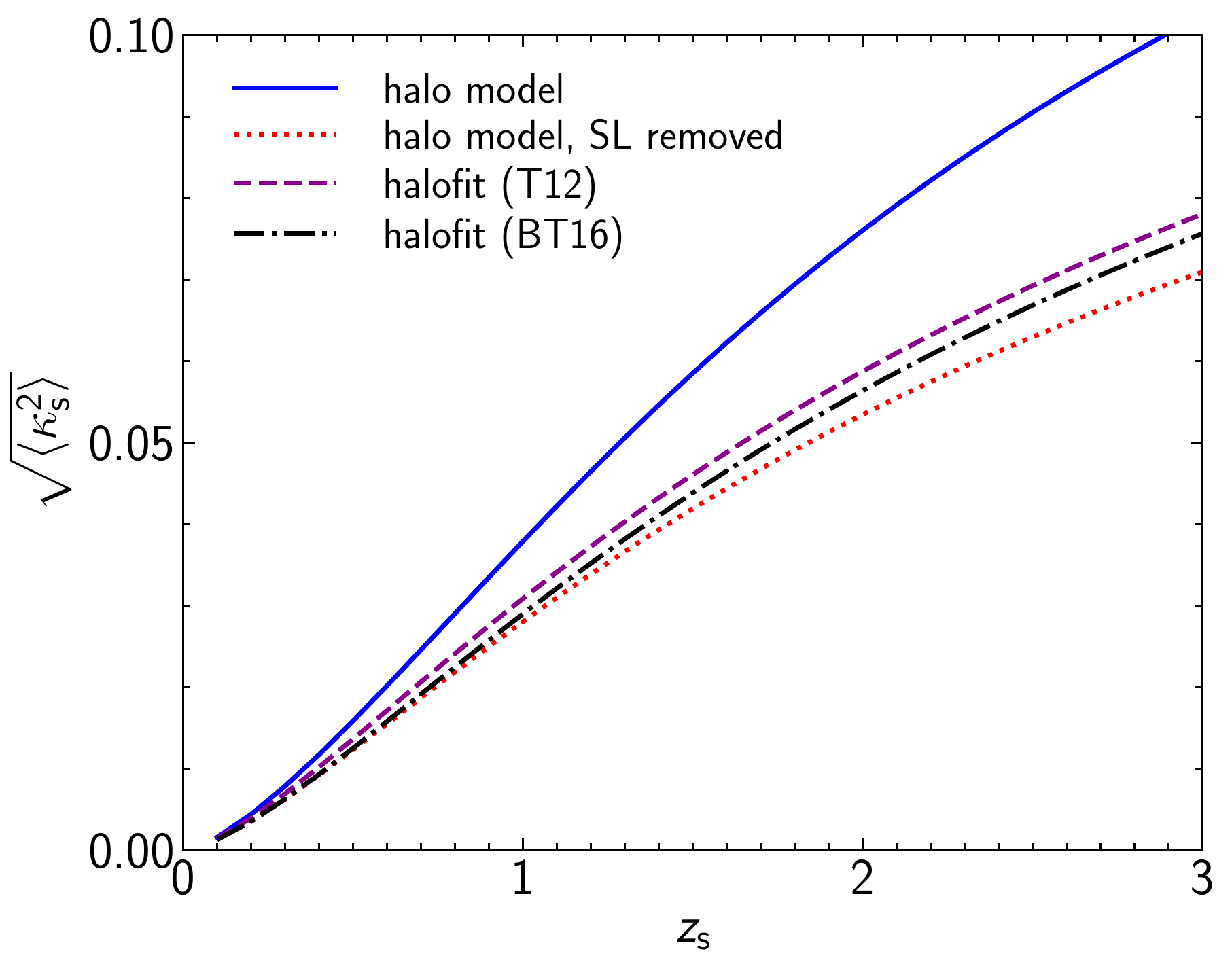}
\end{center}
\caption{Gravitational lensing dispersions
  (equation~\ref{eq:dis_kap_geo}) as a function of the source redshift
  $z_{\rm s}$ for the geometric optics case, assuming a compact source
  size of $\beta_{\rm s}=10^{-3}$~arcsec. We show results using
  our halo model including effects of baryon and subhalos ({\it
    solid}), our halo model but the contribution of strong lensing is
  removed following the prescription in Section~\ref{sec:sl_remove} 
  ({\it dotted}), the halofit model of \citet[][T12]{takahashi12} 
  ({\it dashed}), and the halofit model of 
  \citet[][BT16]{bendayan16} ({\it dash-dotted}). 
\label{fig:var_geo}}
\end{figure}

\subsection{Geometric Optics Case}

Geometric optics provides a good approximation for calculating
gravitational lensing dispersions of traditional astronomical sources
such as supernovae. In this case the dispersion of convergence
smoothed over the angular size $\beta_{\rm s}$ is given by 
\citep[e.g.,][]{takahashi11}
\begin{equation}
\langle\kappa_{\rm s}^2\rangle=\int_0^{\chi_{\rm s}} d\chi\,W^2(\chi)\int
\frac{k\,dk}{2\pi}P(k) W_{\rm s}^2(k\chi\beta_{\rm s}),
\label{eq:dis_kap_geo}
\end{equation}
where $\chi$ is the radial distance, $\chi_{\rm s}$ is the radial
distance to the source, the $W(\chi)$ is a lensing weight
function given by (note that $\bar{\rho}$ here is the comoving matter
density)  
\begin{equation}
W(\chi)=\frac{4\pi G}{c^2} \bar{\rho}a^{-1}\frac{\chi(\chi_{\rm
      s}-\chi)}{\chi_{\rm s}},
\label{eq:wl_weight}
\end{equation}
and $W_{\rm s}(x)$ is a smoothing kernel for which we assume a top-hat
filter 
\begin{equation}
W_{\rm s}(x)=\frac{2J_1(x)}{x}.
\label{eq:source_smooth}
\end{equation}
In most cases, the dispersion of magnification rather than that of
convergence is observed. For a similarly smoothed magnification
$\mu_{\rm s}$, a weak lensing approximation
\begin{equation}
\mu_{\rm s}\approx 1+2\kappa_{\rm s},
\end{equation}
is expected to hold when $|\mu_{\rm s}-1|\ll1$, and in this case we
simply have 
$\langle(\mu_{\rm s}-1)^2\rangle\approx 4\langle\kappa_{\rm s}^2\rangle$.

Figure~\ref{fig:var_geo} shows examples of dispersions of
convergence computed using our halo model as well as the halofit
model. We find that our halo model including baryonic effects predicts
significantly larger dispersions than halofit model predictions for
which baryonic effects are not included. However, as discussed in
Section~\ref{sec:sl_remove}, the significant fraction of the
enhancement by baryonic effects comes from centers of massive galaxies
that produce strong lensing. Once such regions are removed from the
calculation of the matter power spectrum (see
Section~\ref{sec:sl_remove} for more details), we find that
gravitational lensing dispersions from our halo model with baryonic
effects approximately match those from the halofit model for which
baryonic effects are not included. 

\begin{figure*}
\begin{center}
 \includegraphics[width=0.46\hsize]{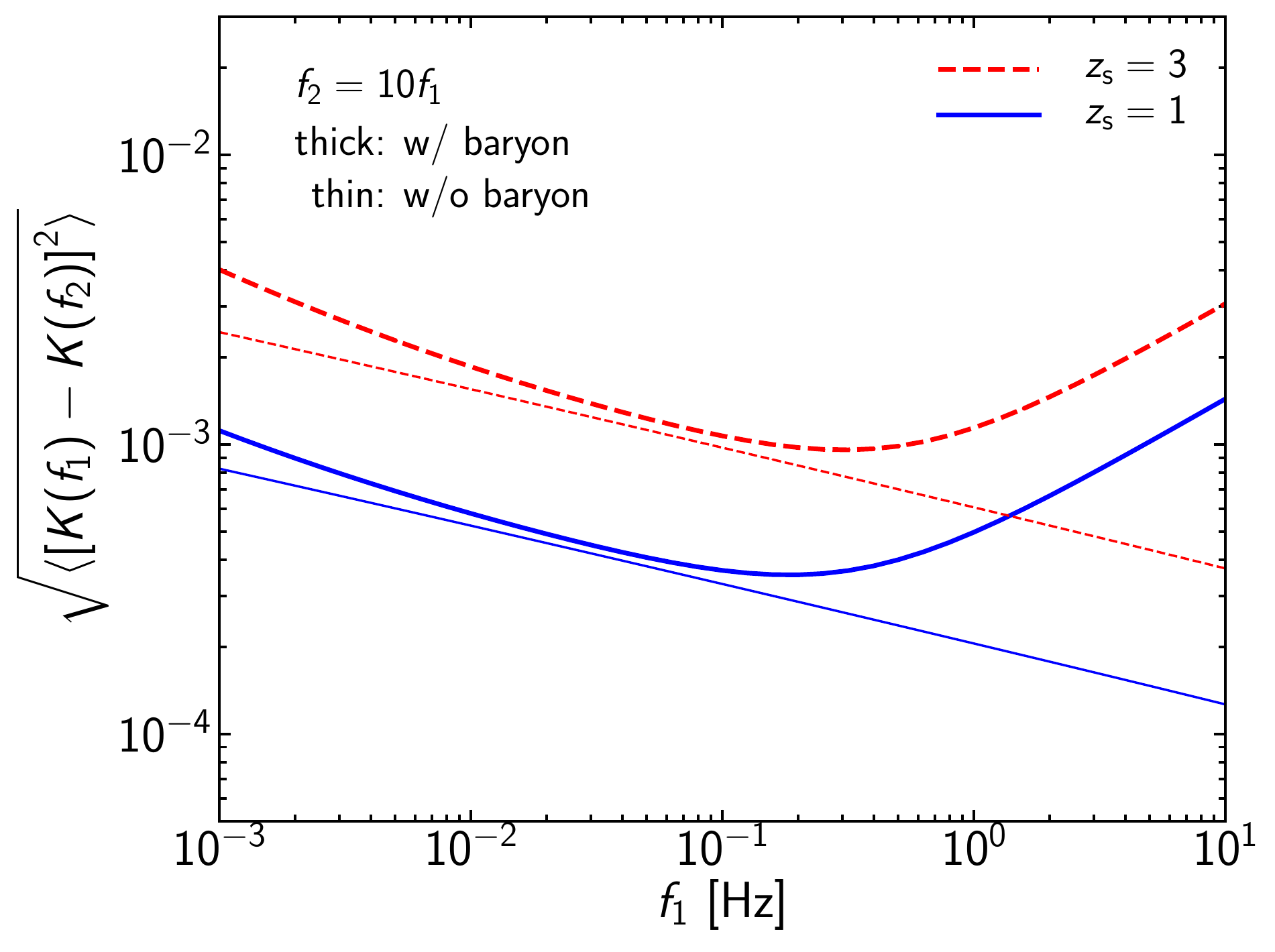}
 \includegraphics[width=0.46\hsize]{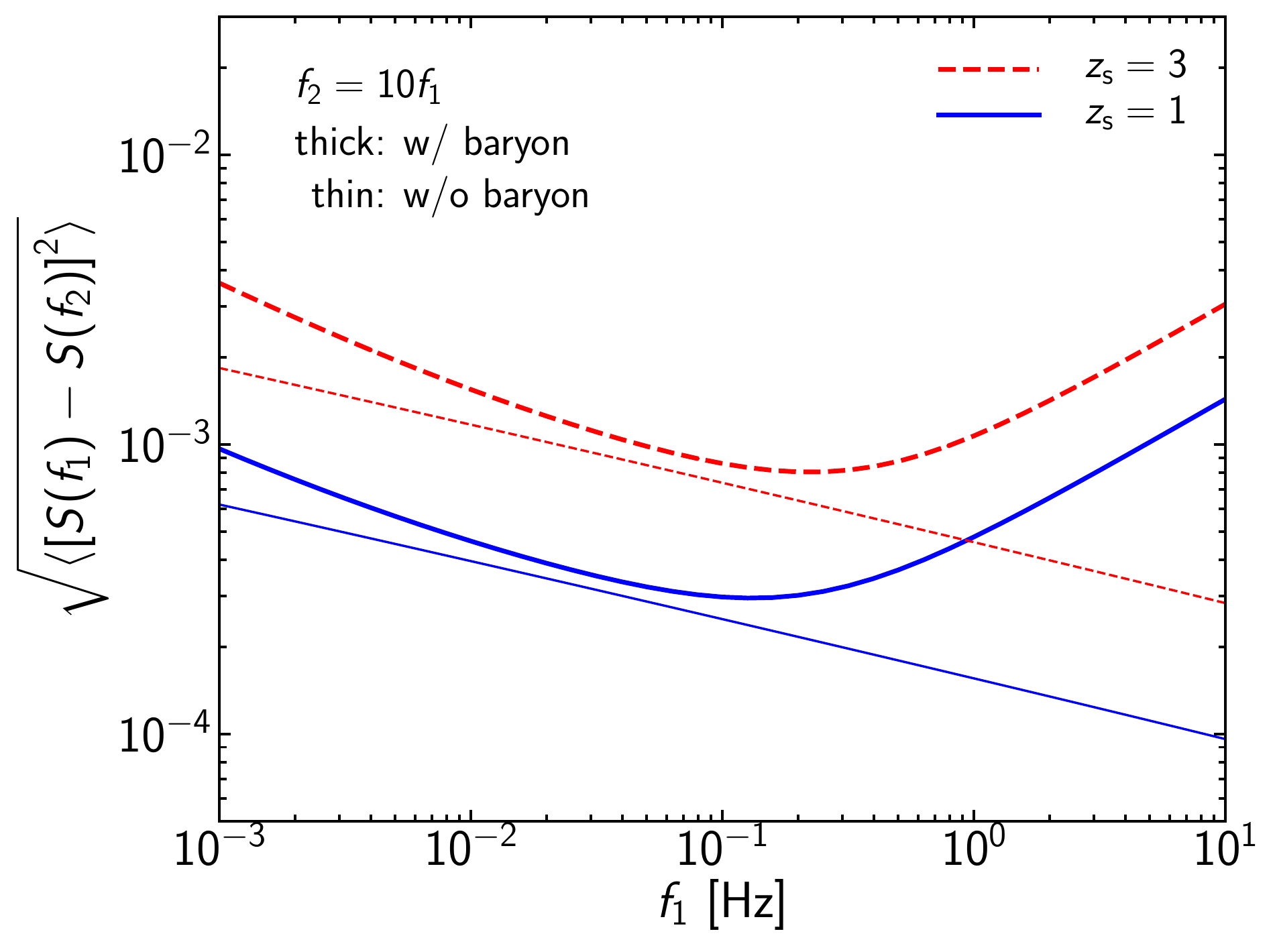}
\end{center}
\caption{{\it Left:} Gravitational lensing dispersion of the
  gravitational wave amplitude (equation~\ref{eq:def_k2_f12}) for
  $z_{\rm s}=1$ ({\it solid}) and $z_{\rm s}=3$ ({\it dashed}) as a
  function of $f_1$.  The frequency $f_2$ is fixed to
  $f_2=10f_1$. Thick and thin lines show calculations with and without
  baryonic effects, respectively.  {\it Right:} Similar to the left panel, but for
  gravitational lensing dispersion of the gravitational wave phase
 (equation~\ref{eq:def_s2_f12}).
\label{fig:var_f}}
\end{figure*}

\subsection{Wave Optics Case}

The propagation of gravitational waves in the inhomogeneous density
field has been studied in the literature, which indicates that density
fluctuations below the Fresnel scale $\sim (f\chi_{\rm s})^{1/2}$ 
\citep{macquart04,takahashi05,takahashi06} are subject to the wave
effect and does not affect the propagation of gravitational waves due
to diffraction effects. We provide detailed calculations in
Appendix~\ref{app:lensing_wave}, and here we give a short summary. 
We denote $\phi^0_{\rm obs}(f)$ as observed gravitational waves at
frequency $f$ in absence of gravitational lensing and 
$\phi_{\rm obs}(f)$ as observed gravitational waves with
gravitational lensing effects. Adopting the weak lensing
approximation, gravitational lensing effects can be 
described by small amplitude and phase shifts $K(f)$ and $S(f)$
\begin{equation}
\frac{\phi_{\rm obs}(f)}{\phi^0_{\rm obs}(f)}\approx 
\left[1+K(f)\right]e^{iS(f)}.
\end{equation}
In the geometric optics limit $K(f)$ and $S(f)$ reduce to convergence
and gravitational time delay, respectively. Dispersions of $K(f)$ and
$S(f)$  are computed as
\begin{equation}
\langle K^2(f)\rangle = \int_0^{\chi_{\rm s}}  d\chi W^2(\chi)\int
\frac{k\,dk}{2\pi}P(k) F_K^2,
\label{eq:def_k2}
\end{equation}
\begin{equation}
F_K=\frac{\sin(r_{\rm F}^2k^2/2)}{r_{\rm F}^2k^2/2},
\label{eq:filter_fk}
\end{equation}
\begin{equation}
\langle S^2(f)\rangle = \int_0^{\chi_{\rm s}}  d\chi W^2(\chi)\int
\frac{k\,dk}{2\pi}P(k) F_S^2,
\label{eq:def_s2}
\end{equation}
\begin{equation}
F_S=\frac{\cos(r_{\rm F}^2k^2/2)-1}{r_{\rm F}^2k^2/2},
\label{eq:filter_fs}
\end{equation}
where $r_{\rm F}$ denotes the Fresnel scale 
\citep{macquart04,takahashi05,takahashi06}
\begin{equation}
r_{\rm F}=\sqrt{\frac{c\chi(\chi_{\rm s}-\chi)}{2\pi f\chi_{\rm s}}}.
\label{eq:r_fresnel}
\end{equation}
Equation~(\ref{eq:def_k2}) suggests that the Fresnel scale can be
interpreted as an effective source size, as is also discussed in
Appendix~\ref{app:shot}.  

As discussed in Appendix~\ref{app:lensing_wave}, $K(f)$ and $S(f)$ can
be measured by comparing inspiral waveforms at different frequencies
\begin{equation}
\left\langle\left[K(f_1)-K(f_2)\right]^2\right\rangle 
= \int_0^{\chi_{\rm s}}  d\chi W^2(\chi)\int
\frac{k\,dk}{2\pi}
P(k)F_{K,12}^2,
\label{eq:def_k2_f12}
\end{equation}
\begin{equation}
F_{K,12}=\frac{\sin(r_{\rm F1}^2k^2/2)}
{r_{\rm F1}^2k^2/2}-
\frac{\sin(r_{\rm F2}^2k^2/2)}
{r_{\rm F2}^2k^2/2},
\label{eq:filter_fk12}
\end{equation}
\begin{equation}
\left\langle\left[S(f_1)-S(f_2)\right]^2\right\rangle 
= \int_0^{\chi_{\rm s}}  d\chi W^2(\chi)\int
\frac{k\,dk}{2\pi}
P(k)F_{S,12}^2,
\label{eq:def_s2_f12}
\end{equation}
\begin{equation}
F_{S,12}=\frac{\cos(r_{\rm F1}^2k^2/2)-1}
{r_{\rm F1}^2k^2/2}
-\frac{\cos(r_{\rm F2}^2k^2/2)-1}
{r_{\rm F2}^2k^2/2},
\label{eq:filter_fs12}
\end{equation}
where $r_{\rm F1}$ and $r_{\rm F2}$ 
denote the Fresnel scales (equation~\ref{eq:r_fresnel}) evaluated at
frequency $f_1$ and $f_2$, respectively. We can also consider the
cross-correlation between $K(f)$ and $S(f)$ as
\begin{eqnarray}
\left\langle\left[K(f_1)-K(f_2)\right]
\left[S(f_1)-S(f_2)\right]\right\rangle &&\nonumber\\
&&\hspace*{-45mm}=\int_0^{\chi_{\rm s}}  d\chi W^2(\chi)\int
\frac{k\,dk}{2\pi}
P(k)F_{K,12}F_{S,12}.
\label{eq:def_ks_f12}
\end{eqnarray}

We show some examples in Figures~\ref{fig:var_f}. Since the Fresnel
scale is proportional to $f^{-1/2}$ (see equation~\ref{eq:r_fresnel}),
at lower frequency the dispersion probe the matter power spectrum
at smaller $k$. Effects of baryon are minimized around the frequency
of $f\sim 0.1$~Hz, where the matter power spectrum at $k\sim 10^6
h{\rm Mpc}^{-1}$ is probed. At frequency higher than $f\sim 0.1$~Hz, 
the dispersion quickly increases because of the shot noise from stars. 
Gravitational wave observations around $f\sim 0.1$~Hz will be
conducted by e.g., B-DECIGO \citep{nakamura16}. However, expected
dispersions are small, $\mathcal{O}(10^{-3})$, suggesting that high
$S/N$ observations for many events are needed to detect the
dispersions. Some discussions on the detectability are given in
Section~\ref{sec:detectability}. 

\begin{figure*}
\begin{center}
 \includegraphics[width=0.46\hsize]{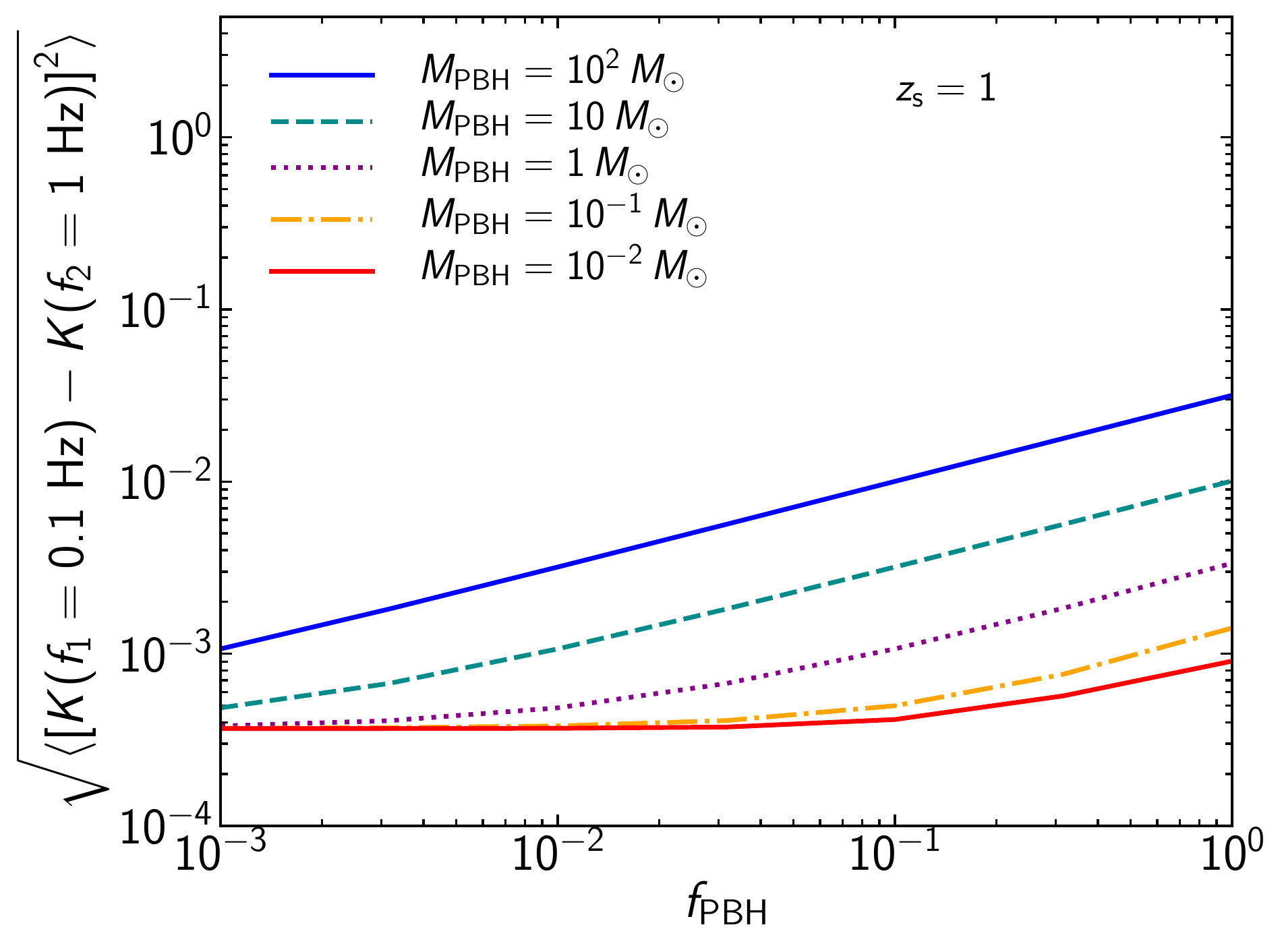}
 \includegraphics[width=0.46\hsize]{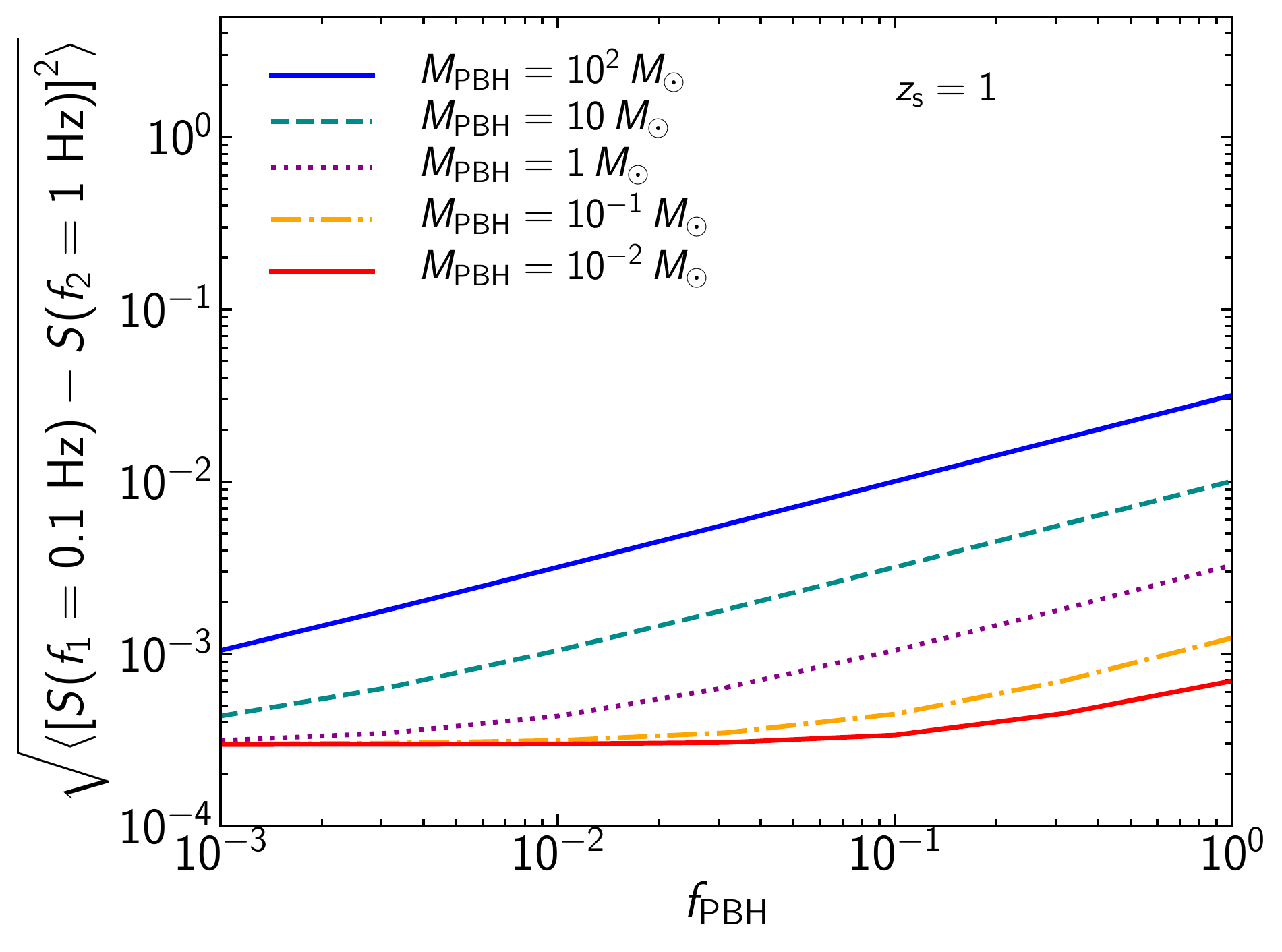}
 \includegraphics[width=0.46\hsize]{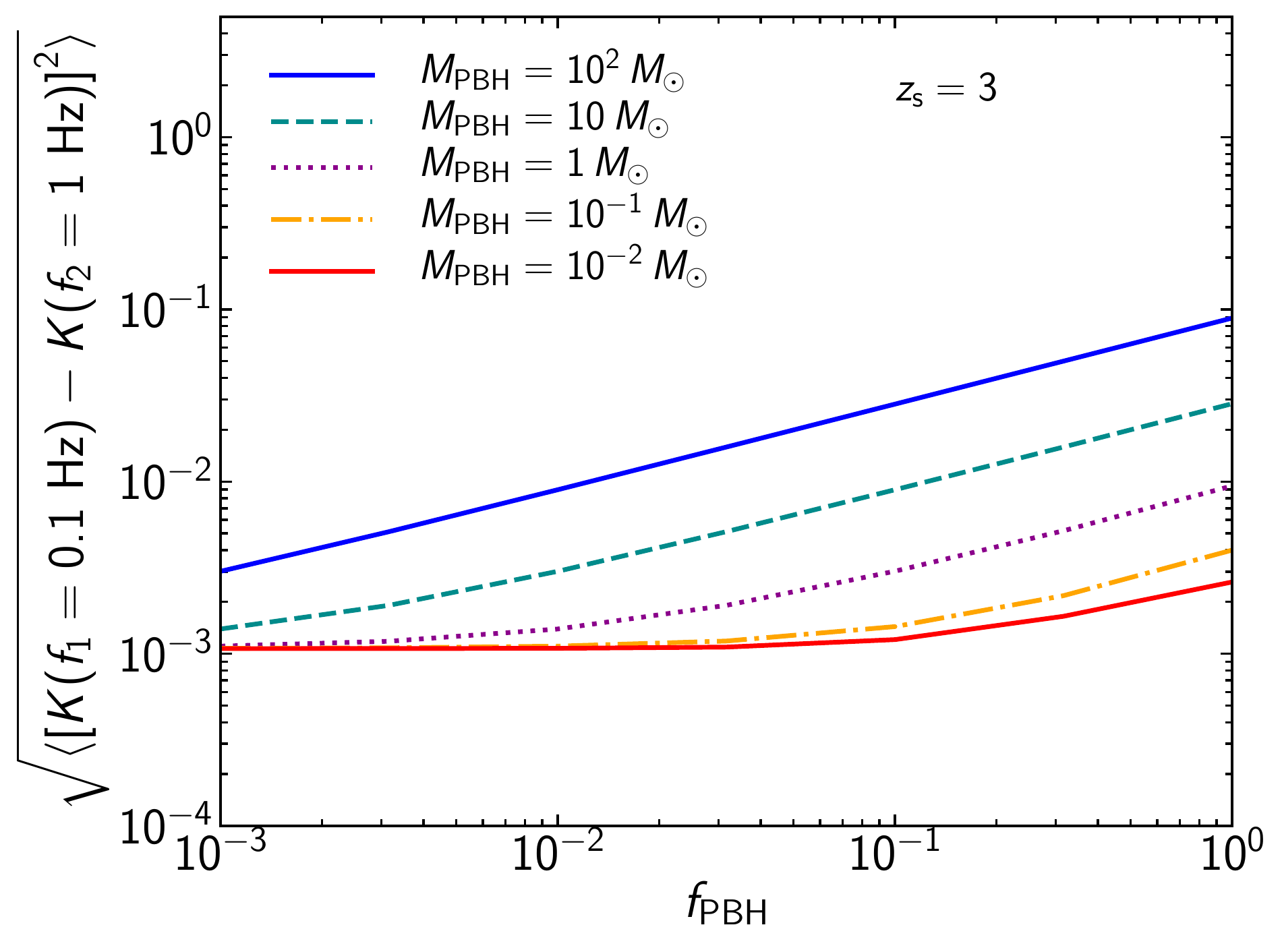}
 \includegraphics[width=0.46\hsize]{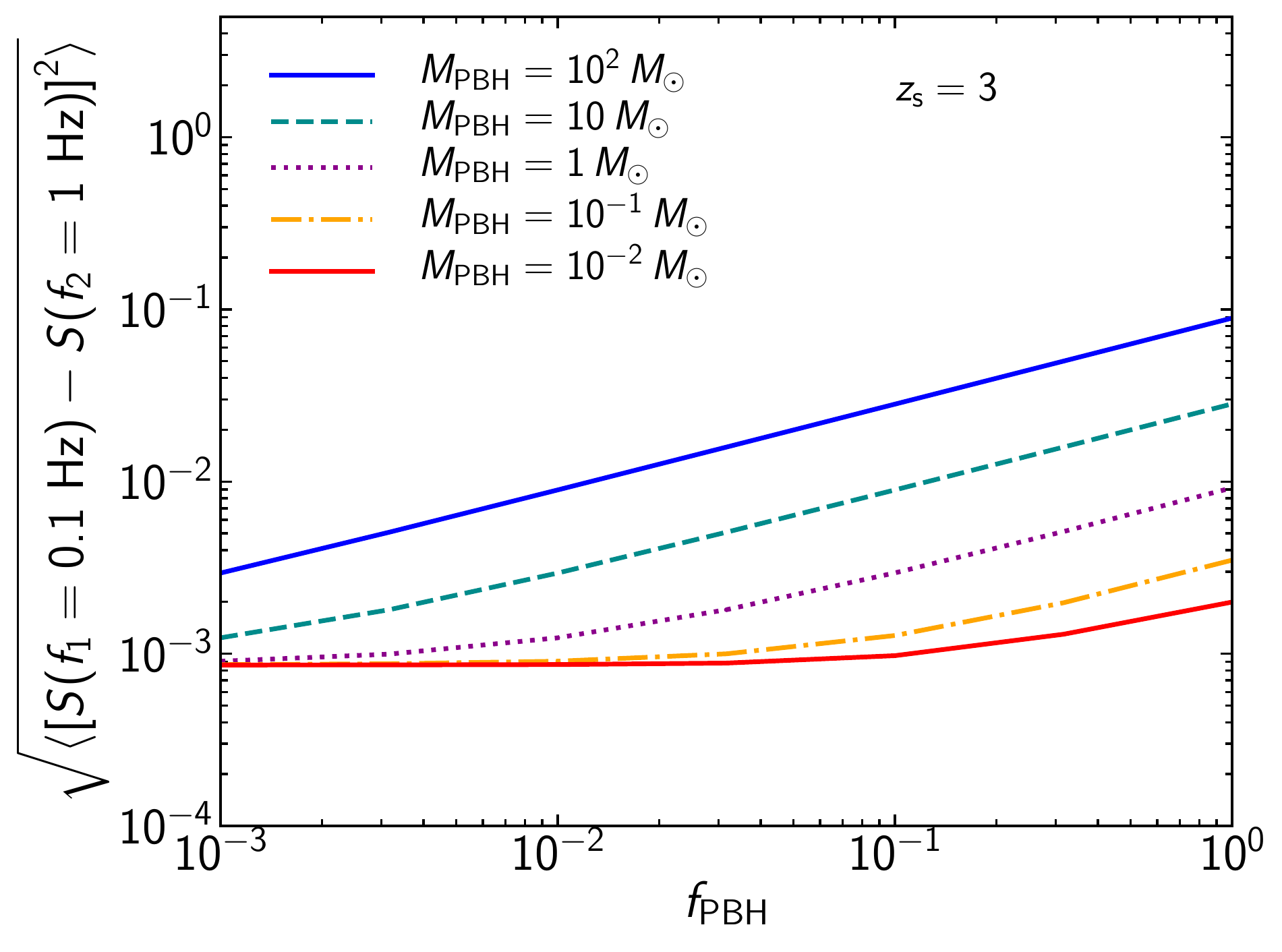}
\end{center}
\caption{Gravitational lensing dispersions of gravitational wave
  amplitude (equation~\ref{eq:def_k2_f12}) and phase
  (equation~\ref{eq:def_s2_f12}) are shown in left and right
  panels, respectively. Upper panels show results at $z_{\rm s}=1$ and
  lower panels show results at $z_{\rm s}=3$. Frequencies are fixed
  to $f_1=0.1~{\rm Hz}$ and $f_2=1~{\rm Hz}$. In each panel,
  gravitational lensing dispersions as a function of PBH fraction
  $f_{\rm PBH}$ for several different PBH masses are shown.
\label{fig:var_f_pbh_f01}}
\end{figure*}

\begin{figure*}
\begin{center}
 \includegraphics[width=0.46\hsize]{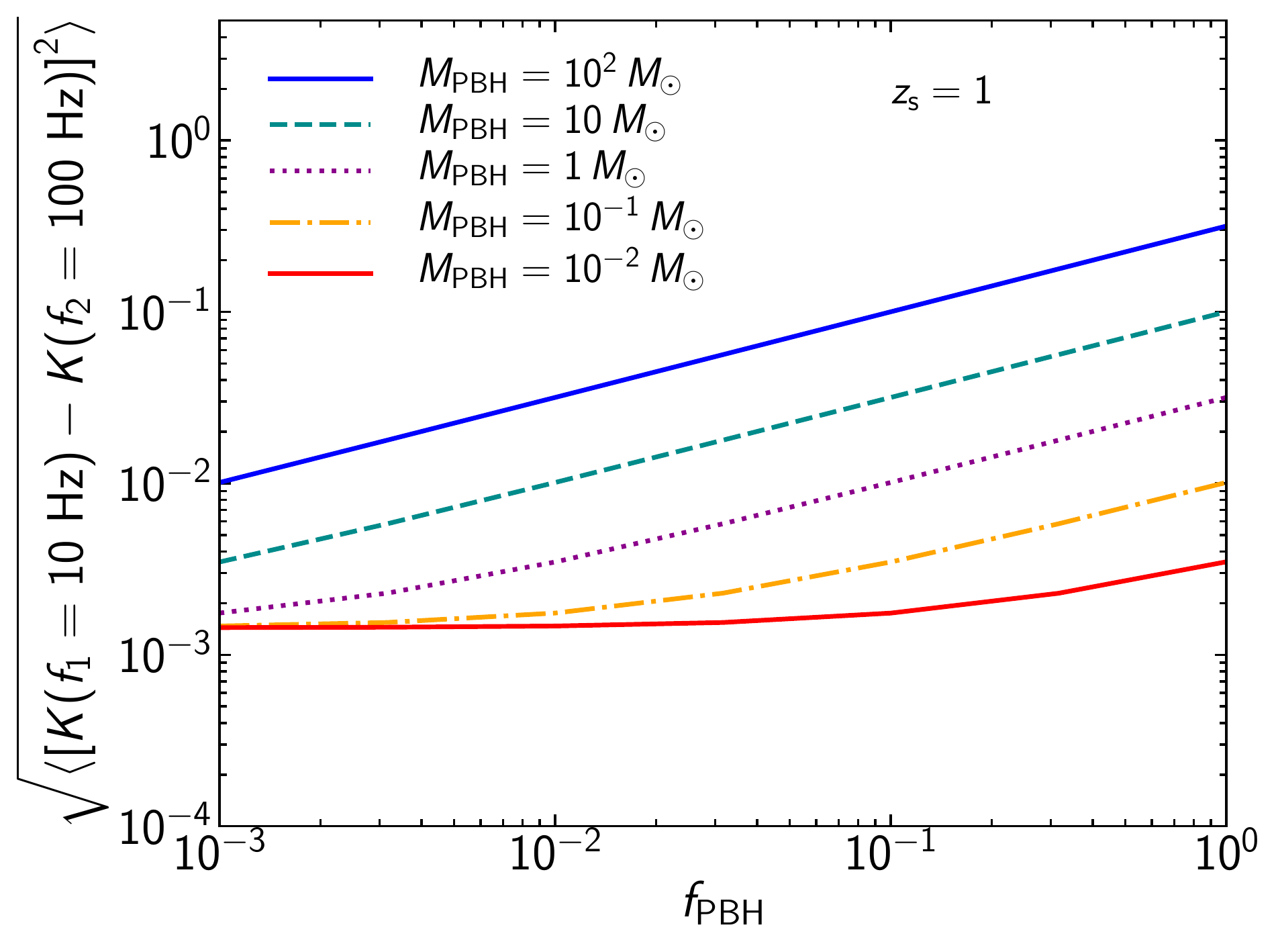}
 \includegraphics[width=0.46\hsize]{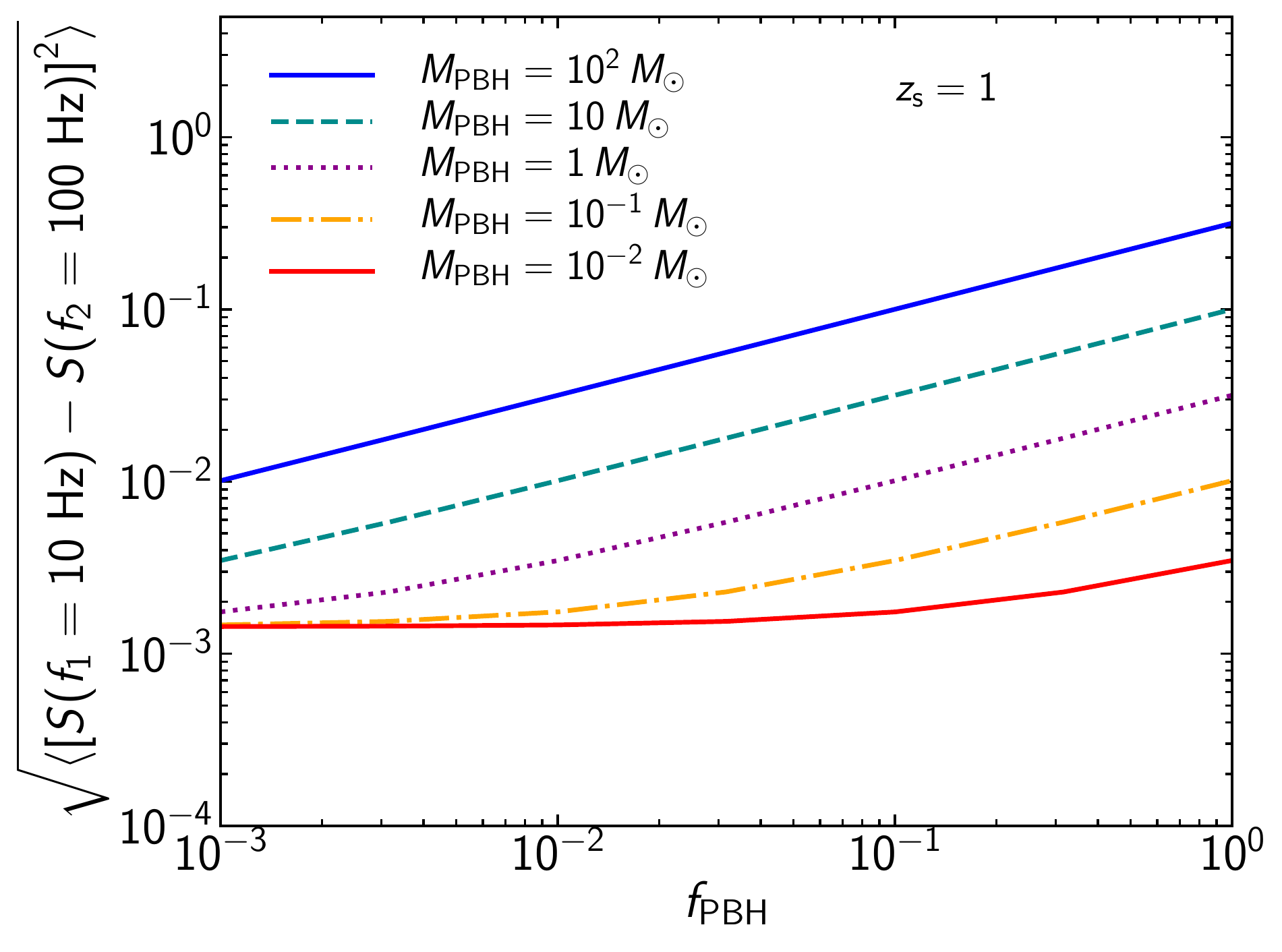}
 \includegraphics[width=0.46\hsize]{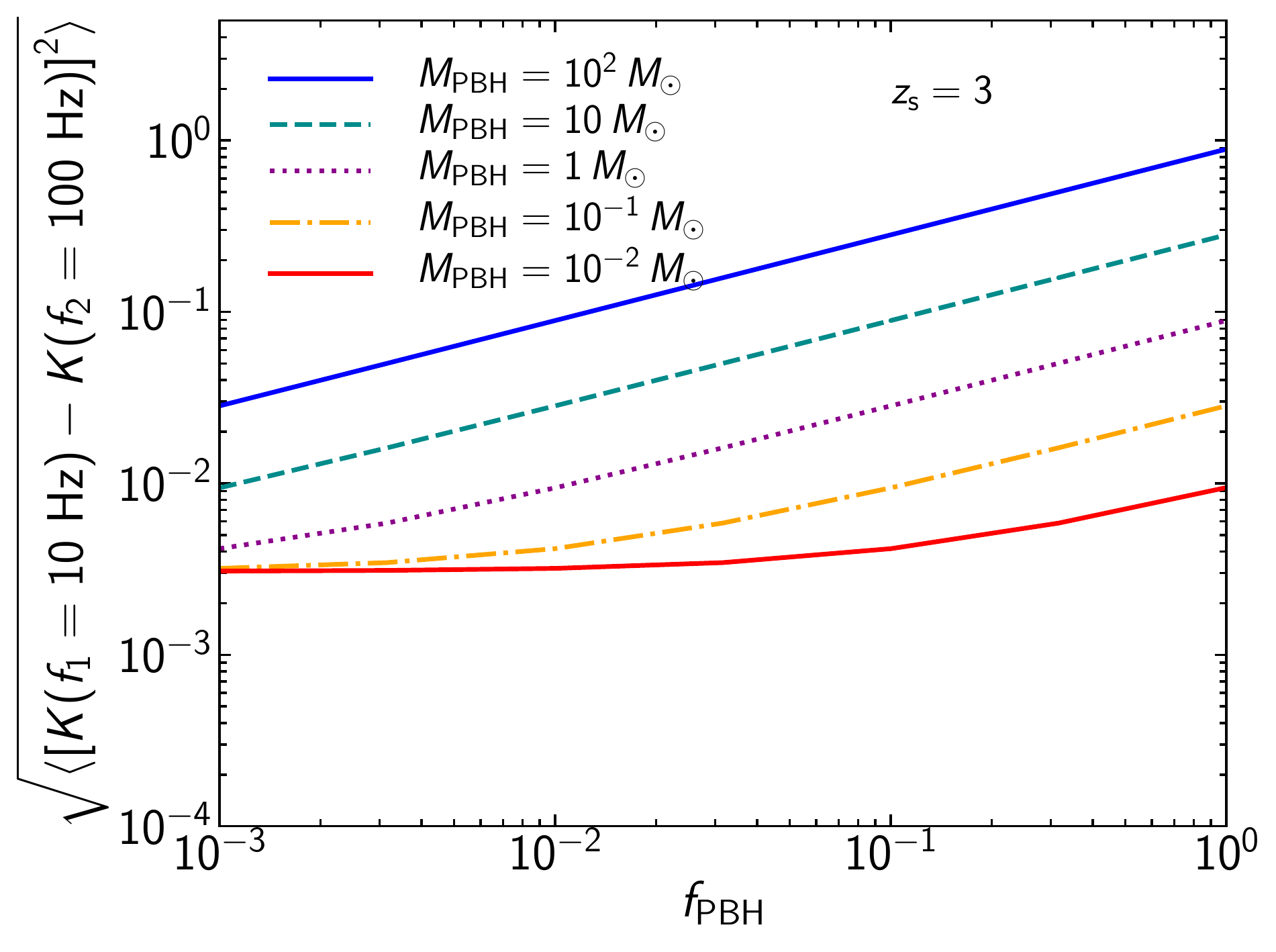}
 \includegraphics[width=0.46\hsize]{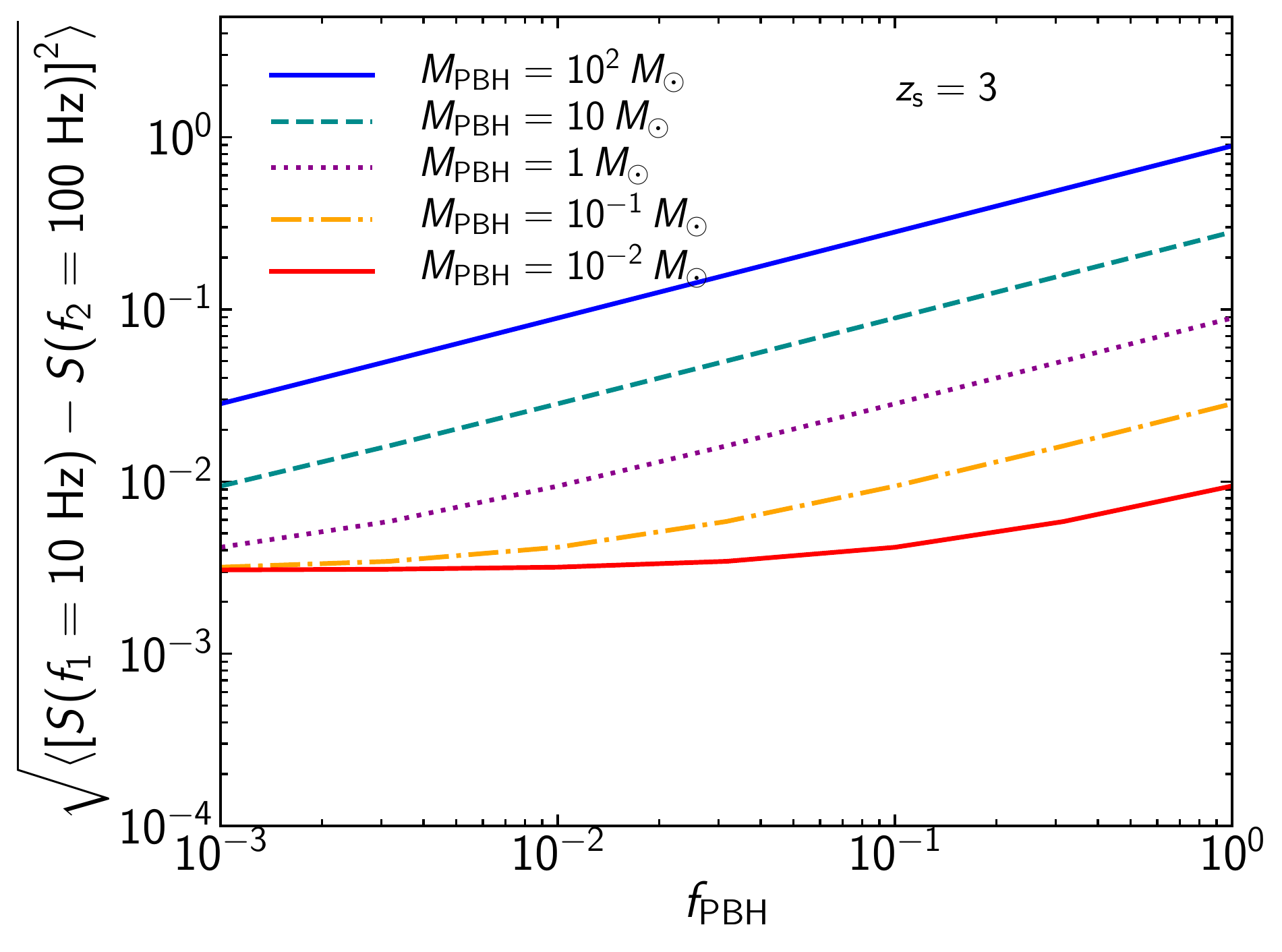}
\end{center}
\caption{Similar to Figure~\ref{fig:var_f_pbh_f01}, but for
  frequencies $f_1=10~{\rm Hz}$ and $f_2=100~{\rm Hz}$.
\label{fig:var_f_pbh_f10}}
\end{figure*}

\subsection{Enhancement due to Primordial Black Holes}

As shown in Section~\ref{sec:pbh}, the presence of PBHs can
significantly enhance the matter power spectrum at high $k$, 
suggesting that observations of gravitational lensing dispersions of
gravitational waves can be used to constrain the abundance of PBHs.
We explore this possibility using our model of the matter power
spectrum presented in Section~\ref{sec:pbh}.

We show results in Figures~\ref{fig:var_f_pbh_f01} and
\ref{fig:var_f_pbh_f10}. Here we consider two combinations of
frequency ranges, from $f_1=0.1~{\rm Hz}$ and $f_2=1~{\rm Hz}$
corresponding to space observations of gravitational waves
(Figure~\ref{fig:var_f_pbh_f01}), and from $f_1=10~{\rm Hz}$ and
$f_2=100~{\rm Hz}$ corresponding to ground observations of
gravitational waves (Figure~\ref{fig:var_f_pbh_f10}). We find that the
enhancement of gravitational lensing dispersion by PBHs is indeed
significant, more than an order of magnitude in some cases. 
The enhancement is particularly large for $f_1=10~{\rm Hz}$ and
$f_2=100~{\rm Hz}$, for which the shot noise from PBHs dominates 
gravitational lensing dispersion (see also Section~\ref{sec:pbh}).
Weak lensing by the shot noise is discussed further in
Section~\ref{sec:discussion} and Appendix~\ref{app:shot}.
Since the shot noise power spectrum (equation~\ref{eq:pk_pbh_shot}) 
is $\Delta P_{\rm shot}(k)=f_{\rm PBH}^2/\bar{n}_{\rm PBH}\propto
f_{\rm PBH} M_{\rm PBH}$ for a fixed $\Omega_{\rm DM}$, gravitational
lensing dispersions at the shot noise dominated region behave as 
$\propto \sqrt{f_{\rm PBH} M_{\rm PBH}}$ as shown in
Figure~\ref{fig:var_f_pbh_f10}. 

\section{Discussions}
\label{sec:discussion}

\subsection{Detectability}
\label{sec:detectability}

For each gravitational wave event, we can measure amplitude and phase
fluctuations with an accuracy of $\sim 1/\rho$, where $\rho$ denotes
the signal-to-noise ratio of the gravitational wave observation 
\citep{lindblom08}. Unless gravitational lensing dispersions are
significantly boosted by PBHs, its typical value is 
$\mathcal{O}(10^{-3})$, indicating that $\rho\gtrsim 10^3$ is needed
to directly measure amplitude and phase shifts due to gravitational
lensing for individual gravitational wave events.  We use the
calculation method described in \citet{oguri18b} to estimate $\rho$
for the chirp mass of $30~M_\odot$ and the redshift of $1$ and find
that $\rho\sim 30$ for B-DECIGO and $\rho\sim 60$ for Einstein
Telescope. Therefore measurements of gravitational lensing dispersions
are likely to be achieved by combining observations of many
gravitational wave events. For instance, by combining $N_{\rm event}$
gravitational wave events, we can measure amplitude and phase
dispersions down to $\sim (2/N_{\rm event})^{1/4}(1/\rho)$, suggesting
that $N_{\rm event}=3\times 10^5$ with $\rho=50$ leads to the
measurement of the dispersion at the level of $10^{-3}$. The required
number is large but can be achieved by next-generation
gravitational wave experiments. Alternatively, observations of the
modest number of events with very sensitive space based gravitational
wave detectors such as DECIGO \citep{seto01} may allow us to measure
gravitational lensing dispersions.

We note that measurements of gravitational lensing dispersions do not
necessarily require measurements of redshifts of individual
gravitational wave events. For instance, by considering the ratio of
strain amplitudes at different frequencies the dependence on the
distance to the gravitational wave source cancels out. The dependence
on antenna pattern functions also cancels out if the frequency
evolution is much faster than the change of antenna pattern functions
with time. The dispersion of phases may also be measured without
knowing the distance to the source, although the degeneracy with
binary model parameters may be in issue. We leave detailed studies of
measurements of gravitational lensing dispersions in a realistic setup
for future work.

\subsection{Validity of Weak Lensing Approximation}
\label{sec:wl_approx}

Since our results rely on the weak lensing approximation, it is
important to check the validity of the approximation. This is partly
done in Section~\ref{sec:sl_remove} in which the matter power spectrum
at $k\sim 10^6 h{\rm Mpc}^{-1}$, which is responsible for gravitational 
lensing dispersions at $f\sim 0.1-1$~Hz,  is shown to be largely
unaffected by removing central regions of galaxies that can produce
strong lensing. This is because the matter power spectrum at 
$k \sim 10^6 h{\rm Mpc}^{-1}$ mostly originates from halos and subhalos
with masses $1h^{-1}M_\odot \la M \la 10^4h^{-1}M_\odot $. Such
low-mass halos do not contain stars, and the convergence $\kappa$ of
such halos computed from the NFW (or BMO) profile is quite low because
it is proportional to 
$\rho_{\rm s}r_{\rm s} \propto r_{\rm vir}c^2/f_{\rm NFW}(c)$. Given
the weak dependence of $c$ on the halo mass, $\kappa$ is a increasing
function of the mass such that $\kappa\ll 1$ even at the very central
region for halos and subhalos with masses 
$1h^{-1}M_\odot \la M \la 10^4h^{-1}M_\odot$. Thus we can safely
adopt the weak lensing approximation for lensing by dark low-mass
halos. 

When the shot noise contribution to the matter power spectrum is
dominant, it is important to make sure that gravitational lensing by
individual stars or PBHs is not strong. This condition is written as
(see also equation~\ref{eq:kappa_s_wave})
\begin{equation}
w=\left(\frac{R_{\rm Ein}}{r_{\rm F}}\right)^2=2\pi f 
(1+z)\frac{4Gm_{\rm p}}{c^3}< 1, 
\end{equation}
where $R_{\rm Ein}$ is the Einstein radius (equation~\ref{eq:r_ein})
and $m_{\rm p}$ refers to either individual mass of stars $m_{\rm
  star}$ or individual mass of PBHs $M_{\rm PBH}$. This condition 
translates into 
$m_{\rm p}<81(1+z)^{-1}(f/100~{\rm Hz})^{-1}M_\odot$, where
$f=100~{\rm Hz}$ is the highest frequency considered in this
paper. Therefore, the weak lensing approximation is reasonable for all
the situations consider in the paper, except for the case with the PBH
mass  $M_{\rm PBH}=100~M_\odot$ and the frequency range $f=10-100$~Hz
for which the weak lensing approximation is partly broken and thus the
result should be taken with caution. 

\subsection{Degree of Non-Gaussianity}
\label{sec:non_gauss}

Another question to ask is how well the distributions of $K$ and $S$
are described by Gaussian. For instance, if the signal is dominated by
a tiny fraction of strong lensing events the resulting distribution 
is quite non-Gaussian as partly discussed in
Section~\ref{sec:sl_remove}. While this can be studied by computing
higher-order statistics such as skewness and kurtosis, here we
make a simple evaluation of the degree of non-Gaussianity based on the
average number of lenses that contribute to the dispersion. 
Discussions in Appendix~\ref{app:shot} indicate that lenses that fall
within the Fresnel scale contribute to the dispersion when the weak
lensing approximation is valid. When the average number of lenses is
much large than unity, the central limit theorem assures that the
distribution is close to the Gaussian distribution, whereas the
average number is small, say less than unity, we expect the
distribution with a significant skewness.  

The average number $\bar{N}_{\rm proj}$ of any lenses with the
comoving number density $\bar{n}$, which is assumed to be independent
of redshift for simplicity, within the Fresnel scale integrated along
the line-of-sight is given by
\begin{equation}
\bar{N}_{\rm proj}=\int_0^{\chi_{\rm s}} d\chi\, 4 r_{\rm F}^2\bar{n}
=V_{\rm F}\bar{n},
\end{equation}
where the origin of the prefactor 4 in the right hand size is
discussed in Appendix~\ref{app:shot}.
We find 
$V_{\rm F}\approx 3.70\times 10^{-9}(f/1~{\rm Hz})^{-1}(h^{-1}{\rm Mpc})^3$ for 
$z_{\rm s}=1$ and 
$V_{\rm F}\approx 1.36\times 10^{-8}(f/1~{\rm Hz})^{-1}(h^{-1}{\rm Mpc})^3$ for 
$z_{\rm s}=3$.
On the other hand, the number density of dark low-mass halos with 
$1h^{-1}M_\odot \la M \la 10^4h^{-1}M_\odot$ is 
$\sim 10^8(h^{-1}{\rm Mpc})^{-3}$. Thus we expect a moderately skewed
distribution for gravitational lensing dispersions at $f\sim 0.1-1$~Hz
by dark low-mass halos at least when $z_{\rm s}\la 3$. Since the 
volume $V_{\rm F}$ increases with increasing source redshift $z_{\rm s}$,
the distribution should approach to the Gaussian distribution for very
high source redshift $z_{\rm s}$.

In the case of gravitational lensing dispersions produced by the shot
noise, we can use e.g., equation~(\ref{eq:nbar_pbh}) to estimate the
average number $\bar{N}_{\rm proj}$, e.g., 
\begin{equation}
\bar{N}_{\rm proj}\approx 270 \left(\frac{f}{1~{\rm Hz}}\right)^{-1}
f_{\rm PBH}\left(\frac{M_{\rm PBH}}{1h^{-1}M_\odot}\right)^{-1},
\end{equation}
for $z_{\rm s}=1$, and
\begin{equation}
\bar{N}_{\rm proj}\approx 980 \left(\frac{f}{1~{\rm Hz}}\right)^{-1}
f_{\rm PBH}\left(\frac{M_{\rm PBH}}{1h^{-1}M_\odot}\right)^{-1},
\end{equation}
for $z_{\rm s}=3$. Thus in the range of parameters we examine in this
paper there are both cases with $\bar{N}_{\rm proj}>1$ and 
$\bar{N}_{\rm proj}<1$. We note that when $\bar{N}_{\rm proj}\ll1$ 
the situation is close to the one considered by 
\citet{zumalacarregui18} in which the non-Gaussian magnification
probability distribution function due to gravitational lensing by
single PBH masses is used to constrain the abundance of PBHs.

For both dark low-mass halos and PBHs, there are cases when single
events dominate the signal depending on frequencies of gravitational
waves and the mass of lenses. In such cases, it may be possible to
detect individual weak lensing events directly by making use of the 
wavelength dependence of the signal. For instance, the amplitude is
affected by weak lensing as Equations~(\ref{eq:kappa_s_wave}) and
(\ref{eq:kappa_s_wave_ano}), which indicates that weak lensing may
be detected via the modulation of the amplitude as a function of
frequency that is proportional to $w\propto f$ (for PBH) if the
signal-to-noise ratio of the gravitational wave observation is
sufficiently high (see also Section~\ref{sec:detectability}). 
This possibility is partly studied in \citet{dai18b} assuming a
singular isothermal sphere as a lens and is worth exploring more in
various setups.

\section{Conclusion}
\label{sec:conclusion}

In this paper we have explored the possibility of using gravitational
lensing dispersions of gravitational waves to probe the matter power
spectrum at very high $k$ i.e., very small scales. For this purpose we
have analyzed the small scale behavior of the matter power spectrum
using the halo model, including effects of baryon and subhalos. We
have confirmed that our halo model predictions agree reasonably well
with results of IllustrisTNG cosmological hydrodynamical simulations.
Using this halo model that is built on well-studied halo properties
and the stellar mass--halo mass relation, we study the matter power
spectrum at $k>10^3h{\rm Mpc}^{-1}$ that has been poorly explored
before. We find that the matter power spectrum at 
$k\sim 10^6h{\rm Mpc}^{-1}$ is relatively insensitive to baryon effects
and is dominated by dark low-mass halos with 
$1h^{-1}M_\odot \la M \la 10^4h^{-1}M_\odot$. We have also found
that the matter power spectrum at $k\ga 10^5h{\rm Mpc}^{-1}$ can be
significantly enhanced by PBHs due to the enhanced halo formation as
well as the shot noise from PBHs.

Using the halo model power spectrum we have computed frequency
dependent gravitational lensing dispersions of gravitational waves. The
frequency dependence originates from the wave optics nature of the
propagation of gravitational waves. We have found that lensing
dispersions of the amplitude and phase of gravitational waves are
$\mathcal{O}(10^{-3})$ in the frequency range of $f\sim
10^{-3}-100$~Hz for source redshifts of $z_{\rm s}\sim 1-3$. 
In particular, the frequency range of $f\sim 0.1-1$~Hz is found to be
a window appropriate for detecting  dark low-mass halos with 
$1h^{-1}M_\odot \la M \la 10^4h^{-1}M_\odot$. PBHs with $M\gtrsim
0.1~M_\odot$ can enhance gravitational lensing dispersions more than
an order of magnitude, when they constitute a significant fraction of
dark matter. At the frequency range of $f\sim 10-100$~Hz, which
corresponds to frequencies of ground observations of gravitational
waves, gravitational lensing dispersions are dominated by the shot
noise from PBHs and therefore serve as a useful probe of PBHs. 

\section*{Acknowledgments}

This work was supported in part by World Premier International
Research Center Initiative (WPI Initiative), MEXT, Japan, and JSPS
KAKENHI Grant Number JP20H04725, JP20H04723, JP18K03693, and
JP17H01131. 

\appendix
\section{Halo Model Calculations}
\label{app:halo_model}

Here we summarize the derivations of the halo model power spectrum
both in the standard case (Section~\ref{sec:halo_standard}) and with
modifications including stellar components and subhalos
(Section~\ref{sec:halo_modified}). 

First, we derive the standard halo model power spectrum. 
We start with Equation~(\ref{eq:def_rhom_1}) and rewrite it as
\begin{equation}
\rho(\boldsymbol{x})=\sum_i \int dM \int d\boldsymbol{x}' 
\delta^{\rm D}(M-M_i)\delta^{\rm D}(\boldsymbol{x}'-\boldsymbol{x}_i)\, M\,
u\left(\boldsymbol{x}-\boldsymbol{x}'|M\right),
\label{eq:def_rhom_2}
\end{equation}
where $\delta^{\rm D}$ denotes the Dirac delta function. 
The halo mass function $dn/dM$ is given by
\begin{equation}
\frac{dn}{dM}=\left\langle 
\sum_i\delta^{\rm D}(M-M_i)\delta^{\rm D}(\boldsymbol{x}'-\boldsymbol{x}_i)
\right\rangle ,
\label{eq:def_massfunc}
\end{equation}
from which it is shown that
\begin{equation}
\left\langle\rho(\boldsymbol{x})\right\rangle
=\int dM\,M\frac{dn}{dM}=\bar{\rho},
\end{equation}
where $\bar{\rho}$ is the mean comoving density of the Universe. We
now consider density fluctuations. Their expressions in real and
Fourier spaces are given as 
\begin{equation}
\delta(\boldsymbol{x})=\frac{\rho(\boldsymbol{x})}{\bar{\rho}}-1,
\end{equation}
\begin{equation}
\delta(\boldsymbol{k})=\int d\boldsymbol{x}\,\delta (\boldsymbol{x})
\,e^{-i\boldsymbol{k}\cdot\boldsymbol{x}}.
\end{equation}
From Equation~(\ref{eq:def_rhom_2}), $\delta(\boldsymbol{k})$ is
calculated as
\begin{equation}
\delta(\boldsymbol{k})=\frac{1}{\bar{\rho}}\sum_i \int dM \int d\boldsymbol{x}\, 
\delta^{\rm D}(M-M_i)\delta^{\rm D}(\boldsymbol{x}-\boldsymbol{x}_i)\, M\,
u\left(\boldsymbol{k}|M\right)\,e^{-i\boldsymbol{k}\cdot\boldsymbol{x}},
\label{eq:def_deltak}
\end{equation}
where $u\left(\boldsymbol{k}|M\right)=u\left(k|M\right)$ assuming a
statistically spherical symmetric halo shape is the Fourier transform
of the normalized density profile $u(\boldsymbol{x}|m)$. From this
expression, we compute the correlation of $\delta(\boldsymbol{k})$
\begin{equation}
\langle\delta(\boldsymbol{k})\delta(\boldsymbol{k}')\rangle
=V\delta^{\rm D}(\boldsymbol{k}+\boldsymbol{k}')P(k),
\label{eq:comp_ps}
\end{equation}
from which the power spectrum is computed as
\begin{eqnarray}
P(k)&=&P^{\rm 1h}(k)+P^{\rm 2h}(k)\nonumber\\
&=&\int dM\frac{dn}{dM}\left(\frac{M}{\bar{\rho}}\right)^2u^2(k|M)
+\int dM_1\frac{dn}{dM_1}\left(\frac{M_1}{\bar{\rho}}\right) u(k|M_1)
\int dM_2\frac{dn}{dM_2}\left(\frac{M_2}{\bar{\rho}}\right) u(k|M_2)
\,P_{\rm hh}(k|M_1, M_2),
\end{eqnarray}
where 
\begin{equation}
\left\langle 
\sum_{i,j}\delta^{\rm D}(M-M_i)(M'-M_j)\delta^{\rm D}(\boldsymbol{x}-\boldsymbol{x}_i)
\delta^{\rm D}(\boldsymbol{x}'-\boldsymbol{x}_j)
\right\rangle=
\frac{dn}{dm_1}\frac{dn}{dm_2}\xi_{\rm hh}(|\boldsymbol{x}-\boldsymbol{x}'||M_1, M_2),
\end{equation}
$\xi_{\rm hh}$ is the halo-halo correlation function, and $P_{\rm hh}$
is its Fourier counterpart. In what follows we simply assume a linear
halo bias 
\begin{equation}
P_{\rm hh}(k|M_1, M_2)=b(M_1)b(M_2)P_{\rm lin}(k),
\end{equation}
where $P_{\rm lin}(k)$ is the linear matter power spectrum.
In this case the 2-halo term reduces to Equation~(\ref{eq:pk_2h}).

Next we consider modifications of 1-halo term. Starting from 
Equation~(\ref{eq:def_rhom_mod}), the Fourier transform of the density
fluctuation is written as
\begin{eqnarray}
\delta(\boldsymbol{k})&=&\frac{1}{\bar{\rho}}\sum_i \int dM \int d\boldsymbol{x}\,
\delta^{\rm D}(M-M_i)\delta^{\rm D}(\boldsymbol{x}-\boldsymbol{x}_i)
(1-f_{\rm s})M\,u_{\rm
  h}\left(\boldsymbol{k}|M\right)\,e^{-i\boldsymbol{k}\cdot\boldsymbol{x}}\nonumber\\
&&+\frac{1}{\bar{\rho}}\sum_{i,j}\int dM \int dm\int d\boldsymbol{x} \int d\boldsymbol{x}' 
\delta^{\rm D}(M-M_i)\delta^{\rm D}(m-m_j)\delta^{\rm D}(\boldsymbol{x}-\boldsymbol{x}_i)
\delta^{\rm D}(\boldsymbol{x}'-\boldsymbol{x}_j)
m\,u_{\rm sub}\left(\boldsymbol{k}|M,m\right)\,e^{-i\boldsymbol{k}\cdot\boldsymbol{x}'},
\label{eq:def_deltak_mod}
\end{eqnarray}
where we ignored the dependence of $u_{\rm sub}$ on the position 
within a halo i.e., 
$u_{\rm sub}(\boldsymbol{x}-\boldsymbol{x}_j|M_i,m_j,\boldsymbol{x}_j-\boldsymbol{x}_i)=
u_{\rm sub} (\boldsymbol{x}-\boldsymbol{x}_j|M_i,m_j)$. 
We also need to specify the subhalo mass function $dN_i/dm$ and their
spatial distribution  $U(\boldsymbol{x}-\boldsymbol{x}_i|M_i,m)$
within the $i$-th halo, which is given in a manner similar to
Equation~(\ref{eq:def_massfunc}) as 
\begin{equation}
\frac{dN_i}{dm}U(\boldsymbol{x}'-\boldsymbol{x}_i|M_i, m)=\left.\left\langle 
\sum_j\delta^{\rm D}(m-m_j)\delta^{\rm D}(\boldsymbol{x}'-\boldsymbol{x}_j)
\right\rangle\right|_i,
\end{equation}
where they satisfy
\begin{equation}
\int dm\,m\frac{dN_i}{dm}=f_{\rm s}M_i,
\end{equation}
\begin{equation}
\int d\boldsymbol{x}\,U(\boldsymbol{x}-\boldsymbol{x}_i|M_i,m)=1.
\end{equation}
From these relations, it can be easily shown that 
$\left\langle\rho(\boldsymbol{x})\right\rangle=\bar{\rho}$ also for
this modified 1-halo case. From Equations~(\ref{eq:comp_ps}) and
(\ref{eq:def_deltak_mod}), we can derive the 1-halo power spectrum 
as Equation~(\ref{eq:pk_1h_mod}).

As discussed in Section~\ref{sec:halo_modified}, the shot noise from
stars can become important at very small scales. The effect is
evaluated by replacing $u_*(\boldsymbol{x}|M)$ as  
\begin{equation}
u_*(\boldsymbol{x}|M) \rightarrow 
u_*^{\rm s}(\boldsymbol{x}|M)=
\frac{1}{N_*}\sum_j 
\delta^{\rm D}(\boldsymbol{x}-\boldsymbol{x}_j),
\end{equation}
where for simplicity we assume that all stars share the same mass
$m_{\rm star}$ and $N_*=f_*M/m_{\rm star}$ denotes the total number of
stars in each halo with mass $M$. The following relation
\begin{equation}
N_*u_*(\boldsymbol{x}|M)=
\left\langle\sum_j 
\delta^{\rm D}(\boldsymbol{x}-\boldsymbol{x}_j)\right\rangle,
\end{equation}
ensures that $\langle u_*^{\rm s}(\boldsymbol{x}|M) \rangle=u_*(\boldsymbol{x}|M)$.
The Fourier transform of $u_*$ is modified as
\begin{equation}
u_*(\boldsymbol{k}|M)\rightarrow 
u_*^{\rm s}(\boldsymbol{k}|M)=
\frac{1}{N_*}\int d\boldsymbol{x}
\sum_j \delta^{\rm D}(\boldsymbol{x}-\boldsymbol{x}_j) 
e^{-i\boldsymbol{k}\cdot\boldsymbol{x}},
\end{equation}
from which we obtain the effect of the shot noise as
\begin{equation}
\left\langle u_*^{\rm s}(\boldsymbol{k}|M)
u_*^{\rm s}(-\boldsymbol{k}|M)\right\rangle 
=u_*(\boldsymbol{k}|M)
u_*(-\boldsymbol{k}|M)+\frac{1}{N_*}.
\label{eq:pk_shot_noise}
\end{equation}
That is, we can simply replace $u_*^2(k|M)$ with 
$u_*^2(k|M)+1/N_*$  to include the shot noise effect.

\section{A Simple Analytic Model of Subhalos}
\label{app:subhalo_model}

Analytic models of subhalos have been proposed in the literature
\citep[e.g.,][]{lee04,oguri04,vandenbosch05,giocoli08a,giocoli08b,han16,jiang16,hiroshima18,ando19},
in which important physical effects such as tidal stripping are taken
into account. Here we present a new analytic model of subhalos 
partly following \citet{oguri04} in which both tidal stripping and
dynamical frictions are taken into account. We keep this model as
simple as possible so that it can easily be computed numerically.

Following previous work we base our analytic model on the extended
Press-Schechter theory \citep{bond91,bower91,lacey93}, which
predicts the number distribution of progenitors with mass $m_{\rm f}$
at redshift $z_{\rm f}$ for a halo with mass $M$ and redshift $z$ as
\begin{equation}
\frac{dN_{\rm EPS}}{dm_{\rm f}}=\frac{M}{m_{\rm f}}P(m_{\rm f}, z_{\rm
  f}|M, z)dm_{\rm f},
\label{eq:n_eps}
\end{equation}
\begin{equation}
P(m_{\rm f}, z_{\rm f}|M, z)=\frac{1}{\sqrt{2\pi}}
\frac{\Delta\omega}
{\Delta S^{3/2}}
\exp\left(-\frac{\Delta\omega^2}{2\Delta S}\right)
\left|\frac{d\Delta S}{dm_{\rm f}}\right|,
\end{equation}
with $\Delta\omega=\delta_{\rm c}(z_{\rm f})-\delta_{\rm c}(z)$ and 
$\Delta S=\sigma^2(m_{\rm f})-\sigma^2(M)$. We adopt 
$\delta_{\rm c}(z)=(3/20)(12\pi)^{2/3}\left\{\Omega_{\rm m}(z)\right\}^{0.0055}/D_+(z)$
\citep{navarro97} with $D_+(z)$ being the linear growth rate. The
square root of the mass variance $\sigma(M)$ is computed in the
standard way by integrating the linear matter power spectrum with a
top-hat filter.  

We evaluate Equation~(\ref{eq:n_eps}) at the median formation time of each
halo. Following \citet{giocoli07}, we derive the median formation time
by solving the following equation
\begin{equation}
\delta_{\rm c}(z_{\rm f})=\delta_{\rm c}(z)+\frac{0.974}{\sqrt{q}}
\sqrt{\sigma^2(f_{\rm f}M)-\sigma^2(M)},
\end{equation}
where $q=0.707$ and $f_{\rm f}=0.5$.

We consider the effect of mass loss due to tidal stripping. We connect
$m_{\rm f}$ and $m$ that refer to subhalo masses before and after
tidal stripping as follows
\begin{equation}
m=m_{\rm f}\frac{f_{\rm BMO}(\tau^{\rm ave})}{f_{\rm NFW}\left(c(m_{\rm f},z_{\rm f})\right)},
\end{equation}
\begin{equation}
\tau^{\rm ave}=\frac{r_{\rm t}^{\rm ave}(m,f_{\rm f}M,z_{\rm
    f})}{r_s(m_{\rm f},c,z_{\rm f})},
\end{equation}
\begin{equation}
r_{\rm t}^{\rm ave}(m,M,z)=\int 4\pi r^2dr\, r_{\rm t}(r,m,M) U(r|M,m) 
=\int 4\pi r^2dr\, r\left[\frac{m}{3M(<r)}\right]^{1/3}U(r|M,m) ,
\end{equation}
where $U(r|M,m)$ is the spatial distribution of subhalos and $M(<r)$
is an enclosed mass of the host halo, both of which we compute using
the BMO profile.

We also take account of the dynamical friction. We adopt the following
crude approximation of the dynamical friction timescale
\citep[e.g.,][]{mo10} 
\begin{equation}
t_{\rm df}(m,M,z)=2\frac{M}{m}\frac{r_{\rm vir}}{V_{\rm vir}(M)},
\end{equation}
where $V_{\rm vir}(M)=\sqrt{GM/r_{\rm vir}}$. The prefactor of 2 is 
introduced to better reproduce the numerical results. 
We assume that the subhalo mass function is suppressed by the
following factor 
\begin{equation}
f_{\rm df}=
\exp\left[-
\left\{\frac{t(z)-t(z_{\rm f})}{t_{\rm df}(m_{\rm f},f_{\rm f}M,z_{\rm f})}\right\}^2\right].
\end{equation}
 
\begin{figure}[t]
\begin{center}
 \includegraphics[width=0.49\hsize]{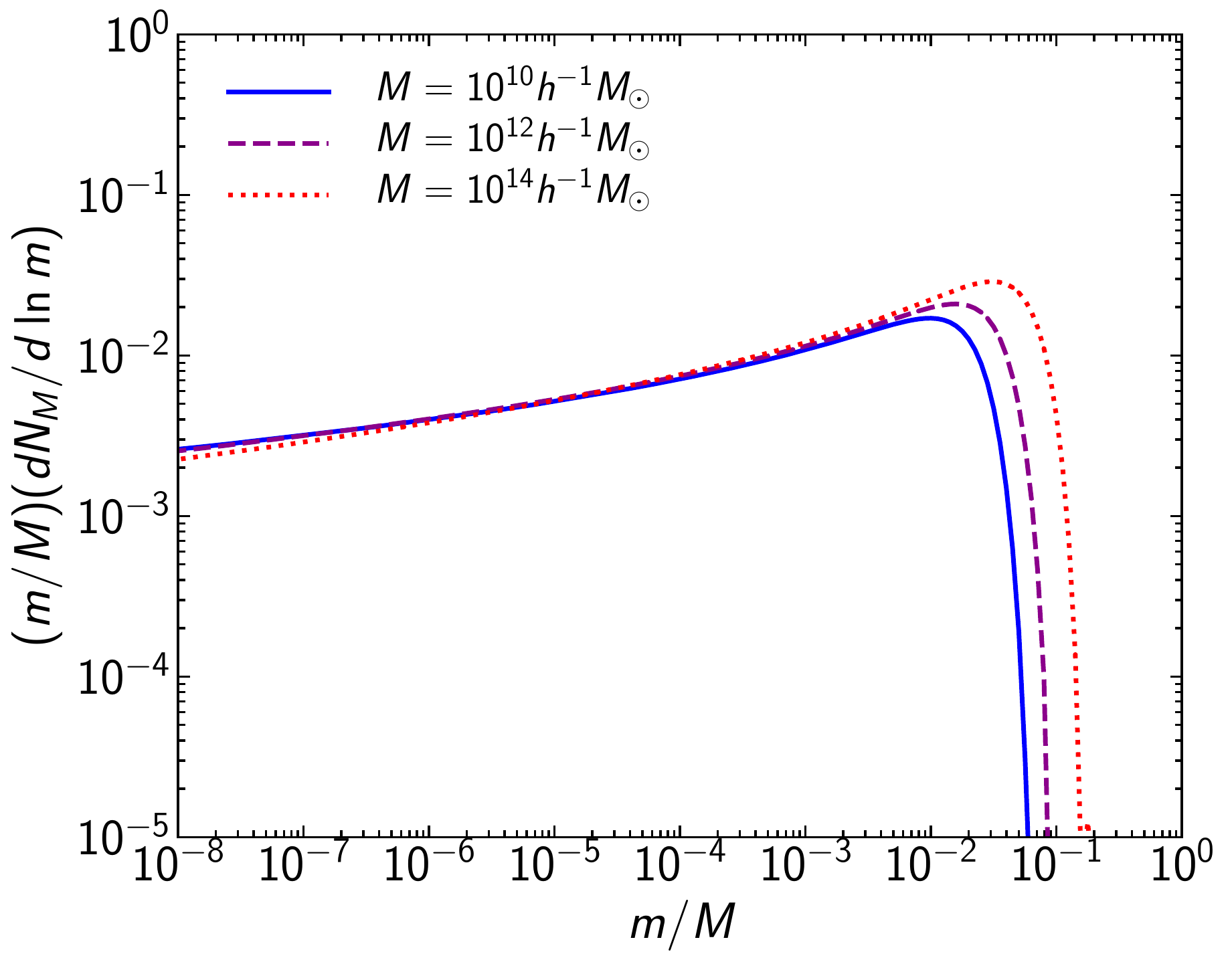}
 \includegraphics[width=0.46\hsize]{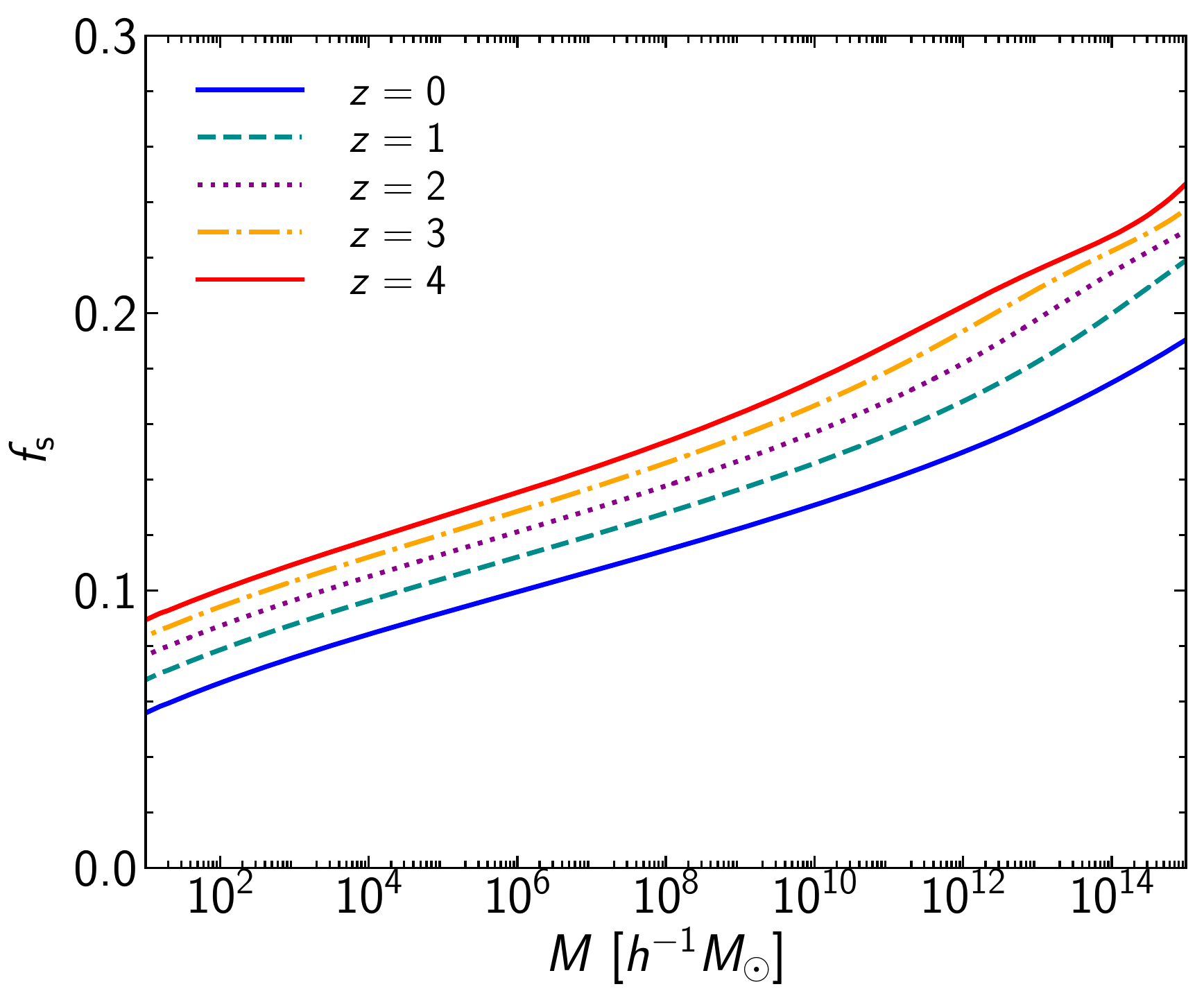}
\end{center}
\caption{{\it Left:} Examples of subhalo mass functions as a function
  of the subhalo mass $m$ at redshift $z=0$ for three host halo masses
  $M$. {\it Right:} Subhalo mass fractions $f_{\rm s}$ as a function
  of the host halo mass $M$ at several different redshifts.
\label{fig:mf_sub}}
\end{figure}

Finally we combine these results to compute the subhalo mass function as
\begin{equation}
\frac{dn_M}{dm}=f_{\rm df}\frac{dN_{\rm EPS}}{dm_{\rm f}}\frac{dm_{\rm f}}{dm}.
\end{equation}
We model the density profile of subhalos by the BMO profile with the
truncation radius $r_{\rm t}^{\rm ave}(m,f_{\rm f}M,z_{\rm f})$ by
that computed above. For an accurate prediction of the scale radius of
each subhalo, we use a fitting form of the concentration
parameter for subhalos $c_{\rm sub}$ by 
\citet[][see also \citealt{moline17,ishiyama20}]{ando19}. In order to
convert $c_{200}$ to $c_{\rm vir}$ we multiply it by
$[200/\Delta_{\rm vir}\Omega_{\rm m}(z)]^{1/3}[H(z)/H_0]^{-2/3}$,
where $\Delta_{\rm vir}$ is the virial overdensity computed from the
spherical collapse model
\citep{nakamura97}. Figure~\ref{fig:mf_sub} shows examples of subhalo
mass functions $dN_M/dm$ as well as subhalo mass fractions $f_{\rm s}$.

\section{Weak Gravitational Lensing in Wave Optics}
\label{app:lensing_wave}

Here we summarize weak gravitational lensing in wave optics, following
and extending work by \citet{takahashi05} and \citet{takahashi06}.
Under the Born approximation, the observed wave at comoving spatial
coordinate $\boldsymbol{x}$ with comoving frequency $f$ in the
presence of gravitational potential 
$\Phi(\boldsymbol{x})$ is 
\begin{equation}
\phi(f,\boldsymbol{x})=\phi^0(f,\boldsymbol{x})+\delta\phi(f,\boldsymbol{x}),
\end{equation}
\begin{equation}
\delta\phi(f,\boldsymbol{x})=-\frac{4\pi f^2}{c^4}\int d\boldsymbol{x}'\frac{e^{2\pi i f|\boldsymbol{x}-\boldsymbol{x}'|/c}}{|\boldsymbol{x}-\boldsymbol{x}'|}\Phi(\boldsymbol{x}')\phi^0(f,\boldsymbol{x}'),
\end{equation}
where $\phi^0(f,\boldsymbol{x})$ denotes the solution with
$\Phi(\boldsymbol{x})=0$. Here we adopt a spherical coordinate in a
flat Universe, $\boldsymbol{x}$=($\chi$, $\boldsymbol{r}$),
centered at the observer with a flat sky approximation, and assume
that a spherical wave is emitted from a source at 
$\boldsymbol{x}_{\rm s}$. Then $\phi^0(f,\boldsymbol{x})$ is given by
\begin{equation}
\phi^0(f,\boldsymbol{x})=
\frac{A e^{2\pi i f|\boldsymbol{x}-\boldsymbol{x}_{\rm s}|/c}}
{|\boldsymbol{x}-\boldsymbol{x}_{\rm s}|}. 
\end{equation}
Setting $\boldsymbol{x}_{\rm s}$=($\chi_{\rm s}$, $0$) and assuming
$|\boldsymbol{r}|\ll\chi$, the observed wave at $\boldsymbol{x}=0$ 
(i.e., $\delta\phi_{\rm obs}(f)=\delta\phi(f,0)$) is
calculated as 
\begin{equation}
\delta\phi_{\rm obs}(f)=-\frac{4\pi f^2 A}{c^4}\int d\chi\int d\boldsymbol{r}
\frac{\Phi(\chi,\boldsymbol{r})}{\chi(\chi_{\rm s}-\chi)}e^{2\pi i f \chi_{\rm s}/c}
e^{2\pi i f \Delta t(\chi, \boldsymbol{r})},
\label{eq:phiobs_calc1}
\end{equation}
\begin{equation}
\Delta t(\chi, \boldsymbol{r})=\frac{\chi_{\rm s}}{c\chi(\chi_{\rm
    s}-\chi)}
\frac{|\boldsymbol{r}|^2}{2}.
\end{equation}
We then consider the Fourier transform of the gravitational potential 
\begin{equation}
\Phi(\chi,\boldsymbol{r})
=\int \frac{dk_\parallel}{2\pi}\int\frac{d\boldsymbol{k}_\perp}{(2\pi)^2}\Phi(k_\parallel,\boldsymbol{k}_\perp)e^{ik_\parallel\chi+i\boldsymbol{k}_\perp\cdot\boldsymbol{r}}.
\end{equation}
Inserting this expression and using 
$\int d\boldsymbol{r}e^{i|\boldsymbol{r}|^2}=i\pi$, we obtain 
\begin{equation}
\frac{\delta\phi_{\rm obs}(f)}{\phi_{\rm obs}^0(f)}=-\frac{4\pi i f}{c^3}
\int d\chi \int \frac{dk_\parallel}{2\pi}\int
\frac{d\boldsymbol{k}_\perp}{(2\pi)^2}\Phi(k_\parallel,\boldsymbol{k}_\perp)
\exp\left[ik_\parallel\chi -i\frac{\chi(\chi_{\rm s}-\chi)}
{4\pi f \chi_{\rm s}/c}|\boldsymbol{k}_\perp|^2\right].
\label{eq:delphi}
\end{equation}
We now consider the high-frequency limit ($f\rightarrow \infty$)  that
corresponds to the geometric optics limit. Using the following
approximation 
\begin{equation}
\exp\left[-i\frac{\chi(\chi_{\rm s}-\chi)}
{4\pi f \chi_{\rm s}/c}|\boldsymbol{k}_\perp|^2\right] 
\simeq
1-i\frac{\chi(\chi_{\rm s}-\chi)}
{4\pi f \chi_{\rm s}/c}|\boldsymbol{k}_\perp|^2,
\end{equation}
we obtain 
\begin{equation}
\frac{\delta\phi_{\rm obs}(f)}{\phi_{\rm obs}^0(f)}\simeq
-\frac{4\pi i f}{c^3} \int d\chi\,\Phi(\chi,0)
+\frac{1}{c^2}\int d\chi \frac{\chi(\chi_{\rm s}-\chi)}{\chi_{\rm s}}
\left.\Delta_{\boldsymbol{r}}\Phi\right|_{\boldsymbol{r}=0}.
\label{eq:phiobs_calc2}
\end{equation}
The first term of Equation~(\ref{eq:phiobs_calc2}) represents a phase
shift due to gravitational time delay 
$\Delta t_{\rm g}=-(2/c^3)\int d\chi\,\Phi(\chi,0)$, whereas the
second term of coincides with convergence $\kappa$. Thus we can
rewrite Equation~(\ref{eq:phiobs_calc2}) as 
\begin{equation}
\frac{\phi_{\rm obs}(f)}{\phi_{\rm obs}^0(f)}\simeq\left(1+2\pi i
f\Delta t_{\rm g}+\kappa\right)
\simeq \left(1+\kappa\right)e^{2\pi i f\Delta t_{\rm
  g}}.
\end{equation}
More generally, if we define
\begin{equation}
K(f)={\rm Re}\left[\frac{\delta\phi_{\rm obs}(f)}{\phi_{\rm obs}^0(f)}\right],
\end{equation}
\begin{equation}
S(f)={\rm Im}\left[\frac{\delta\phi_{\rm obs}(f)}{\phi_{\rm obs}^0(f)}\right],
\end{equation}
we have 
\begin{equation}
\frac{\phi_{\rm obs}(f)}{\phi_{\rm obs}^0(f)}\simeq \left[1+K(f)\right]e^{iS(f)}.
\end{equation}
Note that  we have a freedom to change the origin of time (intrinsic
phase, which is unobservable) at the source such that $\phi_{\rm
  obs}^0(f)\rightarrow \phi_{\rm obs}^0(f)e^{2\pi i f t_0}$ so that
$\Delta t_{\rm g}$ is unobservable. However given a complex
dependence on $f$ in general the effect of $S(f)$ may be observed.

The limit $f\rightarrow \infty$ corresponds to the situation that only
light paths around $\boldsymbol{r}=0$ (solution in the geometric
optics limit given the Born approximation) contribute. To see this,
we Taylor-expand the gravitational potential
\begin{equation}
\Phi(\chi,\boldsymbol{r})=\Phi(\chi,0)+\frac{r_1^2}{2}\left.\frac{\partial^2\Phi}{\partial
  r_1^2}\right|_{\boldsymbol{r}=0}+\frac{r_2^2}{2}\left.\frac{\partial^2\Phi}{\partial
  r_2^2}\right|_{\boldsymbol{r}=0}+\mathcal{O}(r^3),
\end{equation}
where $\boldsymbol{r}=(r_1,\,r_2)$ and terms that disappear after the
integration in Equation~(\ref{eq:phiobs_calc1}) are not
shown. Inserting this expression to Equation~(\ref{eq:phiobs_calc1}),
we obtain
\begin{equation}
\frac{\delta\phi_{\rm obs}(f)}{\phi_{\rm obs}^0(f)}=-\frac{4\pi
  f^2}{c^4}\int d\chi\int d\boldsymbol{r}
\frac{\chi_{\rm s}\Phi(\chi,\boldsymbol{r})}{\chi(\chi_{\rm s}-\chi)}
e^{2\pi i f \Delta t(\chi, \boldsymbol{r})}
= -\frac{4\pi i f}{c^3} \int d\chi\,\Phi(\chi,0)
+\frac{1}{c^2}\int d\chi \frac{\chi(\chi_{\rm s}-\chi)}{\chi_{\rm s}}
\left.\Delta_{\boldsymbol{r}}\Phi\right|_{\boldsymbol{r}=0},
\label{eq:phiobs_calc3}
\end{equation}
where we used $\int_{-\infty}^\infty dx e^{ix^2}=\sqrt{i\pi}$
and $\int_{-\infty}^\infty dx x^2 e^{ix^2}=-\sqrt{i\pi}/(2i)$. 
We see that Equation~(\ref{eq:phiobs_calc3}) is same as
Equation~(\ref{eq:phiobs_calc2}). 

We now consider correlations of $K(f)$ and $S(f)$. Denoting 
$\eta=\delta\phi_{\rm obs}(f)/\phi_{\rm obs}^0(f)$, we have
\begin{equation}
\langle K^2(f)\rangle = \frac{1}{2}\left[\langle \eta\eta^*\rangle+{\rm
  Re}(\langle \eta^2\rangle)\right],
\end{equation}
\begin{equation}
\langle S^2(f)\rangle = \frac{1}{2}\left[\langle \eta\eta^*\rangle-{\rm
  Re}(\langle \eta^2\rangle)\right].
\end{equation}
Also the gravitational potential is related with density fluctuations
by the following Poisson equation 
\begin{equation}
-k^2\Phi(\boldsymbol{k})=
4\pi G\bar{\rho}a^{-1}\delta(\boldsymbol{k}),
\end{equation}
and the matter power spectrum is calculated as 
$\langle \delta(\boldsymbol{k})\delta(\boldsymbol{k}')\rangle=(2\pi)^3\delta^{\rm
  D}(\boldsymbol{k}+\boldsymbol{k}')P(k)$. 
Therefore,
\begin{equation}
\langle \Phi(\boldsymbol{k})\Phi
(\boldsymbol{k}')\rangle=(2\pi)^3\delta^{\rm
  D}(\boldsymbol{k}+\boldsymbol{k}')
\left(\frac{4\pi G\bar{\rho}a^{-1}}{k^2}\right)^2P(k).
\end{equation}
We first compute $\langle \eta\eta^*\rangle$ as
\begin{equation}
\langle \eta\eta^*\rangle =
\left(\frac{4\pi f}{c^3}\right)^2\int d\chi \int d\chi' \int \frac{dk_\parallel}{2\pi}\int
\frac{d\boldsymbol{k}_\perp}{(2\pi)^2}\left(\frac{4\pi G \bar{\rho}a^{-1}}{k^2}\right)^2
P(k)e^{ik_\parallel(\chi-\chi')}
\exp\left[-i\left\{\frac{\chi(\chi_{\rm s}-\chi)}
{4\pi f \chi_{\rm s}/c}-\frac{\chi'(\chi_{\rm s}-\chi')}
{4\pi f \chi_{\rm s}/c}\right\}|\boldsymbol{k}_\perp|^2\right].
\end{equation}
For any $g(k)$ that is a smooth function of $k$, we can use the
following Limber approximation
\begin{equation}
\int \frac{dk_\parallel}{2\pi} g(k) e^{ik_\parallel(\chi-\chi')}\simeq 
\delta^{\rm D}(\chi-\chi')g(|\boldsymbol{k}_\perp|),
\end{equation}
to simplify the expression above as
\begin{equation}
\langle \eta\eta^*\rangle = \left(\frac{4\pi f}{c^3}\right)^2\int d\chi \int
\frac{d\boldsymbol{k}_\perp}{(2\pi)^2}
\left(\frac{4\pi G \bar{\rho}a^{-1}}{|\boldsymbol{k}_\perp|^2}\right)^2
P(|\boldsymbol{k}_\perp|).
\end{equation}
We simplify this expression using the lensing weight function
$W(\chi)$ defined in Equation~(\ref{eq:wl_weight}) and
is the Fresnel scale $r_{\rm F}$ defined in Equation~(\ref{eq:r_fresnel})
\begin{equation}
\langle \eta\eta^*\rangle = \int d\chi W^2(\chi)\int
\frac{d\boldsymbol{k}_\perp}{(2\pi)^2}
\left(\frac{2}{r_{\rm F}^2|\boldsymbol{k}_\perp|^2}\right)^2
P(|\boldsymbol{k}_\perp|).
\end{equation}
Similarly, $\langle \eta^2\rangle$ is evaluated as
\begin{equation}
\langle \eta^2\rangle = -\int d\chi W^2(\chi)\int
\frac{d\boldsymbol{k}_\perp}{(2\pi)^2}
\left(\frac{2}{r_{\rm F}^2|\boldsymbol{k}_\perp|^2}\right)^2
e^{-i r_{\rm F}^2|\boldsymbol{k}_\perp|^2}
P(|\boldsymbol{k}_\perp|).
\end{equation}
Therefore, we obtain
\begin{equation}
\langle K^2(f)\rangle = \int d\chi W^2(\chi)\int
\frac{d\boldsymbol{k}_\perp}{(2\pi)^2}
\left[\frac{\sin(r_{\rm F}^2|\boldsymbol{k}_\perp|^2/2)}
{r_{\rm F}^2|\boldsymbol{k}_\perp|^2/2}\right]^2
P(|\boldsymbol{k}_\perp|),
\label{eq:def_cor_k2}
\end{equation}
\begin{equation}
\langle S^2(f)\rangle = \int d\chi W^2(\chi)\int
\frac{d\boldsymbol{k}_\perp}{(2\pi)^2}
\left[\frac{\cos(r_{\rm F}^2|\boldsymbol{k}_\perp|^2/2)}
{r_{\rm F}^2|\boldsymbol{k}_\perp|^2/2}\right]^2
P(|\boldsymbol{k}_\perp|).
\label{eq:def_cor_s2}
\end{equation}
As discussed above, we have a freedom to shift the origin of time. To
see this point, we subtract the phase shift in the geometric optics
limit ($f\rightarrow \infty$)
\begin{equation}
\hat{\eta}=\eta-2\pi i f \Delta t_{\rm g}
=\eta+\frac{4\pi i f}{c^3} \int d\chi\int \frac{dk_\parallel}{2\pi}
\int\frac{d\boldsymbol{k}_\perp}{(2\pi)^2}\Phi(k_\parallel,\boldsymbol{k}_\perp)e^{ik_\parallel\chi}.
\end{equation}
Specifically, from Equation~(\ref{eq:delphi}) $\hat{\eta}$ is
written as
\begin{equation}
\hat{\eta}=-\frac{4\pi i f}{c^3} \int
d\chi \int \frac{dk_\parallel}{2\pi}\int
\frac{d\boldsymbol{k}_\perp}{(2\pi)^2}
\Phi(k_\parallel,\boldsymbol{k}_\perp)
e^{ik_\parallel\chi}\left(e^{-ir_{\rm F}^2|\boldsymbol{k}_\perp|^2/2}-1\right).
\label{eq:delphi_hat}
\end{equation}
In this case,
\begin{equation}
\langle \hat{\eta}\hat{\eta}^*\rangle = \int d\chi W^2(\chi)\int
\frac{d\boldsymbol{k}_\perp}{(2\pi)^2}
\left(\frac{2}{r_{\rm F}^2|\boldsymbol{k}_\perp|^2}\right)^2P(|\boldsymbol{k}_\perp|)\nonumber\\
\left[2-2\cos(r_{\rm F}^2|\boldsymbol{k}_\perp|^2/2)\right],
\end{equation}
\begin{equation}
{\rm Re}(\langle \hat{\eta}^2\rangle) = -\int d\chi W^2(\chi)\int
\frac{d\boldsymbol{k}_\perp}{(2\pi)^2}
\left(\frac{2}{r_{\rm F}^2|\boldsymbol{k}_\perp|^2}\right)^2
P(|\boldsymbol{k}_\perp|)\left[\cos(r_{\rm F}^2|\boldsymbol{k}_\perp|^2)
+1-2\cos(r_{\rm F}^2|\boldsymbol{k}_\perp|^2/2)\right].
\end{equation}
Thus $\langle K^2(f)\rangle$ is unchanged from Equation~(\ref{eq:def_cor_k2})
but $\langle S^2(f)\rangle$ is modified to
\begin{equation}
\langle S^2(f)\rangle = \int d\chi W^2(\chi)\int
\frac{d\boldsymbol{k}_\perp}{(2\pi)^2}
\left[\frac{\cos(r_{\rm F}^2|\boldsymbol{k}_\perp|^2/2)-1}
{r_{\rm F}^2|\boldsymbol{k}_\perp|^2/2}\right]^2
P(|\boldsymbol{k}_\perp|).
\end{equation}

\begin{figure}
\begin{center}
 \includegraphics[width=0.32\hsize]{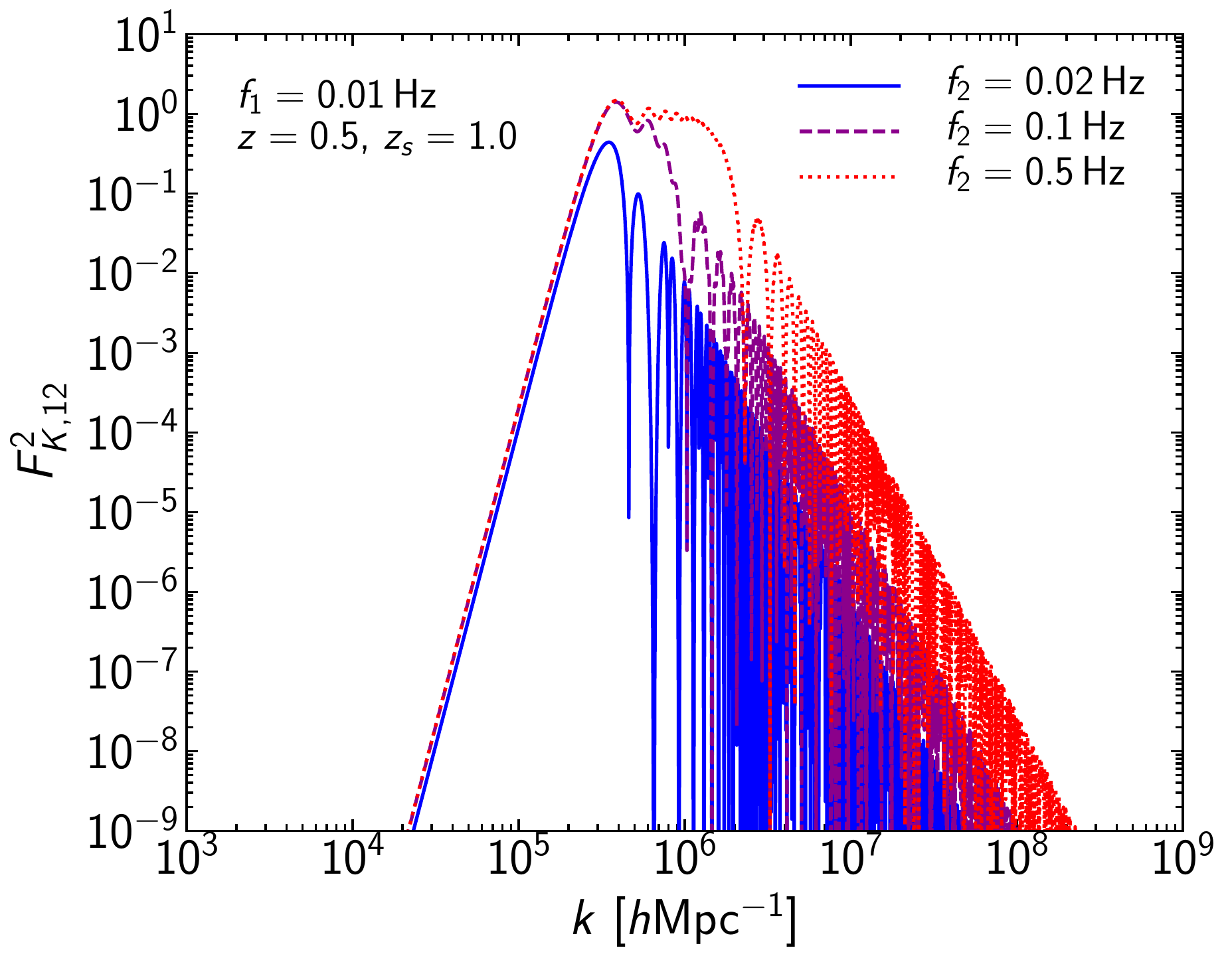}
 \includegraphics[width=0.32\hsize]{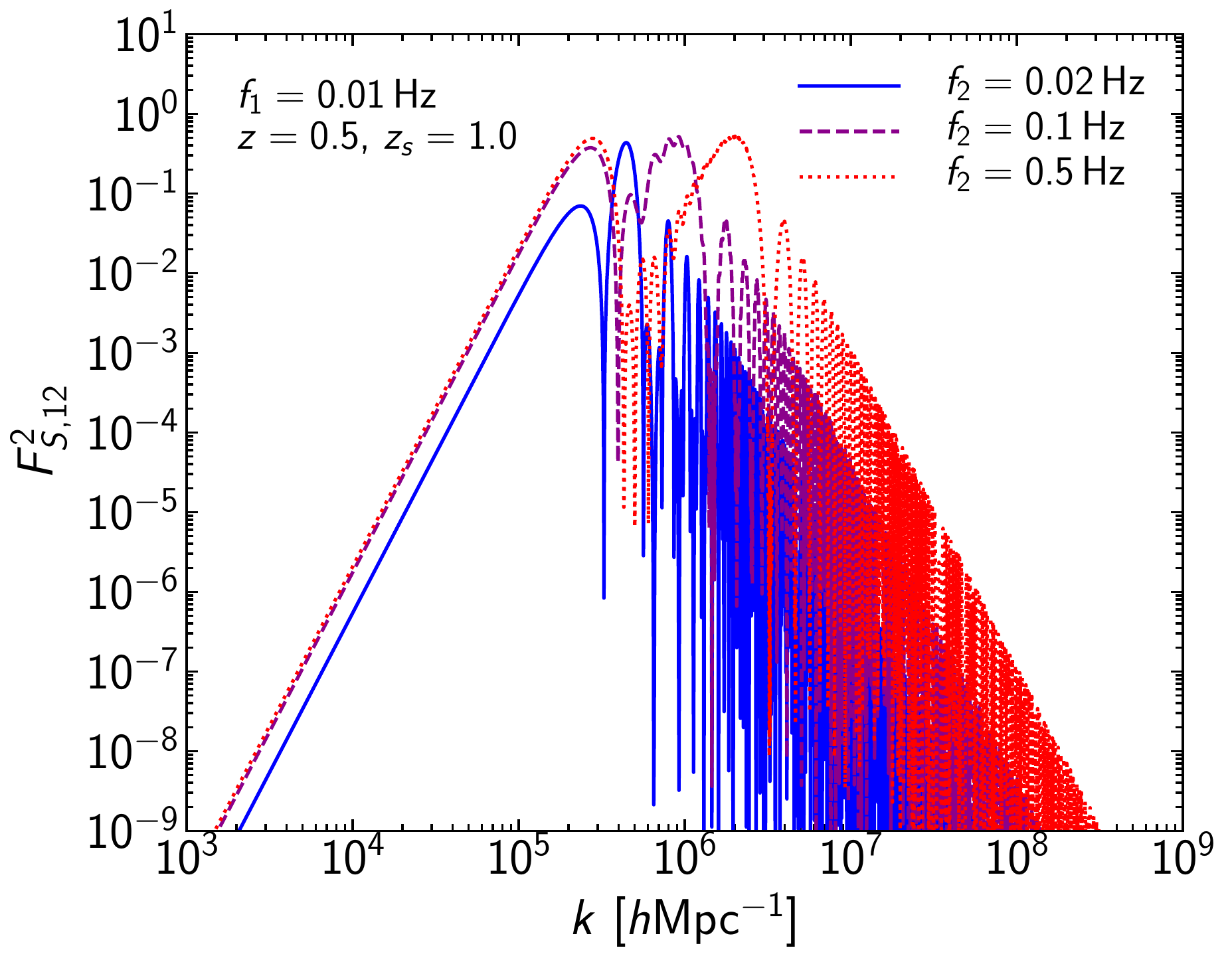}
 \includegraphics[width=0.32\hsize]{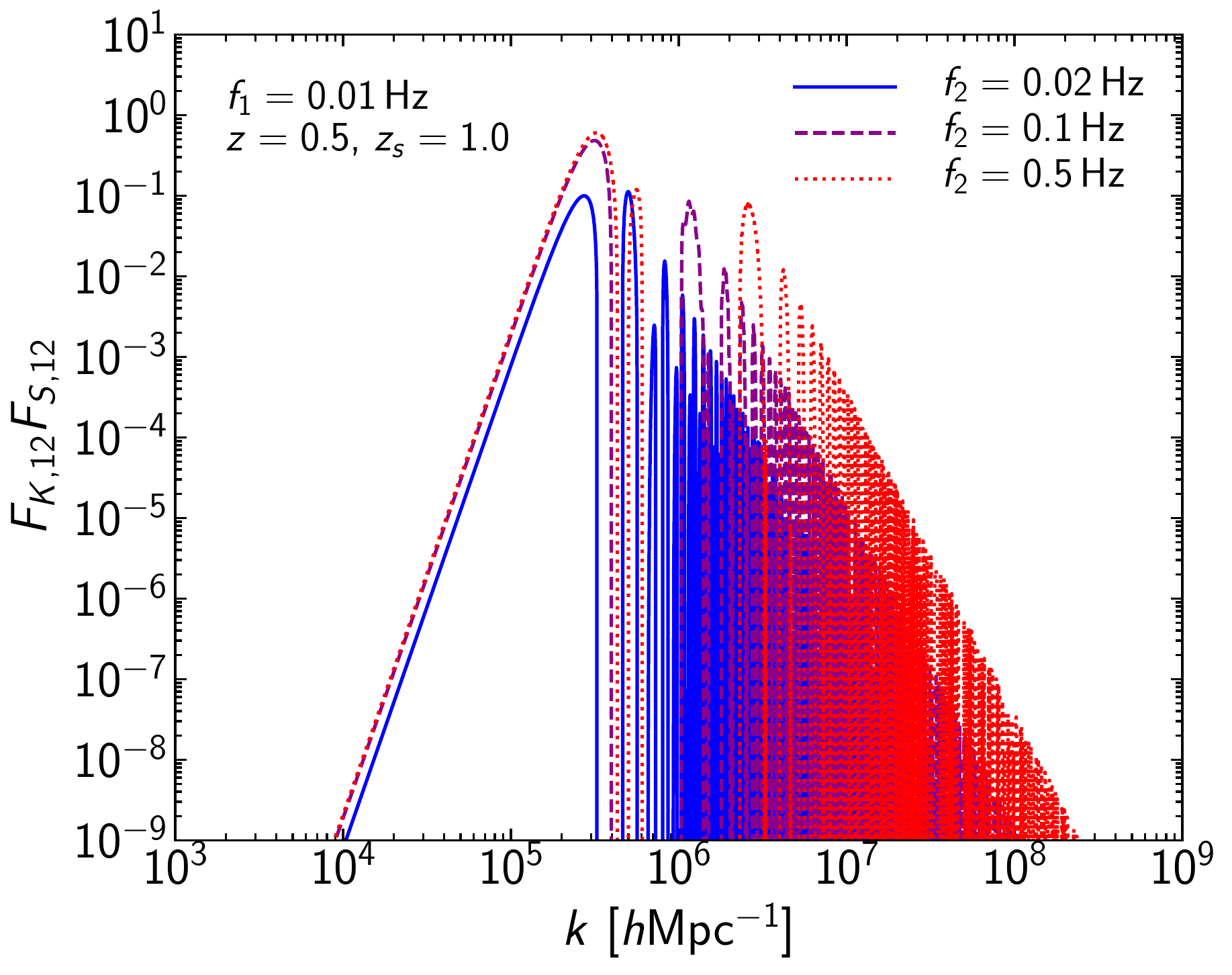}
\end{center}
\caption{Filter functions $F_{K,12}^2$ ({\it left}), $F_{S,12}^2$
  ({\it middle}), and $F_{K,12}F_{S,12}$  ({\it right}) used for 
 calculations of auto and cross lensing dispersions
 (equations~\ref{eq:def_k2_f12}, \ref{eq:def_s2_f12}, \ref{eq:def_ks_f12}).
 $F_{K,12}$ and $F_{S,12}$ are defined in Equations~(\ref{eq:filter_fk12})
 and (\ref{eq:filter_fs12}), respectively. In all panels, we fix
 redshift $z=0.5$ and $z_{\rm s}=1$, and frequency $f_1=0.01$~Hz, and
 consider three different frequency $f_2$, $f_2=0.02$~Hz 
  ({\it solid}), $0.1$~Hz ({\it dashed}), and $0.5$~Hz ({\it dotted}).
\label{fig:filter}}
\end{figure}

In practice, $K(f)$ and $S(f)$ have to be measured by comparing
signals at different frequencies. For example, in an ideal case where
the model waveform $h^{\rm model}(f)$ without gravitational lensing
effects is completely known the following ratio between frequencies
$f_1$ and $f_2$ reduces to
\begin{equation}
\frac{h(f_1)/h^{\rm model}(f_1)}{h(f_2)/h^{\rm model}(f_2)}\approx
1+K(f_1)-K(f_2)+i\left[S(f_1)-S(f_2)\right],
\end{equation}
from which we can measure auto and cross correlations of the
differences of $K(f)$ and $S(f)$. After some calculations, 
they are found to be given by Equations~(\ref{eq:def_k2_f12}), 
(\ref{eq:def_s2_f12}), and (\ref{eq:def_ks_f12}).
We show examples of filter functions used for these calculations 
in Figure~\ref{fig:filter}.

\section{Convergence and Magnification by the Shot Noise}
\label{app:shot}

We discuss the connection between lensing effects by individual
point mass lenses and the variance of the convergence computed by the
shot noise power spectrum. Specifically we consider a population of
point mass lenses with their individual mass $m_{\rm p}$, the comoving
number density $\bar{n}_{\rm p}$, and the mass fraction 
$m_{\rm p}\bar{n}_{\rm p}/\bar{\rho}=f_{\rm p}$.
The shot noise contribution to the convergence variance smoothed over
a circle with radius  $\beta_{\rm s}$ is
\begin{equation}
\langle\kappa^2_{\rm shot}\rangle=\int d\chi W^2(\chi)\int
\frac{k\,dk}{2\pi}\Delta P_{\rm shot}(k) W_{\rm s}^2(k\chi\beta_{\rm s}),
\end{equation}
where the lensing weight function $W(\chi)$ is defined in
Equation~(\ref{eq:wl_weight}) and $W_{\rm s}(x)$ is a smoothing kernel
defined in Equation~(\ref{eq:source_smooth}). The shot noise
contribution to the matter power spectrum is given by  
\begin{equation}
\Delta P_{\rm shot}(k)=\frac{f_{\rm p}^2}{\bar{n}_{\rm p}}.
\end{equation}
The lensing weight function is also rewritten as
\begin{equation}
W(\chi)
=\frac{4\pi G\bar{\rho}a^{-1}\chi(\chi_{\rm s}-\chi)}{c^2\chi_{\rm s}}
=\frac{\bar{\rho}a^{-2}}{\Sigma_{\rm crit}}
=\pi R_{\rm Ein}^2\frac{\bar{n}_{\rm p}}{f_{\rm p}},
\end{equation}
where the comoving Einstein radius $R_{\rm Ein}$ for a point mass lens
is given by
\begin{equation}
R_{\rm Ein}=\chi\theta_{\rm Ein}
=\frac{1}{a}\sqrt{\frac{m_{\rm p}}{\pi\Sigma_{\rm crit}}}.
\label{eq:r_ein}
\end{equation}
By combining these calculations, we obtain
\begin{equation}
\langle\kappa^2_{\rm shot}\rangle=\int d\chi \frac{(\pi R_{\rm
    Ein}^2)^2\bar{n}_{\rm p}}{\pi(\chi\beta_{\rm s})^2}=\sum \kappa_{\rm
  s}^2(\beta_{\rm s})\langle N_{\rm p}^2\rangle,
\end{equation}
where 
\begin{equation}
\kappa_{\rm s}(\beta_{\rm s})=\left(\frac{R_{\rm Ein}}{\chi\beta_{\rm s}}\right)^2,
\label{eq:kappa_smooth}
\end{equation}
is the smoothed convergence for each point mass lens and
\begin{equation}
\langle N_{\rm p}^2\rangle = \pi(\chi\beta_{\rm s})^2\bar{n}_{\rm p}\Delta\chi
\end{equation}
is the variance of the number of point mass lenses within 
$\beta_{\rm s}$, which are assumed to be randomly distributed, within
the radial distance slice $\Delta\chi$.  

The discussion above suggests that the weak lensing approximation
($\mu_{\rm s}\approx 1+2\kappa_{\rm s}$) breaks down at
$\chi\beta_{\rm s}\la R_{\rm Ein}$, where 
$R_{\rm Ein}\sim 10^{-8}$~Mpc for $m_{\rm p}=1~M_\odot$, $z_{\rm s}=1$
and $z=0.5$. 

For gravitational wave sources, the Fresnel scale $r_{\rm F}$ 
(equation~\ref{eq:r_fresnel}) should be interpreted as the effective
size of the source. Given the difference of the top-hat filter used
above and the filter used to define the convergence of gravitational
waves in Equation~(\ref{eq:def_k2}), we connect $\beta_{\rm s}$ with
$r_{\rm F}$ as 
\begin{equation}
\int\frac{k\,dk}{2\pi}F_K^2=\int\frac{k\,dk}{2\pi}W_{\rm s}^2(k\chi\beta_{\rm s}),
\end{equation}
where $F_K$ is defined in Equation~(\ref{eq:filter_fk}). By solving this
equation we find $(\chi\beta_{\rm s})^2/r_{\rm F}^2=4/\pi$. Thus the
smoothed convergence (equation~\ref{eq:kappa_smooth}) is rewritten as
\begin{equation}
\kappa_{\rm s}=\frac{\pi}{4}\left(\frac{R_{\rm Ein}}{r_{\rm F}}\right)^2
=\frac{\pi}{4}2\pi f (1+z)\frac{4Gm_{\rm p}}{c^3}=\frac{\pi}{4}w,
\label{eq:kappa_s_wave}
\end{equation}
where $w$ denotes the dimensionless parameter that controls the
wave optics effect in gravitational lensing \citep[see e.g.,][]{oguri19}.
The smoothed convergence can also be derived by directly evaluating 
$\delta\phi_{\rm obs}(f)/\phi_{\rm obs}^0(f)$ in the Born
approximation for the case of a point mass lens \citep{takahashi05}
\begin{equation}
\frac{\delta\phi_{\rm obs}(f)}{\phi_{\rm
    obs}^0(f)}=\frac{i}{2}w\left[{\rm Ci}
  \left(w\frac{R_\perp^2}{2R_{\rm Ein}^2}\right)+i\left\{{\rm Si}\left(w\frac{R_\perp^2}{2R_{\rm Ein}^2}\right)-\frac{\pi}{2}\right\}\right],
\end{equation}
where $R_\perp$ denote the comoving impact parameter at the lens. 
From this expression we can read off the smoothed convergence as
\begin{equation}
\kappa_{\rm s}={\rm Re}\left[\frac{\delta\phi_{\rm obs}(f)}{\phi_{\rm
    obs}^0(f)}\right]\approx \frac{\pi}{4}w \;\;\;(R_\perp\la r_{\rm
  F}),
\label{eq:kappa_s_wave_ano}
\end{equation}
which is consistent with Equation~(\ref{eq:kappa_s_wave}).

On the other hand, using the $w$ parameter, the magnification factor
of a point mass lens with the comoving impact parameter $R_\perp$ 
in the wave optics limit is described as \citep{deguchi86a,deguchi86b}
\begin{equation}
  \mu_{\rm s}=\frac{\pi w}{1-e^{-\pi w}}
\left|{}_1F_1\left(\frac{i}{2}w, 1;
\frac{i}{2}w\frac{R_\perp^2}{R_{\rm Ein}^2}\right)\right|^2
\approx 1+\frac{\pi}{2}w \;\;\;(w\ll 1,\,R_\perp\la r_{\rm F}),
\end{equation}
where ${}_1F_1$ is the confluent hypergeometric function. Hence the
weak lensing relation $\mu_{\rm s}\approx 1+2\kappa_{\rm s}$
holds also for the case of the wave optics lensing by a point mass
lens.

\bibliography{ref} 

\begin{thebibliography}{}
\expandafter\ifx\csname natexlab\endcsname\relax\def\natexlab#1{#1}\fi
\providecommand{\url}[1]{\href{#1}{#1}}
\providecommand{\dodoi}[1]{doi:~\href{http://doi.org/#1}{\nolinkurl{#1}}}
\providecommand{\doeprint}[1]{\href{http://ascl.net/#1}{\nolinkurl{http://ascl.net/#1}}}
\providecommand{\doarXiv}[1]{\href{https://arxiv.org/abs/#1}{\nolinkurl{https://arxiv.org/abs/#1}}}

\bibitem[{{Abbott} {et~al.}(2016){Abbott}, {Abbott}, {Abbott}, {Abernathy},
  {Acernese}, {Ackley}, {Adams}, {Adams}, {Addesso}, {Adhikari}, {Adya},
  {Affeldt}, {Agathos}, {Agatsuma}, {Aggarwal}, {Aguiar}, {Aiello}, {Ain},
  {Ajith}, {Allen}, {Allocca}, {Altin}, {Anderson}, {Anderson}, {Arai},
  {Arain}, {Araya}, {Arceneaux}, {Areeda}, {Arnaud}, {Arun}, {Ascenzi},
  {Ashton}, {Ast}, {Aston}, {Astone}, {Aufmuth}, {Aulbert}, {Babak}, {Bacon},
  {Bader}, {Baker}, {Baldaccini}, {Ballardin}, {Ballmer}, {Barayoga},
  {Barclay}, {Barish}, {Barker}, {Barone}, {Barr}, {Barsotti}, {Barsuglia},
  {Barta}, {Bartlett}, {Barton}, {Bartos}, {Bassiri}, {Basti}, {Batch},
  {Baune}, {Bavigadda}, {Bazzan}, {Behnke}, {Bejger}, {Belczynski}, {Bell},
  {Bell}, {Berger}, {Bergman}, {Bergmann}, {Berry}, {Bersanetti}, {Bertolini},
  {Betzwieser}, {Bhagwat}, {Bhandare}, {Bilenko}, {Billingsley}, {Birch},
  {Birney}, {Birnholtz}, {Biscans}, {Bisht}, {Bitossi}, {Biwer}, {Bizouard},
  {Blackburn}, {Blair}, {Blair}, {Blair}, {Bloemen}, {Bock}, {Bodiya}, {Boer},
  {Bogaert}, {Bogan}, {Bohe}, {Bojtos}, {Bond}, {Bondu}, {Bonnand}, {Boom},
  {Bork}, {Boschi}, {Bose}, {Bouffanais}, {Bozzi}, {Bradaschia}, {Brady},
  {Braginsky}, {Branchesi}, {Brau}, {Briant}, {Brillet}, {Brinkmann},
  {Brisson}, {Brockill}, {Brooks}, {Brown}, {Brown}, {Brown}, {Buchanan},
  {Buikema}, {Bulik}, {Bulten}, {Buonanno}, {Buskulic}, {Buy}, {Byer},
  {Cabero}, {Cadonati}, {Cagnoli}, {Cahillane}, {Bustillo}, {Callister},
  {Calloni}, {Camp}, {Cannon}, {Cao}, {Capano}, {Capocasa}, {Carbognani},
  {Caride}, {Casanueva Diaz}, {Casentini}, {Caudill}, {Cavagli{\`a}},
  {Cavalier}, {Cavalieri}, {Cella}, {Cepeda}, {Baiardi}, {Cerretani},
  {Cesarini}, {Chakraborty}, {Chalermsongsak}, {Chamberlin}, {Chan}, {Chao},
  {Charlton}, {Chassand e-Mottin}, {Chen}, {Chen}, {Cheng}, {Chincarini},
  {Chiummo}, {Cho}, {Cho}, {Chow}, {Christensen}, {Chu}, {Chua}, {Chung},
  {Ciani}, {Clara}, {Clark}, {Cleva}, {Coccia}, {Cohadon}, {Colla}, {Collette},
  {Cominsky}, {Constancio}, {Conte}, {Conti}, {Cook}, {Corbitt}, {Cornish},
  {Corsi}, {Cortese}, {Costa}, {Coughlin}, {Coughlin}, {Coulon}, {Countryman},
  {Couvares}, {Cowan}, {Coward}, {Cowart}, {Coyne}, {Coyne}, {Craig},
  {Creighton}, {Creighton}, {Cripe}, {Crowder}, {Cruise}, {Cumming},
  {Cunningham}, {Cuoco}, {Dal Canton}, {Danilishin}, {D'Antonio}, {Danzmann},
  {Darman}, {Da Silva Costa}, {Dattilo}, {Dave}, {Daveloza}, {Davier},
  {Davies}, {Daw}, {Day}, {De}, {DeBra}, {Debreczeni}, {Degallaix}, {De
  Laurentis}, {Del{\'e}glise}, {Del Pozzo}, {Denker}, {Dent}, {Dereli},
  {Dergachev}, {DeRosa}, {De Rosa}, {DeSalvo}, {Dhurandhar}, {D{\'\i}az}, {Di
  Fiore}, {Di Giovanni}, {Di Lieto}, {Di Pace}, {Di Palma}, {Di Virgilio},
  {Dojcinoski}, {Dolique}, {Donovan}, {Dooley}, {Doravari}, {Douglas},
  {Downes}, {Drago}, {Drever}, {Driggers}, {Du}, {Ducrot}, {Dwyer}, {Edo},
  {Edwards}, {Effler}, {Eggenstein}, {Ehrens}, {Eichholz}, {Eikenberry},
  {Engels}, {Essick}, {Etzel}, {Evans}, {Evans}, {Everett}, {Factourovich},
  {Fafone}, {Fair}, {Fairhurst}, {Fan}, {Fang}, {Farinon}, {Farr}, {Farr},
  {Favata}, {Fays}, {Fehrmann}, {Fejer}, {Feldbaum}, {Ferrante}, {Ferreira},
  {Ferrini}, {Fidecaro}, {Finn}, {Fiori}, {Fiorucci}, {Fisher}, {Flaminio},
  {Fletcher}, {Fong}, {Fournier}, {Franco}, {Frasca}, {Frasconi}, {Frede},
  {Frei}, {Freise}, {Frey}, {Frey}, {Fricke}, {Fritschel}, {Frolov}, {Fulda},
  {Fyffe}, {Gabbard}, {Gair}, {Gammaitoni}, {Gaonkar}, {Garufi}, {Gatto},
  {Gaur}, {Gehrels}, {Gemme}, {Gendre}, {Genin}, {Gennai}, {George}, {Gergely},
  {Germain}, {Ghosh}, {Ghosh}, {Ghosh}, {Giaime}, {Giardina}, {Giazotto},
  {Gill}, {Glaefke}, {Gleason}, {Goetz}, {Goetz}, {Gondan}, {Gonz{\'a}lez},
  {Castro}, {Gopakumar}, {Gordon}, {Gorodetsky}, {Gossan}, {Gosselin},
  {Gouaty}, {Graef}, {Graff}, {Granata}, {Grant}, {Gras}, {Gray}, {Greco},
  {Green}, {Greenhalgh}, {Groot}, {Grote}, {Grunewald}, {Guidi}, {Guo},
  {Gupta}, {Gupta}, {Gushwa}, {Gustafson}, {Gustafson}, {Hacker}, {Hall},
  {Hall}, {Hammond}, {Haney}, {Hanke}, {Hanks}, {Hanna}, {Hannam}, {Hanson},
  {Hardwick}, {Harms}, {Harry}, {Harry}, {Hart}, {Hartman}, {Haster},
  {Haughian}, {Healy}, {Heefner}, {Heidmann}, {Heintze}, {Heinzel}, {Heitmann},
  {Hello}, {Hemming}, {Hendry}, {Heng}, {Hennig}, {Heptonstall}, {Heurs},
  {Hild}, {Hoak}, {Hodge}, {Hofman}, {Hollitt}, {Holt}, {Holz}, {Hopkins},
  {Hosken}, {Hough}, {Houston}, {Howell}, {Hu}, {Huang}, {Huerta}, {Huet},
  {Hughey}, {Husa}, {Huttner}, {Huynh-Dinh}, {Idrisy}, {Indik}, {Ingram},
  {Inta}, {Isa}, {Isac}, {Isi}, {Islas}, {Isogai}, {Iyer}, {Izumi}, {Jacobson},
  {Jacqmin}, {Jang}, {Jani}, {Jaranowski}, {Jawahar}, {Jim{\'e}nez-Forteza},
  {Johnson}, {Johnson-McDaniel}, {Jones}, {Jones}, {Jonker}, {Ju}, {Haris},
  {Kalaghatgi}, {Kalogera}, {Kandhasamy}, {Kang}, {Kanner}, {Karki},
  {Kasprzack}, {Katsavounidis}, {Katzman}, {Kaufer}, {Kaur}, {Kawabe},
  {Kawazoe}, {K{\'e}f{\'e}lian}, {Kehl}, {Keitel}, {Kelley}, {Kells},
  {Kennedy}, {Keppel}, {Key}, {Khalaidovski}, {Khalili}, {Khan}, {Khan},
  {Khan}, {Khazanov}, {Kijbunchoo}, {Kim}, {Kim}, {Kim}, {Kim}, {Kim}, {Kim},
  {King}, {King}, {Kinzel}, {Kissel}, {Kleybolte}, {Klimenko}, {Koehlenbeck},
  {Kokeyama}, {Koley}, {Kondrashov}, {Kontos}, {Koranda}, {Korobko}, {Korth},
  {Kowalska}, {Kozak}, {Kringel}, {Krishnan}, {Kr{\'o}lak}, {Krueger}, {Kuehn},
  {Kumar}, {Kumar}, {Kuo}, {Kutynia}, {Kwee}, {Lackey}, {Landry}, {Lange},
  {Lantz}, {Lasky}, {Lazzarini}, {Lazzaro}, {Leaci}, {Leavey}, {Lebigot},
  {Lee}, {Lee}, {Lee}, {Lee}, {Lenon}, {Leonardi}, {Leong}, {Leroy},
  {Letendre}, {Levin}, {Levine}, {Li}, {Libson}, {Littenberg}, {Lockerbie},
  {Logue}, {Lombardi}, {London}, {Lord}, {Lorenzini}, {Loriette}, {Lormand},
  {Losurdo}, {Lough}, {Lousto}, {Lovelace}, {L{\"u}ck}, {Lundgren}, {Luo},
  {Lynch}, {Ma}, {MacDonald}, {Machenschalk}, {MacInnis}, {Macleod},
  {Maga{\~n}a-Sandoval}, {Magee}, {Mageswaran}, {Majorana}, {Maksimovic},
  {Malvezzi}, {Man}, {Mandel}, {Mandic}, {Mangano}, {Mansell}, {Manske},
  {Mantovani}, {Marchesoni}, {Marion}, {M{\'a}rka}, {M{\'a}rka}, {Markosyan},
  {Maros}, {Martelli}, {Martellini}, {Martin}, {Martin}, {Martynov}, {Marx},
  {Mason}, {Masserot}, {Massinger}, {Masso-Reid}, {Matichard}, {Matone},
  {Mavalvala}, {Mazumder}, {Mazzolo}, {McCarthy}, {McClelland}, {McCormick},
  {McGuire}, {McIntyre}, {McIver}, {McManus}, {McWilliams}, {Meacher},
  {Meadors}, {Meidam}, {Melatos}, {Mendell}, {Mendoza-Gandara}, {Mercer},
  {Merilh}, {Merzougui}, {Meshkov}, {Messenger}, {Messick}, {Meyers},
  {Mezzani}, {Miao}, {Michel}, {Middleton}, {Mikhailov}, {Milano}, {Miller},
  {Millhouse}, {Minenkov}, {Ming}, {Mirshekari}, {Mishra}, {Mitra},
  {Mitrofanov}, {Mitselmakher}, {Mittleman}, {Moggi}, {Mohan}, {Mohapatra},
  {Montani}, {Moore}, {Moore}, {Moraru}, {Moreno}, {Morriss}, {Mossavi},
  {Mours}, {Mow-Lowry}, {Mueller}, {Mueller}, {Muir}, {Mukherjee}, {Mukherjee},
  {Mukherjee}, {Mukund}, {Mullavey}, {Munch}, {Murphy}, {Murray}, {Mytidis},
  {Nardecchia}, {Naticchioni}, {Nayak}, {Necula}, {Nedkova}, {Nelemans},
  {Neri}, {Neunzert}, {Newton}, {Nguyen}, {Nielsen}, {Nissanke}, {Nitz},
  {Nocera}, {Nolting}, {Normandin}, {Nuttall}, {Oberling}, {Ochsner}, {O'Dell},
  {Oelker}, {Ogin}, {Oh}, {Oh}, {Ohme}, {Oliver}, {Oppermann}, {Oram},
  {O'Reilly}, {O'Shaughnessy}, {Ott}, {Ottaway}, {Ottens}, {Overmier}, {Owen},
  {Pai}, {Pai}, {Palamos}, {Palashov}, {Palomba}, {Pal-Singh}, {Pan}, {Pan},
  {Pankow}, {Pannarale}, {Pant}, {Paoletti}, {Paoli}, {Papa}, {Paris},
  {Parker}, {Pascucci}, {Pasqualetti}, {Passaquieti}, {Passuello},
  {Patricelli}, {Patrick}, {Pearlstone}, {Pedraza}, {Pedurand }, {Pekowsky},
  {Pele}, {Penn}, {Perreca}, {Pfeiffer}, {Phelps}, {Piccinni}, {Pichot},
  {Pickenpack}, {Piergiovanni}, {Pierro}, {Pillant}, {Pinard}, {Pinto},
  {Pitkin}, {Poeld}, {Poggiani}, {Popolizio}, {Post}, {Powell}, {Prasad},
  {Predoi}, {Premachandra}, {Prestegard}, {Price}, {Prijatelj}, {Principe},
  {Privitera}, {Prix}, {Prodi}, {Prokhorov}, {Puncken}, {Punturo}, {Puppo},
  {P{\"u}rrer}, {Qi}, {Qin}, {Quetschke}, {Quintero}, {Quitzow-James}, {Raab},
  {Rabeling}, {Radkins}, {Raffai}, {Raja}, {Rakhmanov}, {Ramet}, {Rapagnani},
  {Raymond}, {Razzano}, {Re}, {Read}, {Reed}, {Regimbau}, {Rei}, {Reid},
  {Reitze}, {Rew}, {Reyes}, {Ricci}, {Riles}, {Robertson}, {Robie}, {Robinet},
  {Rocchi}, {Rolland}, {Rollins}, {Roma}, {Romano}, {Romano}, {Romanov},
  {Romie}, {Rosi{\'n}ska}, {Rowan}, {R{\"u}diger}, {Ruggi}, {Ryan}, {Sachdev},
  {Sadecki}, {Sadeghian}, {Salconi}, {Saleem}, {Salemi}, {Samajdar}, {Sammut},
  {Sampson}, {Sanchez}, {Sandberg}, {Sandeen}, {Sand ers}, {Sanders},
  {Sassolas}, {Sathyaprakash}, {Saulson}, {Sauter}, {Savage}, {Sawadsky},
  {Schale}, {Schilling}, {Schmidt}, {Schmidt}, {Schnabel}, {Schofield},
  {Sch{\"o}nbeck}, {Schreiber}, {Schuette}, {Schutz}, {Scott}, {Scott},
  {Sellers}, {Sengupta}, {Sentenac}, {Sequino}, {Sergeev}, {Serna},
  {Setyawati}, {Sevigny}, {Shaddock}, {Shaffer}, {Shah}, {Shahriar}, {Shaltev},
  {Shao}, {Shapiro}, {Shawhan}, {Sheperd}, {Shoemaker}, {Shoemaker}, {Siellez},
  {Siemens}, {Sigg}, {Silva}, {Simakov}, {Singer}, {Singer}, {Singh}, {Singh},
  {Singhal}, {Sintes}, {Slagmolen}, {Smith}, {Smith}, {Smith}, {Smith}, {Son},
  {Sorazu}, {Sorrentino}, {Souradeep}, {Srivastava}, {Staley}, {Steinke},
  {Steinlechner}, {Steinlechner}, {Steinmeyer}, {Stephens}, {Stevenson},
  {Stone}, {Strain}, {Straniero}, {Stratta}, {Strauss}, {Strigin}, {Sturani},
  {Stuver}, {Summerscales}, {Sun}, {Sutton}, {Swinkels}, {Szczepa{\'n}czyk},
  {Tacca}, {Talukder}, {Tanner}, {T{\'a}pai}, {Tarabrin}, {Taracchini},
  {Taylor}, {Theeg}, {Thirugnanasambandam}, {Thomas}, {Thomas}, {Thomas},
  {Thorne}, {Thorne}, {Thrane}, {Tiwari}, {Tiwari}, {Tokmakov}, {Tomlinson},
  {Tonelli}, {Torres}, {Torrie}, {T{\"o}yr{\"a}}, {Travasso}, {Traylor},
  {Trifir{\`o}}, {Tringali}, {Trozzo}, {Tse}, {Turconi}, {Tuyenbayev},
  {Ugolini}, {Unnikrishnan}, {Urban}, {Usman}, {Vahlbruch}, {Vajente},
  {Valdes}, {Vallisneri}, {van Bakel}, {van Beuzekom}, {van den Brand}, {Van
  Den Broeck}, {Vand er-Hyde}, {van der Schaaf}, {van Heijningen}, {van
  Veggel}, {Vardaro}, {Vass}, {Vas{\'u}th}, {Vaulin}, {Vecchio}, {Vedovato},
  {Veitch}, {Veitch}, {Venkateswara}, {Verkindt}, {Vetrano}, {Vicer{\'e}},
  {Vinciguerra}, {Vine}, {Vinet}, {Vitale}, {Vo}, {Vocca}, {Vorvick}, {Voss},
  {Vousden}, {Vyatchanin}, {Wade}, {Wade}, {Wade}, {Waldman}, {Walker},
  {Wallace}, {Walsh}, {Wang}, {Wang}, {Wang}, {Wang}, {Wang}, {Ward}, {Ward},
  {Warner}, {Was}, {Weaver}, {Wei}, {Weinert}, {Weinstein}, {Weiss}, {Welborn},
  {Wen}, {We{\ss}els}, {Westphal}, {Wette}, {Whelan}, {Whitcomb}, {White},
  {Whiting}, {Wiesner}, {Wilkinson}, {Willems}, {Williams}, {Williams},
  {Williamson}, {Willis}, {Willke}, {Wimmer}, {Winkelmann}, {Winkler}, {Wipf},
  {Wiseman}, {Wittel}, {Woan}, {Worden}, {Wright}, {Wu}, {Yablon}, {Yakushin},
  {Yam}, {Yamamoto}, {Yancey}, {Yap}, {Yu}, {Yvert}, {Zadro{\.Z}ny},
  {Zangrando}, {Zanolin}, {Zendri}, {Zevin}, {Zhang}, {Zhang}, {Zhang},
  {Zhang}, {Zhao}, {Zhou}, {Zhou}, {Zhu}, {Zucker}, {Zuraw}, {Zweizig}, {LIGO
  Scientific Collaboration}, \& {Virgo Collaboration}}]{ligo16}
{Abbott}, B.~P., {Abbott}, R., {Abbott}, T.~D., {et~al.} 2016, \prl, 116,
  061102, \dodoi{10.1103/PhysRevLett.116.061102}

\bibitem[{{Afshordi} {et~al.}(2003){Afshordi}, {McDonald}, \&
  {Spergel}}]{afshordi03}
{Afshordi}, N., {McDonald}, P., \& {Spergel}, D.~N. 2003, \apjl, 594, L71,
  \dodoi{10.1086/378763}

\bibitem[{{Agrawal} {et~al.}(2019){Agrawal}, {Okumura}, \&
  {Futamase}}]{agrawal19}
{Agrawal}, A., {Okumura}, T., \& {Futamase}, T. 2019, \prd, 100, 063534,
  \dodoi{10.1103/PhysRevD.100.063534}

\bibitem[{{Alam} {et~al.}(2017){Alam}, {Ata}, {Bailey}, {Beutler}, {Bizyaev},
  {Blazek}, {Bolton}, {Brownstein}, {Burden}, {Chuang}, {Comparat}, {Cuesta},
  {Dawson}, {Eisenstein}, {Escoffier}, {Gil-Mar{\'\i}n}, {Grieb}, {Hand}, {Ho},
  {Kinemuchi}, {Kirkby}, {Kitaura}, {Malanushenko}, {Malanushenko}, {Maraston},
  {McBride}, {Nichol}, {Olmstead}, {Oravetz}, {Padmanabhan},
  {Palanque-Delabrouille}, {Pan}, {Pellejero-Ibanez}, {Percival}, {Petitjean},
  {Prada}, {Price-Whelan}, {Reid}, {Rodr{\'\i}guez-Torres}, {Roe}, {Ross},
  {Ross}, {Rossi}, {Rubi{\~n}o-Mart{\'\i}n}, {Saito}, {Salazar-Albornoz},
  {Samushia}, {S{\'a}nchez}, {Satpathy}, {Schlegel}, {Schneider},
  {Sc{\'o}ccola}, {Seo}, {Sheldon}, {Simmons}, {Slosar}, {Strauss}, {Swanson},
  {Thomas}, {Tinker}, {Tojeiro}, {Maga{\~n}a}, {Vazquez}, {Verde}, {Wake},
  {Wang}, {Weinberg}, {White}, {Wood-Vasey}, {Y{\`e}che}, {Zehavi}, {Zhai}, \&
  {Zhao}}]{boss17}
{Alam}, S., {Ata}, M., {Bailey}, S., {et~al.} 2017, \mnras, 470, 2617,
  \dodoi{10.1093/mnras/stx721}

\bibitem[{{Ando} {et~al.}(2019){Ando}, {Ishiyama}, \& {Hiroshima}}]{ando19}
{Ando}, S., {Ishiyama}, T., \& {Hiroshima}, N. 2019, Galaxies, 7, 68,
  \dodoi{10.3390/galaxies7030068}

\bibitem[{{Baltz} {et~al.}(2009){Baltz}, {Marshall}, \& {Oguri}}]{baltz09}
{Baltz}, E.~A., {Marshall}, P., \& {Oguri}, M. 2009, \jcap, 2009, 015,
  \dodoi{10.1088/1475-7516/2009/01/015}

\bibitem[{{Banik} {et~al.}(2019){Banik}, {Bovy}, {Bertone}, {Erkal}, \& {de
  Boer}}]{banik19}
{Banik}, N., {Bovy}, J., {Bertone}, G., {Erkal}, D., \& {de Boer}, T.~J.~L.
  2019, arXiv e-prints, arXiv:1911.02663.
\newblock \doarXiv{1911.02663}

\bibitem[{{Battaglieri} {et~al.}(2017){Battaglieri}, {Belloni}, {Chou},
  {Cushman}, {Echenard}, {Essig}, {Estrada}, {Feng}, {Flaugher}, {Fox},
  {Graham}, {Hall}, {Harnik}, {Hewett}, {Incandela}, {Izaguirre}, {McKinsey},
  {Pyle}, {Roe}, {Rybka}, {Sikivie}, {Tait}, {Toro}, {Van De Water}, {Weiner},
  {Zurek}, {Adelberger}, {Afanasev}, {Alexander}, {Alexander}, {Cristian
  Antochi}, {Asner}, {Baer}, {Banerjee}, {Baracchini}, {Barbeau}, {Barrow},
  {Bastidon}, {Battat}, {Benson}, {Berlin}, {Bird}, {Blinov}, {Boddy}, {Bondi},
  {Bonivento}, {Boulay}, {Boyce}, {Brodeur}, {Broussard}, {Budnik}, {Bunting},
  {Caffee}, {Caiazza}, {Campbell}, {Cao}, {Carosi}, {Carpinelli}, {Cavoto},
  {Celentano}, {Hyeok Chang}, {Chattopadhyay}, {Chavarria}, {Chen}, {Clark},
  {Clarke}, {Colegrove}, {Coleman}, {Cooke}, {Cooper}, {Crisler}, {Crivelli},
  {D'Eramo}, {D'Urso}, {Dahl}, {Dawson}, {De Napoli}, {De Vita},
  {DeNiverville}, {Derenzo}, {Di Crescenzo}, {Di Marco}, {Dienes}, {Diwan},
  {Hand iipondola Dongwi}, {Drlica-Wagner}, {Ellis}, {Chigbo Ezeribe},
  {Farrar}, {Ferrer}, {Figueroa-Feliciano}, {Filippi}, {Fiorillo}, {Fornal},
  {Freyberger}, {Frugiuele}, {Galbiati}, {Galon}, {Gardner}, {Geraci},
  {Gerbier}, {Graham}, {Gschwendtner}, {Hearty}, {Heise}, {Henning}, {Hill},
  {Hitlin}, {Hochberg}, {Hogan}, {Holtrop}, {Hong}, {Hossbach}, {Humensky},
  {Ilten}, {Irwin}, {Jaros}, {Johnson}, {Jones}, {Kahn}, {Kalantarians},
  {Kaplinghat}, {Khatiwada}, {Knapen}, {Kohl}, {Kouvaris}, {Kozaczuk},
  {Krnjaic}, {Kubarovsky}, {Kuflik}, {Kusenko}, {Lang}, {Leach}, {Lin},
  {Lisanti}, {Liu}, {Liu}, {Liu}, {Loomba}, {Lykken}, {Mack}, {Mans}, {Maris},
  {Markiewicz}, {Marsicano}, {Martoff}, {Mazzitelli}, {McCabe}, {McDermott},
  {McDonald}, {McKinnon}, {Mei}, {Melia}, {Miller}, {Miuchi}, {Nazeer},
  {Moreno}, {Morozov}, {Mouton}, {Mueller}, {Murphy}, {Neilson}, {Nelson},
  {Neu}, {Nosochkov}, {O'Hare}, {Oblath}, {Orrell}, {Ouellet}, {Pastore},
  {Paul}, {Perelstein}, {Peter}, {Phan}, {Phinney}, {Pivovaroff}, {Pocar},
  {Pospelov}, {Pradler}, {Privitera}, {Profumo}, {Raggi}, {Rajendran}, {Rand
  azzo}, {Raubenheimer}, {Regenfus}, {Renshaw}, {Ritz}, {Rizzo}, {Rosenberg},
  {Rubbia}, {Rybolt}, {Saab}, {Safdi}, {Santopinto}, {Scarff}, {Schneider},
  {Schuster}, {Seidel}, {Sekiya}, {Seong}, {Simi}, {Sipala}, {Slatyer},
  {Slone}, {Smith}, {Smolinsky}, {Snowden-Ifft}, {Solt}, {Sonnenschein},
  {Sorensen}, {Spooner}, {Srivastava}, {Stancu}, {Strigari}, {Strube},
  {Sushkov}, {Szydagis}, {Tanedo}, {Tanner}, {Tayloe}, {Terrano}, {Thaler},
  {Thomas}, {Thorpe}, {Thorpe}, {Tiffenberg}, {Tran}, {Trovato}, {Tully},
  {Tyson}, {Vachaspati}, {Vahsen}, {van Bibber}, {Vand enbroucke}, {Villano},
  {Volansky}, {Wang}, {Ward}, {Wester}, {Whitbeck}, {Williams}, {Wing},
  {Winslow}, {Wojtsekhowski}, {Yu}, {Yu}, {Yu}, {Zhang}, {Zhao}, \&
  {Zhong}}]{uscv17}
{Battaglieri}, M., {Belloni}, A., {Chou}, A., {et~al.} 2017, arXiv e-prints,
  arXiv:1707.04591.
\newblock \doarXiv{1707.04591}

\bibitem[{{Behroozi} {et~al.}(2019){Behroozi}, {Wechsler}, {Hearin}, \&
  {Conroy}}]{behroozi19}
{Behroozi}, P., {Wechsler}, R.~H., {Hearin}, A.~P., \& {Conroy}, C. 2019,
  \mnras, 488, 3143, \dodoi{10.1093/mnras/stz1182}

\bibitem[{{Ben-Dayan} \& {Kalaydzhyan}(2014)}]{bendayan14}
{Ben-Dayan}, I., \& {Kalaydzhyan}, T. 2014, \prd, 90, 083509,
  \dodoi{10.1103/PhysRevD.90.083509}

\bibitem[{{Ben-Dayan} \& {Takahashi}(2016)}]{bendayan16}
{Ben-Dayan}, I., \& {Takahashi}, R. 2016, \mnras, 455, 552,
  \dodoi{10.1093/mnras/stv2356}

\bibitem[{{Bernardeau} {et~al.}(1997){Bernardeau}, {van Waerbeke}, \&
  {Mellier}}]{bernardeau97}
{Bernardeau}, F., {van Waerbeke}, L., \& {Mellier}, Y. 1997, \aap, 322, 1.
\newblock \doarXiv{astro-ph/9609122}

\bibitem[{{Blumenthal} {et~al.}(1984){Blumenthal}, {Faber}, {Primack}, \&
  {Rees}}]{blumenthal84}
{Blumenthal}, G.~R., {Faber}, S.~M., {Primack}, J.~R., \& {Rees}, M.~J. 1984,
  \nat, 311, 517, \dodoi{10.1038/311517a0}

\bibitem[{{Bond} {et~al.}(1991){Bond}, {Cole}, {Efstathiou}, \&
  {Kaiser}}]{bond91}
{Bond}, J.~R., {Cole}, S., {Efstathiou}, G., \& {Kaiser}, N. 1991, \apj, 379,
  440, \dodoi{10.1086/170520}

\bibitem[{{Bovy} {et~al.}(2017){Bovy}, {Erkal}, \& {Sanders}}]{bovy17}
{Bovy}, J., {Erkal}, D., \& {Sanders}, J.~L. 2017, \mnras, 466, 628,
  \dodoi{10.1093/mnras/stw3067}

\bibitem[{{Bower}(1991)}]{bower91}
{Bower}, R.~G. 1991, \mnras, 248, 332, \dodoi{10.1093/mnras/248.2.332}

\bibitem[{{Bullock} \& {Boylan-Kolchin}(2017)}]{bullock17}
{Bullock}, J.~S., \& {Boylan-Kolchin}, M. 2017, \araa, 55, 343,
  \dodoi{10.1146/annurev-astro-091916-055313}

\bibitem[{{Carr} {et~al.}(2020){Carr}, {Kohri}, {Sendouda}, \&
  {Yokoyama}}]{carr20}
{Carr}, B., {Kohri}, K., {Sendouda}, Y., \& {Yokoyama}, J. 2020, arXiv
  e-prints, arXiv:2002.12778.
\newblock \doarXiv{2002.12778}

\bibitem[{{Chisari} {et~al.}(2019){Chisari}, {Mead}, {Joudaki}, {Ferreira},
  {Schneider}, {Mohr}, {Tr{\"o}ster}, {Alonso}, {McCarthy}, {Martin-Alvarez},
  {Devriendt}, {Slyz}, \& {van Daalen}}]{chisari19}
{Chisari}, N.~E., {Mead}, A.~J., {Joudaki}, S., {et~al.} 2019, The Open Journal
  of Astrophysics, 2, 4, \dodoi{10.21105/astro.1905.06082}

\bibitem[{{Cooray} \& {Sheth}(2002)}]{cooray02}
{Cooray}, A., \& {Sheth}, R. 2002, \physrep, 372, 1,
  \dodoi{10.1016/S0370-1573(02)00276-4}

\bibitem[{{Dai} {et~al.}(2018{\natexlab{a}}){Dai}, {Li}, {Zackay}, {Mao}, \&
  {Lu}}]{dai18b}
{Dai}, L., {Li}, S.-S., {Zackay}, B., {Mao}, S., \& {Lu}, Y.
  2018{\natexlab{a}}, \prd, 98, 104029, \dodoi{10.1103/PhysRevD.98.104029}

\bibitem[{{Dai} \& {Miralda-Escud{\'e}}(2020)}]{dai20}
{Dai}, L., \& {Miralda-Escud{\'e}}, J. 2020, \aj, 159, 49,
  \dodoi{10.3847/1538-3881/ab5e83}

\bibitem[{{Dai} {et~al.}(2018{\natexlab{b}}){Dai}, {Venumadhav}, {Kaurov}, \&
  {Miralda-Escud}}]{dai18a}
{Dai}, L., {Venumadhav}, T., {Kaurov}, A.~A., \& {Miralda-Escud}, J.
  2018{\natexlab{b}}, \apj, 867, 24, \dodoi{10.3847/1538-4357/aae478}

\bibitem[{{Davis} {et~al.}(1985){Davis}, {Efstathiou}, {Frenk}, \&
  {White}}]{davis85}
{Davis}, M., {Efstathiou}, G., {Frenk}, C.~S., \& {White}, S.~D.~M. 1985, \apj,
  292, 371, \dodoi{10.1086/163168}

\bibitem[{{Debackere} {et~al.}(2020){Debackere}, {Schaye}, \&
  {Hoekstra}}]{debackere20}
{Debackere}, S. N.~B., {Schaye}, J., \& {Hoekstra}, H. 2020, \mnras, 492, 2285,
  \dodoi{10.1093/mnras/stz3446}

\bibitem[{{Deguchi} \& {Watson}(1986{\natexlab{a}})}]{deguchi86a}
{Deguchi}, S., \& {Watson}, W.~D. 1986{\natexlab{a}}, \apj, 307, 30,
  \dodoi{10.1086/164389}

\bibitem[{{Deguchi} \& {Watson}(1986{\natexlab{b}})}]{deguchi86b}
---. 1986{\natexlab{b}}, \prd, 34, 1708, \dodoi{10.1103/PhysRevD.34.1708}

\bibitem[{{Diemer} \& {Joyce}(2019)}]{diemer19}
{Diemer}, B., \& {Joyce}, M. 2019, \apj, 871, 168,
  \dodoi{10.3847/1538-4357/aafad6}

\bibitem[{{Diemer} \& {Kravtsov}(2015)}]{diemer15}
{Diemer}, B., \& {Kravtsov}, A.~V. 2015, \apj, 799, 108,
  \dodoi{10.1088/0004-637X/799/1/108}

\bibitem[{{Dodelson} \& {Vallinotto}(2006)}]{dodelson06}
{Dodelson}, S., \& {Vallinotto}, A. 2006, \prd, 74, 063515,
  \dodoi{10.1103/PhysRevD.74.063515}

\bibitem[{{Dror} {et~al.}(2019){Dror}, {Ramani}, {Trickle}, \&
  {Zurek}}]{dror19}
{Dror}, J.~A., {Ramani}, H., {Trickle}, T., \& {Zurek}, K.~M. 2019, \prd, 100,
  023003, \dodoi{10.1103/PhysRevD.100.023003}

\bibitem[{{Fedeli}(2014)}]{fedeli14a}
{Fedeli}, C. 2014, \jcap, 2014, 028, \dodoi{10.1088/1475-7516/2014/04/028}

\bibitem[{{Fedeli} \& {Moscardini}(2014)}]{fedeli14}
{Fedeli}, C., \& {Moscardini}, L. 2014, \mnras, 442, 2659,
  \dodoi{10.1093/mnras/stu1043}

\bibitem[{{Fedeli} {et~al.}(2014){Fedeli}, {Semboloni}, {Velliscig}, {Van
  Daalen}, {Schaye}, \& {Hoekstra}}]{fedeli14b}
{Fedeli}, C., {Semboloni}, E., {Velliscig}, M., {et~al.} 2014, \jcap, 2014,
  028, \dodoi{10.1088/1475-7516/2014/08/028}

\bibitem[{{Gilman} {et~al.}(2020){Gilman}, {Birrer}, {Nierenberg}, {Treu},
  {Du}, \& {Benson}}]{gilman20}
{Gilman}, D., {Birrer}, S., {Nierenberg}, A., {et~al.} 2020, \mnras, 491, 6077,
  \dodoi{10.1093/mnras/stz3480}

\bibitem[{{Giocoli} {et~al.}(2010){Giocoli}, {Bartelmann}, {Sheth}, \&
  {Cacciato}}]{giocoli10}
{Giocoli}, C., {Bartelmann}, M., {Sheth}, R.~K., \& {Cacciato}, M. 2010,
  \mnras, 408, 300, \dodoi{10.1111/j.1365-2966.2010.17108.x}

\bibitem[{{Giocoli} {et~al.}(2007){Giocoli}, {Moreno}, {Sheth}, \&
  {Tormen}}]{giocoli07}
{Giocoli}, C., {Moreno}, J., {Sheth}, R.~K., \& {Tormen}, G. 2007, \mnras, 376,
  977, \dodoi{10.1111/j.1365-2966.2007.11520.x}

\bibitem[{{Giocoli} {et~al.}(2008{\natexlab{a}}){Giocoli}, {Pieri}, \&
  {Tormen}}]{giocoli08b}
{Giocoli}, C., {Pieri}, L., \& {Tormen}, G. 2008{\natexlab{a}}, \mnras, 387,
  689, \dodoi{10.1111/j.1365-2966.2008.13283.x}

\bibitem[{{Giocoli} {et~al.}(2008{\natexlab{b}}){Giocoli}, {Tormen}, \& {van
  den Bosch}}]{giocoli08a}
{Giocoli}, C., {Tormen}, G., \& {van den Bosch}, F.~C. 2008{\natexlab{b}},
  \mnras, 386, 2135, \dodoi{10.1111/j.1365-2966.2008.13182.x}

\bibitem[{{Gong} \& {Kitajima}(2017)}]{gong17}
{Gong}, J.-O., \& {Kitajima}, N. 2017, \jcap, 2017, 017,
  \dodoi{10.1088/1475-7516/2017/08/017}

\bibitem[{{Hada} \& {Futamase}(2016)}]{hada16}
{Hada}, R., \& {Futamase}, T. 2016, \apj, 828, 112,
  \dodoi{10.3847/0004-637X/828/2/112}

\bibitem[{{Hada} \& {Futamase}(2019)}]{hada19}
---. 2019, \jcap, 2019, 033, \dodoi{10.1088/1475-7516/2019/06/033}

\bibitem[{{Hamana} \& {Futamase}(2000)}]{hamana00}
{Hamana}, T., \& {Futamase}, T. 2000, \apj, 534, 29, \dodoi{10.1086/308758}

\bibitem[{{Han} {et~al.}(2016){Han}, {Cole}, {Frenk}, \& {Jing}}]{han16}
{Han}, J., {Cole}, S., {Frenk}, C.~S., \& {Jing}, Y. 2016, \mnras, 457, 1208,
  \dodoi{10.1093/mnras/stv2900}

\bibitem[{{Hernquist}(1990)}]{hernquist90}
{Hernquist}, L. 1990, \apj, 356, 359, \dodoi{10.1086/168845}

\bibitem[{{Hiroshima} {et~al.}(2018){Hiroshima}, {Ando}, \&
  {Ishiyama}}]{hiroshima18}
{Hiroshima}, N., {Ando}, S., \& {Ishiyama}, T. 2018, \prd, 97, 123002,
  \dodoi{10.1103/PhysRevD.97.123002}

\bibitem[{{Holz} \& {Hughes}(2005)}]{holz05}
{Holz}, D.~E., \& {Hughes}, S.~A. 2005, \apj, 629, 15, \dodoi{10.1086/431341}

\bibitem[{{Huang} {et~al.}(2017){Huang}, {Fall}, {Ferguson}, {van der Wel},
  {Grogin}, {Koekemoer}, {Lee}, {P{\'e}rez-Gonz{\'a}lez}, \& {Wuyts}}]{huang17}
{Huang}, K.-H., {Fall}, S.~M., {Ferguson}, H.~C., {et~al.} 2017, \apj, 838, 6,
  \dodoi{10.3847/1538-4357/aa62a6}

\bibitem[{{Ibata} {et~al.}(2002){Ibata}, {Lewis}, {Irwin}, \&
  {Quinn}}]{ibata02}
{Ibata}, R.~A., {Lewis}, G.~F., {Irwin}, M.~J., \& {Quinn}, T. 2002, \mnras,
  332, 915, \dodoi{10.1046/j.1365-8711.2002.05358.x}

\bibitem[{{Inman} \& {Ali-Ha{\"\i}moud}(2019)}]{inman19}
{Inman}, D., \& {Ali-Ha{\"\i}moud}, Y. 2019, \prd, 100, 083528,
  \dodoi{10.1103/PhysRevD.100.083528}

\bibitem[{{Inoue} \& {Chiba}(2003)}]{inoue03}
{Inoue}, K.~T., \& {Chiba}, M. 2003, \apjl, 591, L83, \dodoi{10.1086/377247}

\bibitem[{{Inoue} {et~al.}(2015){Inoue}, {Takahashi}, {Takahashi}, \&
  {Ishiyama}}]{inoue15}
{Inoue}, K.~T., {Takahashi}, R., {Takahashi}, T., \& {Ishiyama}, T. 2015,
  \mnras, 448, 2704, \dodoi{10.1093/mnras/stv194}

\bibitem[{{Ishiyama} \& {Ando}(2020)}]{ishiyama20}
{Ishiyama}, T., \& {Ando}, S. 2020, \mnras, 492, 3662,
  \dodoi{10.1093/mnras/staa069}

\bibitem[{{Jiang} \& {van den Bosch}(2016)}]{jiang16}
{Jiang}, F., \& {van den Bosch}, F.~C. 2016, \mnras, 458, 2848,
  \dodoi{10.1093/mnras/stw439}

\bibitem[{{J{\"o}nsson} {et~al.}(2007){J{\"o}nsson}, {Dahl{\'e}n}, {Goobar},
  {M{\"o}rtsell}, \& {Riess}}]{jonsson07}
{J{\"o}nsson}, J., {Dahl{\'e}n}, T., {Goobar}, A., {M{\"o}rtsell}, E., \&
  {Riess}, A. 2007, \jcap, 2007, 002, \dodoi{10.1088/1475-7516/2007/06/002}

\bibitem[{{J{\"o}nsson} {et~al.}(2010){J{\"o}nsson}, {Sullivan}, {Hook},
  {Basa}, {Carlberg}, {Conley}, {Fouchez}, {Howell}, {Perrett}, \&
  {Pritchet}}]{jonsson10}
{J{\"o}nsson}, J., {Sullivan}, M., {Hook}, I., {et~al.} 2010, \mnras, 405, 535,
  \dodoi{10.1111/j.1365-2966.2010.16467.x}

\bibitem[{{Karpenka} {et~al.}(2013){Karpenka}, {March}, {Feroz}, \&
  {Hobson}}]{karpenka13}
{Karpenka}, N.~V., {March}, M.~C., {Feroz}, F., \& {Hobson}, M.~P. 2013,
  \mnras, 433, 2693, \dodoi{10.1093/mnras/sts700}

\bibitem[{{Kashiyama} \& {Oguri}(2018)}]{kashiyama18}
{Kashiyama}, K., \& {Oguri}, M. 2018, arXiv e-prints, arXiv:1801.07847.
\newblock \doarXiv{1801.07847}

\bibitem[{{Kawamata} {et~al.}(2015){Kawamata}, {Ishigaki}, {Shimasaku},
  {Oguri}, \& {Ouchi}}]{kawamata15}
{Kawamata}, R., {Ishigaki}, M., {Shimasaku}, K., {Oguri}, M., \& {Ouchi}, M.
  2015, \apj, 804, 103, \dodoi{10.1088/0004-637X/804/2/103}

\bibitem[{{Kawamata} {et~al.}(2018){Kawamata}, {Ishigaki}, {Shimasaku},
  {Oguri}, {Ouchi}, \& {Tanigawa}}]{kawamata18}
{Kawamata}, R., {Ishigaki}, M., {Shimasaku}, K., {et~al.} 2018, \apj, 855, 4,
  \dodoi{10.3847/1538-4357/aaa6cf}

\bibitem[{{Kelly} {et~al.}(2018){Kelly}, {Diego}, {Rodney}, {Kaiser},
  {Broadhurst}, {Zitrin}, {Treu}, {P{\'e}rez-Gonz{\'a}lez}, {Morishita},
  {Jauzac}, {Selsing}, {Oguri}, {Pueyo}, {Ross}, {Filippenko}, {Smith},
  {Hjorth}, {Cenko}, {Wang}, {Howell}, {Richard}, {Frye}, {Jha}, {Foley},
  {Norman}, {Bradac}, {Zheng}, {Brammer}, {Benito}, {Cava}, {Christensen}, {de
  Mink}, {Graur}, {Grillo}, {Kawamata}, {Kneib}, {Matheson}, {McCully},
  {Nonino}, {P{\'e}rez-Fournon}, {Riess}, {Rosati}, {Schmidt}, {Sharon}, \&
  {Weiner}}]{kelly18}
{Kelly}, P.~L., {Diego}, J.~M., {Rodney}, S., {et~al.} 2018, Nature Astronomy,
  2, 334, \dodoi{10.1038/s41550-018-0430-3}

\bibitem[{{Koopmans}(2005)}]{koopmans05}
{Koopmans}, L.~V.~E. 2005, \mnras, 363, 1136,
  \dodoi{10.1111/j.1365-2966.2005.09523.x}

\bibitem[{{Kravtsov}(2013)}]{kravtsov13}
{Kravtsov}, A.~V. 2013, \apjl, 764, L31, \dodoi{10.1088/2041-8205/764/2/L31}

\bibitem[{{Kravtsov} {et~al.}(2018){Kravtsov}, {Vikhlinin}, \&
  {Meshcheryakov}}]{kravtsov18}
{Kravtsov}, A.~V., {Vikhlinin}, A.~A., \& {Meshcheryakov}, A.~V. 2018,
  Astronomy Letters, 44, 8, \dodoi{10.1134/S1063773717120015}

\bibitem[{{Kronborg} {et~al.}(2010){Kronborg}, {Hardin}, {Guy}, {Astier},
  {Balland }, {Basa}, {Carlberg}, {Conley}, {Fouchez}, {Hook}, {Howell},
  {J{\"o}nsson}, {Pain}, {Pedersen}, {Perrett}, {Pritchet}, {Regnault}, {Rich},
  {Sullivan}, {Palanque-Delabrouille}, \& {Ruhlmann-Kleider}}]{kronborg10}
{Kronborg}, T., {Hardin}, D., {Guy}, J., {et~al.} 2010, \aap, 514, A44,
  \dodoi{10.1051/0004-6361/200913618}

\bibitem[{{Lacey} \& {Cole}(1993)}]{lacey93}
{Lacey}, C., \& {Cole}, S. 1993, \mnras, 262, 627,
  \dodoi{10.1093/mnras/262.3.627}

\bibitem[{{Lee}(2004)}]{lee04}
{Lee}, J. 2004, \apjl, 604, L73, \dodoi{10.1086/386304}

\bibitem[{{Lindblom} {et~al.}(2008){Lindblom}, {Owen}, \& {Brown}}]{lindblom08}
{Lindblom}, L., {Owen}, B.~J., \& {Brown}, D.~A. 2008, \prd, 78, 124020,
  \dodoi{10.1103/PhysRevD.78.124020}

\bibitem[{{Macquart}(2004)}]{macquart04}
{Macquart}, J.~P. 2004, \aap, 422, 761, \dodoi{10.1051/0004-6361:20034512}

\bibitem[{{Mao} \& {Schneider}(1998)}]{mao98}
{Mao}, S., \& {Schneider}, P. 1998, \mnras, 295, 587,
  \dodoi{10.1046/j.1365-8711.1998.01319.x}

\bibitem[{{Marinacci} {et~al.}(2018){Marinacci}, {Vogelsberger}, {Pakmor},
  {Torrey}, {Springel}, {Hernquist}, {Nelson}, {Weinberger}, {Pillepich},
  {Naiman}, \& {Genel}}]{marinacci18}
{Marinacci}, F., {Vogelsberger}, M., {Pakmor}, R., {et~al.} 2018, \mnras, 480,
  5113, \dodoi{10.1093/mnras/sty2206}

\bibitem[{{Mediavilla} {et~al.}(2017){Mediavilla}, {Jim{\'e}nez-Vicente},
  {Mu{\~n}oz}, {Vives-Arias}, \& {Calder{\'o}n-Infante}}]{mediavilla17}
{Mediavilla}, E., {Jim{\'e}nez-Vicente}, J., {Mu{\~n}oz}, J.~A., {Vives-Arias},
  H., \& {Calder{\'o}n-Infante}, J. 2017, \apjl, 836, L18,
  \dodoi{10.3847/2041-8213/aa5dab}

\bibitem[{{Metcalf}(1999)}]{metcalf99}
{Metcalf}, R.~B. 1999, \mnras, 305, 746,
  \dodoi{10.1046/j.1365-8711.1999.02382.x}

\bibitem[{{Mo} {et~al.}(2010){Mo}, {van den Bosch}, \& {White}}]{mo10}
{Mo}, H., {van den Bosch}, F.~C., \& {White}, S. 2010, {Galaxy Formation and
  Evolution}

\bibitem[{{Molin{\'e}} {et~al.}(2017){Molin{\'e}}, {S{\'a}nchez-Conde},
  {Palomares-Ruiz}, \& {Prada}}]{moline17}
{Molin{\'e}}, {\'A}., {S{\'a}nchez-Conde}, M.~A., {Palomares-Ruiz}, S., \&
  {Prada}, F. 2017, \mnras, 466, 4974, \dodoi{10.1093/mnras/stx026}

\bibitem[{{Mondino} {et~al.}(2020){Mondino}, {Taki}, {Van Tilburg}, \&
  {Weiner}}]{mondino20}
{Mondino}, C., {Taki}, A.-M., {Van Tilburg}, K., \& {Weiner}, N. 2020, arXiv
  e-prints, arXiv:2002.01938.
\newblock \doarXiv{2002.01938}

\bibitem[{{Naiman} {et~al.}(2018){Naiman}, {Pillepich}, {Springel},
  {Ramirez-Ruiz}, {Torrey}, {Vogelsberger}, {Pakmor}, {Nelson}, {Marinacci},
  {Hernquist}, {Weinberger}, \& {Genel}}]{naiman18}
{Naiman}, J.~P., {Pillepich}, A., {Springel}, V., {et~al.} 2018, \mnras, 477,
  1206, \dodoi{10.1093/mnras/sty618}

\bibitem[{{Nakamura} {et~al.}(2016){Nakamura}, {Ando}, {Kinugawa}, {Nakano},
  {Eda}, {Sato}, {Musha}, {Akutsu}, {Tanaka}, {Seto}, {Kanda}, \&
  {Itoh}}]{nakamura16}
{Nakamura}, T., {Ando}, M., {Kinugawa}, T., {et~al.} 2016, Progress of
  Theoretical and Experimental Physics, 2016, 093E01,
  \dodoi{10.1093/ptep/ptw127}

\bibitem[{{Nakamura} \& {Deguchi}(1999)}]{nakamura99}
{Nakamura}, T.~T., \& {Deguchi}, S. 1999, Progress of Theoretical Physics
  Supplement, 133, 137, \dodoi{10.1143/PTPS.133.137}

\bibitem[{{Nakamura} \& {Suto}(1997)}]{nakamura97}
{Nakamura}, T.~T., \& {Suto}, Y. 1997, Progress of Theoretical Physics, 97, 49,
  \dodoi{10.1143/PTP.97.49}

\bibitem[{{Navarro} {et~al.}(1997){Navarro}, {Frenk}, \& {White}}]{navarro97}
{Navarro}, J.~F., {Frenk}, C.~S., \& {White}, S. D.~M. 1997, \apj, 490, 493,
  \dodoi{10.1086/304888}

\bibitem[{{Nelson} {et~al.}(2018){Nelson}, {Pillepich}, {Springel},
  {Weinberger}, {Hernquist}, {Pakmor}, {Genel}, {Torrey}, {Vogelsberger},
  {Kauffmann}, {Marinacci}, \& {Naiman}}]{nelson18}
{Nelson}, D., {Pillepich}, A., {Springel}, V., {et~al.} 2018, \mnras, 475, 624,
  \dodoi{10.1093/mnras/stx3040}

\bibitem[{{Oguri}(2018)}]{oguri18b}
{Oguri}, M. 2018, \mnras, 480, 3842, \dodoi{10.1093/mnras/sty2145}

\bibitem[{{Oguri}(2019)}]{oguri19}
---. 2019, Reports on Progress in Physics, 82, 126901,
  \dodoi{10.1088/1361-6633/ab4fc5}

\bibitem[{{Oguri} {et~al.}(2018){Oguri}, {Diego}, {Kaiser}, {Kelly}, \&
  {Broadhurst}}]{oguri18}
{Oguri}, M., {Diego}, J.~M., {Kaiser}, N., {Kelly}, P.~L., \& {Broadhurst}, T.
  2018, \prd, 97, 023518, \dodoi{10.1103/PhysRevD.97.023518}

\bibitem[{{Oguri} \& {Hamana}(2011)}]{oguri11}
{Oguri}, M., \& {Hamana}, T. 2011, \mnras, 414, 1851,
  \dodoi{10.1111/j.1365-2966.2011.18481.x}

\bibitem[{{Oguri} \& {Lee}(2004)}]{oguri04}
{Oguri}, M., \& {Lee}, J. 2004, \mnras, 355, 120,
  \dodoi{10.1111/j.1365-2966.2004.08304.x}

\bibitem[{{Peebles}(1982)}]{peebles82}
{Peebles}, P.~J.~E. 1982, \apjl, 263, L1, \dodoi{10.1086/183911}

\bibitem[{{Pillepich} {et~al.}(2018){Pillepich}, {Nelson}, {Hernquist},
  {Springel}, {Pakmor}, {Torrey}, {Weinberger}, {Genel}, {Naiman}, {Marinacci},
  \& {Vogelsberger}}]{pillepich18}
{Pillepich}, A., {Nelson}, D., {Hernquist}, L., {et~al.} 2018, \mnras, 475,
  648, \dodoi{10.1093/mnras/stx3112}

\bibitem[{{Planck Collaboration} {et~al.}(2016){Planck Collaboration}, {Ade},
  {Aghanim}, {Arnaud}, {Ashdown}, {Aumont}, {Baccigalupi}, {Banday},
  {Barreiro}, {Bartlett}, {Bartolo}, {Battaner}, {Battye}, {Benabed},
  {Beno{\^\i}t}, {Benoit-L{\'e}vy}, {Bernard}, {Bersanelli}, {Bielewicz},
  {Bock}, {Bonaldi}, {Bonavera}, {Bond}, {Borrill}, {Bouchet}, {Boulanger},
  {Bucher}, {Burigana}, {Butler}, {Calabrese}, {Cardoso}, {Catalano},
  {Challinor}, {Chamballu}, {Chary}, {Chiang}, {Chluba}, {Christensen},
  {Church}, {Clements}, {Colombi}, {Colombo}, {Combet}, {Coulais}, {Crill},
  {Curto}, {Cuttaia}, {Danese}, {Davies}, {Davis}, {de Bernardis}, {de Rosa},
  {de Zotti}, {Delabrouille}, {D{\'e}sert}, {Di Valentino}, {Dickinson},
  {Diego}, {Dolag}, {Dole}, {Donzelli}, {Dor{\'e}}, {Douspis}, {Ducout},
  {Dunkley}, {Dupac}, {Efstathiou}, {Elsner}, {En{\ss}lin}, {Eriksen},
  {Farhang}, {Fergusson}, {Finelli}, {Forni}, {Frailis}, {Fraisse},
  {Franceschi}, {Frejsel}, {Galeotta}, {Galli}, {Ganga}, {Gauthier}, {Gerbino},
  {Ghosh}, {Giard}, {Giraud-H{\'e}raud}, {Giusarma}, {Gjerl{\o}w},
  {Gonz{\'a}lez-Nuevo}, {G{\'o}rski}, {Gratton}, {Gregorio}, {Gruppuso},
  {Gudmundsson}, {Hamann}, {Hansen}, {Hanson}, {Harrison}, {Helou},
  {Henrot-Versill{\'e}}, {Hern{\'a}ndez-Monteagudo}, {Herranz}, {Hildebrand t},
  {Hivon}, {Hobson}, {Holmes}, {Hornstrup}, {Hovest}, {Huang}, {Huffenberger},
  {Hurier}, {Jaffe}, {Jaffe}, {Jones}, {Juvela}, {Keih{\"a}nen}, {Keskitalo},
  {Kisner}, {Kneissl}, {Knoche}, {Knox}, {Kunz}, {Kurki-Suonio}, {Lagache},
  {L{\"a}hteenm{\"a}ki}, {Lamarre}, {Lasenby}, {Lattanzi}, {Lawrence}, {Leahy},
  {Leonardi}, {Lesgourgues}, {Levrier}, {Lewis}, {Liguori}, {Lilje},
  {Linden-V{\o}rnle}, {L{\'o}pez-Caniego}, {Lubin}, {Mac{\'\i}as-P{\'e}rez},
  {Maggio}, {Maino}, {Mandolesi}, {Mangilli}, {Marchini}, {Maris}, {Martin},
  {Martinelli}, {Mart{\'\i}nez-Gonz{\'a}lez}, {Masi}, {Matarrese}, {McGehee},
  {Meinhold}, {Melchiorri}, {Melin}, {Mendes}, {Mennella}, {Migliaccio},
  {Millea}, {Mitra}, {Miville-Desch{\^e}nes}, {Moneti}, {Montier}, {Morgante},
  {Mortlock}, {Moss}, {Munshi}, {Murphy}, {Naselsky}, {Nati}, {Natoli},
  {Netterfield}, {N{\o}rgaard-Nielsen}, {Noviello}, {Novikov}, {Novikov},
  {Oxborrow}, {Paci}, {Pagano}, {Pajot}, {Paladini}, {Paoletti}, {Partridge},
  {Pasian}, {Patanchon}, {Pearson}, {Perdereau}, {Perotto}, {Perrotta},
  {Pettorino}, {Piacentini}, {Piat}, {Pierpaoli}, {Pietrobon}, {Plaszczynski},
  {Pointecouteau}, {Polenta}, {Popa}, {Pratt}, {Pr{\'e}zeau}, {Prunet},
  {Puget}, {Rachen}, {Reach}, {Rebolo}, {Reinecke}, {Remazeilles}, {Renault},
  {Renzi}, {Ristorcelli}, {Rocha}, {Rosset}, {Rossetti}, {Roudier},
  {Rouill{\'e} d'Orfeuil}, {Rowan-Robinson}, {Rubi{\~n}o-Mart{\'\i}n},
  {Rusholme}, {Said}, {Salvatelli}, {Salvati}, {Sandri}, {Santos},
  {Savelainen}, {Savini}, {Scott}, {Seiffert}, {Serra}, {Shellard}, {Spencer},
  {Spinelli}, {Stolyarov}, {Stompor}, {Sudiwala}, {Sunyaev}, {Sutton},
  {Suur-Uski}, {Sygnet}, {Tauber}, {Terenzi}, {Toffolatti}, {Tomasi},
  {Tristram}, {Trombetti}, {Tucci}, {Tuovinen}, {T{\"u}rler}, {Umana},
  {Valenziano}, {Valiviita}, {Van Tent}, {Vielva}, {Villa}, {Wade}, {Wandelt},
  {Wehus}, {White}, {White}, {Wilkinson}, {Yvon}, {Zacchei}, \&
  {Zonca}}]{planck16}
{Planck Collaboration}, {Ade}, P.~A.~R., {Aghanim}, N., {et~al.} 2016, \aap,
  594, A13, \dodoi{10.1051/0004-6361/201525830}

\bibitem[{{Quartin} {et~al.}(2014){Quartin}, {Marra}, \&
  {Amendola}}]{quartin14}
{Quartin}, M., {Marra}, V., \& {Amendola}, L. 2014, \prd, 89, 023009,
  \dodoi{10.1103/PhysRevD.89.023009}

\bibitem[{{Quimby} {et~al.}(2014){Quimby}, {Oguri}, {More}, {More}, {Moriya},
  {Werner}, {Tanaka}, {Folatelli}, {Bersten}, {Maeda}, \& {Nomoto}}]{quimby14}
{Quimby}, R.~M., {Oguri}, M., {More}, A., {et~al.} 2014, Science, 344, 396,
  \dodoi{10.1126/science.1250903}

\bibitem[{{Ritondale} {et~al.}(2019){Ritondale}, {Vegetti}, {Despali}, {Auger},
  {Koopmans}, \& {McKean}}]{ritondale19}
{Ritondale}, E., {Vegetti}, S., {Despali}, G., {et~al.} 2019, \mnras, 485,
  2179, \dodoi{10.1093/mnras/stz464}

\bibitem[{{Sasaki} {et~al.}(2018){Sasaki}, {Suyama}, {Tanaka}, \&
  {Yokoyama}}]{sasaki18}
{Sasaki}, M., {Suyama}, T., {Tanaka}, T., \& {Yokoyama}, S. 2018, Classical and
  Quantum Gravity, 35, 063001, \dodoi{10.1088/1361-6382/aaa7b4}

\bibitem[{{Schutz}(1986)}]{schutz86}
{Schutz}, B.~F. 1986, \nat, 323, 310, \dodoi{10.1038/323310a0}

\bibitem[{{Semboloni} {et~al.}(2011){Semboloni}, {Hoekstra}, {Schaye}, {van
  Daalen}, \& {McCarthy}}]{semboloni11}
{Semboloni}, E., {Hoekstra}, H., {Schaye}, J., {van Daalen}, M.~P., \&
  {McCarthy}, I.~G. 2011, \mnras, 417, 2020,
  \dodoi{10.1111/j.1365-2966.2011.19385.x}

\bibitem[{{Seto} {et~al.}(2001){Seto}, {Kawamura}, \& {Nakamura}}]{seto01}
{Seto}, N., {Kawamura}, S., \& {Nakamura}, T. 2001, \prl, 87, 221103,
  \dodoi{10.1103/PhysRevLett.87.221103}

\bibitem[{{Sheth} \& {Tormen}(1999)}]{sheth99}
{Sheth}, R.~K., \& {Tormen}, G. 1999, \mnras, 308, 119,
  \dodoi{10.1046/j.1365-8711.1999.02692.x}

\bibitem[{{Smith} {et~al.}(2014){Smith}, {Bacon}, {Nichol}, {Campbell},
  {Clarkson}, {Maartens}, {D'Andrea}, {Bassett}, {Cinabro}, {Finley},
  {Frieman}, {Galbany}, {Garnavich}, {Olmstead}, {Schneider}, {Shapiro}, \&
  {Sollerman}}]{smith14}
{Smith}, M., {Bacon}, D.~J., {Nichol}, R.~C., {et~al.} 2014, \apj, 780, 24,
  \dodoi{10.1088/0004-637X/780/1/24}

\bibitem[{{Smith} {et~al.}(2003){Smith}, {Peacock}, {Jenkins}, {White},
  {Frenk}, {Pearce}, {Thomas}, {Efstathiou}, \& {Couchman}}]{smith03}
{Smith}, R.~E., {Peacock}, J.~A., {Jenkins}, A., {et~al.} 2003, \mnras, 341,
  1311, \dodoi{10.1046/j.1365-8711.2003.06503.x}

\bibitem[{{Springel} {et~al.}(2018){Springel}, {Pakmor}, {Pillepich},
  {Weinberger}, {Nelson}, {Hernquist}, {Vogelsberger}, {Genel}, {Torrey},
  {Marinacci}, \& {Naiman}}]{springel18}
{Springel}, V., {Pakmor}, R., {Pillepich}, A., {et~al.} 2018, \mnras, 475, 676,
  \dodoi{10.1093/mnras/stx3304}

\bibitem[{{Takahashi}(2006)}]{takahashi06}
{Takahashi}, R. 2006, \apj, 644, 80, \dodoi{10.1086/503323}

\bibitem[{{Takahashi} {et~al.}(2011){Takahashi}, {Oguri}, {Sato}, \&
  {Hamana}}]{takahashi11}
{Takahashi}, R., {Oguri}, M., {Sato}, M., \& {Hamana}, T. 2011, \apj, 742, 15,
  \dodoi{10.1088/0004-637X/742/1/15}

\bibitem[{{Takahashi} {et~al.}(2012){Takahashi}, {Sato}, {Nishimichi},
  {Taruya}, \& {Oguri}}]{takahashi12}
{Takahashi}, R., {Sato}, M., {Nishimichi}, T., {Taruya}, A., \& {Oguri}, M.
  2012, \apj, 761, 152, \dodoi{10.1088/0004-637X/761/2/152}

\bibitem[{{Takahashi} {et~al.}(2005){Takahashi}, {Suyama}, \&
  {Michikoshi}}]{takahashi05}
{Takahashi}, R., {Suyama}, T., \& {Michikoshi}, S. 2005, \aap, 438, L5,
  \dodoi{10.1051/0004-6361:200500140}

\bibitem[{{van den Bosch} {et~al.}(2005){van den Bosch}, {Tormen}, \&
  {Giocoli}}]{vandenbosch05}
{van den Bosch}, F.~C., {Tormen}, G., \& {Giocoli}, C. 2005, \mnras, 359, 1029,
  \dodoi{10.1111/j.1365-2966.2005.08964.x}

\bibitem[{{Van Tilburg} {et~al.}(2018){Van Tilburg}, {Taki}, \&
  {Weiner}}]{vantilburg18}
{Van Tilburg}, K., {Taki}, A.-M., \& {Weiner}, N. 2018, \jcap, 2018, 041,
  \dodoi{10.1088/1475-7516/2018/07/041}

\bibitem[{{Vegetti} {et~al.}(2012){Vegetti}, {Lagattuta}, {McKean}, {Auger},
  {Fassnacht}, \& {Koopmans}}]{vegetti12}
{Vegetti}, S., {Lagattuta}, D.~J., {McKean}, J.~P., {et~al.} 2012, \nat, 481,
  341, \dodoi{10.1038/nature10669}

\bibitem[{{White}(2004)}]{white04}
{White}, M. 2004, Astroparticle Physics, 22, 211,
  \dodoi{10.1016/j.astropartphys.2004.06.001}

\bibitem[{{Zanisi} {et~al.}(2020){Zanisi}, {Shankar}, {Lapi}, {Menci},
  {Bernardi}, {Duckworth}, {Huertas-Company}, {Grylls}, \&
  {Salucci}}]{zanisi20}
{Zanisi}, L., {Shankar}, F., {Lapi}, A., {et~al.} 2020, \mnras, 492, 1671,
  \dodoi{10.1093/mnras/stz3516}

\bibitem[{{Zhan} \& {Knox}(2004)}]{zhan04}
{Zhan}, H., \& {Knox}, L. 2004, \apjl, 616, L75, \dodoi{10.1086/426712}

\bibitem[{{Zumalac{\'a}rregui} \& {Seljak}(2018)}]{zumalacarregui18}
{Zumalac{\'a}rregui}, M., \& {Seljak}, U. 2018, \prl, 121, 141101,
  \dodoi{10.1103/PhysRevLett.121.141101}

\end{thebibliography}

\end{document}